\documentclass [12pt]{article}
\usepackage{latexsym}
\usepackage{amsfonts}
\usepackage{amsmath}
\usepackage{acronym}
\newcommand{\be}{\begin{equation}}
\newcommand{\ee}{\end{equation}}
\newcommand{\bea}{\begin{eqnarray}}
\newcommand{\eea}{\end{eqnarray}}

\makeatletter
\@addtoreset{equation}{section}
\def\theequation{\thesection.\arabic{equation}}
 \makeatother
\begin{document}
\normalsize
\title{Topological Theory of Helium 4.}
\author
{{\bf L.~D.~Lantsman}\\  
Tel.  049-0381-799-07-24,\\
llantsman@freenet.de}
\maketitle
\begin {abstract} In this paper we attempt to construct the topological theory of superfluid helium $4$ in the framework of the (rigid) $U(1)$ model in which the initial $U(1)$ group is destroyed with appearance of (topologically nontrivial) domains separated by domain walls treated as step voltages between domains (e.g. with neighboring topological numbers). This can explain the superfluid properties in a  helium $4$ specimen as well as the appearance of topologically nontrivial vortices therein. 
\end{abstract}
\noindent PACS: 74.00 12.38.AW 14.80.Hv  11.15.-q  \newline
Keywords: Superfluidity, vortices, superconductivity, QCD, monopoles, Yang-
Mills.
\newpage
 \tableofcontents
\newpage
\section{Introduction. About Topological Defects.}
 The analysis of superfluid properties of a liquid helium led L. D. Landau, in 1941-1948 y. y.,   to the creation
 of the \it two-component  superfluid liquid  phenomenological model \rm involving the potential superfluid liquid and quasiparticle excitations belonging to the normal component in the given liquid helium specimen. 

Bogolubov et al \cite{N.N.} created the quantum version of the superfluidity theory, found giving an exhaustive explanation of the liquid helium phenomenology by Landau \cite{Landau}, including manifest superfluid properties of liquid helium at extremely low temperatures. 

With the example of the model Hamiltonian  (see e.g. \S 5 of  Part 6, Chapter 1 in   \cite{Levich} or the paper \cite{Nels}) they showed how the  quantum theory can result  the energy spectrum near to experimental data about liquids possessing superfluidity. 

That  theory was based on  extracting  the c-number Bose condensate  (the quantum analogue of the superfluid component \cite{Landau})  and Bogolubov canonical transformation \cite{N.N.,Levich,Smir} allowing to calculate the  quasiparticle spectrum by  diagonalizing   the model Hamiltonian (claiming herewith the stability of the considered theory). \par
On the other hand, to complete the phenomenological description of liquid helium, one would take account of the same more kind of collective excitations besides phonons and rotons.
There are so-called {\it quantum vortices} \cite{Halatnikov}. How we do argue in the present study, these are a direct manifestation of non-trivial topological properties naturally inherent in the liquid helium $4$ model.  \par
Although these excitations posses a little statistical weight (as we shall demonstrate below \cite{Halatnikov,Landau52}, this statistical weight is suppressed by the Planck constant $\hbar$), they play an essential role in hydrodynamics of liquid helium. \par
The simplest case of such vortices are rectilinear (infinitely thin) vortices, interspersing with the superfluid as well as the normal components in a liquid  helium II specimen \cite{Landau52}.  They possess the natural topology of a cylinder and thus can serve as patterns of (quantum) rotary motions occurred in a liquid  helium, even when it is rested.

Thus our goal in the present article is to reveal the topological content of such nontrivial dynamics, indeed proving to be inherent in a  helium model  \cite{N.N., Halatnikov, Landau52}.\par \bigskip
In the present work we propose a fundamentally new (accordind to the author little or no investigated today) way violating the initial  $U(1)\cong S^1$ (global) symmetry in such a wise that the circle $S^1$ breaks up into its topological domains (which is related directly to the appropriate fundamental group $\pi_1 S^1 =n$, $n\in {\bf Z}$) separated by domain walls.

And now a quite reasonable question arises. Which is the physical nature of such domain walls? As it is well known ( (see e.g. \S 7.2 in \cite{Linde} or the paper \cite{Ph.tr}), the width of a domain (or \it Bloch\rm, in the terminology \cite{Ph.tr}) wall is roughly proportional to the inverse of the lowest mass of all the physical particles in the (gauge) model considered. For instance, in the Minkowskian Higgs-Yang-Mills (HYM) models,  the typical such scale is the (effective) Higgs mass $m/\sqrt\lambda$ (with $m$ and $\lambda$ being the Higgs mass and self-interaction constant, respectively). This is so since Yang-Mills field is massless. \par
Such a domain walls structure must give its nontrivial
contribution in the appropriate model Lagrangian (Hamiltonian). It's an obvious thing!  In magnetism, for instance, we have examples of such domain walls as   transitions between different magnetic moments. More precisely, a domain wall is a gradual reorientation of individual moments across a finite distance. {\it The energy of a domain wall herewith is simply the difference between the magnetic moments before and after the domain wall was created} (with  correct accounting of the physical units).

A simple example of such domain walls structure is a {\it Bloch wall}. It is a narrow transition region at the boundary between magnetic domains, over which the magnetization changes from its value in one domain to that in the next. The magnetization rotates through the plane of the domain wall. Bloch domain walls appear in bulk materials, i.e. when sizes of magnetic material are considerably larger than domain wall width.

Alike mechanism transition between magnetic domains was suggested by L. Landau and E. Lifshitz  in wich the magnetization vector $M$ turns in the domain wall plane changing its direction to the opposite.
Thus an important point here is just a {\it nontrivial contribution} into the Lagrangian (Hamiltonian) of the magnetization theory.

\medskip The next example of domain walls tells us the notion of {\it step voltage}. It is, actually, the difference of potentials $\phi_1(r)-\phi_2(r)$.  If these potentials possess two "neighboring" topologies: say, $\phi_1 (n_1,r)$  and  $\phi_2 (n_1+1,r)$, then such difference is $\phi_2 (n_1+1,r)-\phi_1 (n_1,r)$. It is just the {\it potential energy} associated with the domain wall between two domains "neighboring on the topology"  if {\it we identify  surfaces swept by these two potentials with such topological domain}.

When it comes to the YMH theory, this very interesting conception was analyzed recently for the so-called {\it Dirac quantization scheme} \cite{Dir} designed to description of the \cite{BPS} BPS monopole "superfluidity" phenomena \cite {LP1,LP2, rem1, rem2} as well as topologically nontrivial dynamics  \cite {rem3} inside the YMH vacuum. Unfortunately, the negative answer was gotten about the possibility of such domain wall in the mentioned model. Now we want to tell our readers this story and explain why such negative answer takes the place.

The base of the model  \cite {LP1,LP2, rem1,rem2, rem3} is resolving the {\it Gauss law constraint}
\be \label {Gau} \partial W/\partial A_0=0\ee
(with $W$ being the action functional of the considered gauge theory). Solving Eq. (\ref{Gau}) one expresses  temporal components $A_0$ of gauge
fields $A$ through their  spatial components; by  that the nondynamical components $A_0$ \footnote{The (pure) YM Lagrangian can be always recast to the shape (cf. \cite{Gitman})
$$  L= \frac 1 4(\dot A_i+ D_i A_0)^2- \frac 1 4 F_{ik}^2,$$
with $D_i$ being the YM covariant derivative; hence the canonical momentum $\partial L /\partial \dot A_0=0$. As a consequence, also the temporal components $A_0$ of the YM fields become nondynamical in the YM Hamiltonian $H$.
} are indeed ruled out from the appropriate Hamiltonians. 

In the (topologically trivial) BPS monopole background $\Phi^{(0)}$, the Gauss law constraint (\ref{Gau}) comes \cite{disc} to the homogeneous PDE \be \label{homo}   [D^2_i(\Phi ^{(0)})]^{ac} A_{0c}=0        \ee (with latin indices $a,b,c=1,2,3$ referring to the $SU(2)$ gauge group) 
permitting, in the Minkowskian Higgs model with vacuum BPS monopoles quantized by Dirac \cite{Dir}, the family of zero mode   solutions \cite{Pervush1, Pervush2}, \be \label{zero}  A_0^c(t,{\bf x})= {\dot N}(t) \Phi_{(0)}^c ({\bf x})\equiv Z^c,
   \ee
implicating the topological variable $\dot N(t)$ and Higgs (topologically trivial) vacuum Higgs BPS monopole modes $\Phi_0^a ({\bf x})$.

In turn, the topological variables $\dot N(t)$ (respectively, $N(t)$), may be specified \cite{LP1,LP2, rem3,Pervush2} via the  relation 
\bea
\label{winding num.}
\nu[A_0,\Phi^{(0)}]&=&\frac{g^2}{16\pi^2}\int\limits_{t_{\rm in} }^{t_{\rm out} }
dt  \int d^3x F^a_{\mu\nu} \widetilde{F}^{a\mu \nu}=\frac{\alpha_s}{4\pi}  \int d^3x F^a_{i0}B_i^a(\Phi^{(0)})[N(t_{\rm out}) -N(t_{\rm in})]\nonumber \\
 &&  =N(t_{\rm out}) -N(t_{\rm in})= \int\limits_{t_{\rm in} }^{t_{\rm out} } dt \dot N(t), 
 \eea 
taking account of the natural duality between the  tensors $ F^a_{i0}$ and $ F^a_{ij}$ in the $SU(2)$ gauge model and the normalization \cite{Pervush2} $$
 \frac{g^2}{16 \pi^2}\int d^3 x D_i^{ab}({\Phi})\Phi^b_0 B_i^a({\Phi})=1; \quad \alpha_s\equiv g^2/4\pi.
 $$ As an useful information, Most likely, the topological variable $N(t)$ arose originally in the work \cite{Niemi} by L. D. Faddeev and A. J. Niemi. 

In turn, this breeds a nontrivial (vacuum) topological dynamics \cite{rem3} in the YMH theory with BPS monopole backgrounds (in the both YM and Higgs sectors) quantized by Dirac \cite{Dir}. This comes actually to the specific {\it Josephson effect} \cite{Josephson, Pervush3} \footnote{This Josephson effect, coming to the non-stop topological rotations inside the YMH Josephson, is valid at the absolute zero  temperature $T=0$ of environment. At $T\neq 0$, a "friction" arises between the YMH and this  environment. This allows, as it is easy to see, to multiply the topological (winding) variable $N(t)$, (\ref{winding num.}), onto the $\cos (T/k)$ factor (with $k$ being the Boltzmann constant): $N(t)\rightarrow N'(t)= N(t)\cos (T/k)$. This is consistent logically with the periodical temperature behavior of bosonic fields of the  $\phi(0)=\phi (T)$ in a state of thermodynamic equilibrium at the temperature $T$.  This implies the same quantum (probabilistic) description in the given bosonic theory at the absolute zero as well as at the fixed temperature $T$.  Expanding $\cos(k T)$ into the Maclaurin series, we have
$$ \cos( T/k)=1-\frac{(T/k)^2}{2!} \pm \dots$$
This expansion creates the basis for the perturbation YMH theory by the degrees of the temperature $T$: in an infinitesimal neighborhood of the absolute zero as well as at an arbitrary large temperature 
 $T$.}.

As its is well known (see e.g. the monograph \cite{Al.S.}),  such nontrivial topological dynamics inside the YMH vacuum manifold is a manifestation of {\it thread topological deffects} inside this manifold. From the thermodynamics point of view,  this comes down to violating the thermodynamics equilibrium over a smooth curve or over an (infinitesimal) small neighbourhood of this curve (see \S $\Phi1$ in \cite{Al.S.}; Fig. 5 ibid). 

This immediately rises the question: what kind should the YMH vacuum manifold look like in order to justify the nontrivial topological dynamics \cite{rem3}
inside it. The answer has been given in the papers \cite{disc, fund}. It turns out that this must take the shape
\be \label{RYM} R_{\rm YM}  \equiv G/H\equiv SU(2)/ U(1)\simeq G_0/[U_0 \otimes {\bf Z}],\ee
with the appropriate first fundamental groups
\be \label{fgr1} \pi_1(U_0)= \pi_1(G_0)=0  \ee
while
$$ SU(2)\equiv G_0; \quad U(1)\equiv U.$$ As always in a gauge theory, $G$ is the complete gauge group and $H$ is the remaining after the (spontaneous) symmetry breaking.

The most important point  here is including the {\it discrete} factor $\bf Z$, the group of integers. This idea originated actually from the work \cite{Pervush1} where it was indicated onto the natural isomorphism (actually, between the homotopical classes of loops of the circle $S^1$ and the  group $\bf Z$ of integers; we recomend our readers the monograph \cite{Postn4}, see Lecture 3 therein, for the proof of this isomorphism) 
\be \label{top1} U(1)\cong S^1\cong {\bf Z}\ee 

Then it can be checked \cite{disc} that \be
 \label{fact1}
U(1) \simeq U_0 \otimes {\bf Z},
 \ee
On the other hand, there is a natural isomorphism $$ SU(2)\cong S^3,$$
and since $\pi_1 S^3=0$  (see Lecture 26 in \cite{Postn4}), {\it $SU(2)$ is a one-connected group}: $$ \pi_1 [SU(2)]=0.$$  

Let us explain 
Eq. (\ref{fact1} ). This isomorphism for the Abelian group $U(1)$ follows from the (manifestly unitary) representation of  an element of $U(1)$ as $\exp(in\theta)=\exp(i0\theta)\exp(in\theta)$ [$\exp(i0\theta)=1$].  This gives sketch to the proof of (\ref{fact1}); however, the complete proof should use the "loop framework" from the cited Lecture 3 by \cite{Postn4}.  More precisely, if $U_0$ consists of trivial loops, i.e. those loops $u$ which are homotopical to the constant loop $e_{p_0}$  ($p_0$ is the center of the circle $S^1$) \footnote{Formally, it is the map $e: [0, 1] \rightarrow X$ such that $e(t) = x_0$ for all $t \in [0, 1]$.}, then any loop homotopical class comes to the 
\be \label{cll} [u]^n =[e_{p_0}]\circ\tau^n,\ee 
with $\tau$ being the homotopical class of loops which just once uniformly running around the circle $S^1$ and herewith counterclockwise;  $n\in {\bf Z}$.

Having proved Eq.  (\ref{fact1}),  Eq. (\ref{RYM}) follows automatically taking into account our reasoning about the initial gauge symmetry $SU(2)$. 

\medskip The representations (\ref{fact1}) for the group $U(1)$  allow to divide the appropriate gauge transformations into the "small", topologically trivial, $U_0$, and topologically nontrivial, with $n\neq 0$. Such a classification of gauge transformations most likely was given first by L. D. Faddeev and R. Jackiw \cite{Fadd2}.

\medskip
Now we can analyze the properties of the vacuum manifold $R_{\rm YM}$. First of all, following \cite{disc}, we represent the complete gauge group $G$ as
\be 
\label{sxema1} 
G=H \oplus G/H \equiv H \oplus R_{vac},\ee
where $ R_{vac}$ is the appropriate {\it degeneration space or  vacuum manifold} (see e.g. Eq. (10.69) in \cite{Ryder}).

 Following this framework, we substitute in (\ref{sxema1}) the "discrete" representations (\ref{fact1}) for the appropriate gauge symmetries groups:
\be
 \label{trick}
  G_0= {\bf Z}\otimes U_0 +  G_0/(U_0\otimes {\bf Z}).
 \ee  
This just proves Eq. (\ref{RYM}). 

Eq. (\ref{fact1}) implies \cite {Al.S.} 
\be\label{dl1} \pi_0 [U_0 \otimes {\bf Z}]=\pi_0 ({\bf Z})= {\bf Z}\ee 
since, in definition,
\be
 \label{oneconnect} 
 \pi_0 G_0 = \pi_0 U_0 =0
\ee  and $\bf Z$ is a discrete set.

In other words, $G_0=SU(2)$ and $U_0$ are  (maximal) connected components \rm \cite {Al.S.} in their gauge groups (respectively, $G$ and $U$).

There is an enough simple explanation of 
Eq.  (\ref{dl1}). When considering $ U_0 \otimes \bf{Z}$
, each element $ n \in \bf{Z} $ can be thought of as indexing a copy of $ U(1)
$ (in the (\ref{cll}) wise). This means that for each fixed 
$n$, there is a corresponding such copy of  $U(1)$  in the space 
$U_0 \otimes \bf{Z}$.  Further, each copy of $ U(1) $ at a fixed $ n $ represents a distinct homotopy class of loops in the $ U(1) $ group space (see Lecture 3 in \cite{Postn4}). Specifically, the loop that winds $n$ times around $ U(1) $ corresponds to the element $ n  \in  \bf{Z}$. Therefore, the “copies” at fixed $n$ are directly related to the homotopy classes of loops in $ U(1)$.

 As a consequence, also $$\pi_0 (G_0/U_0)=0$$ for the "small"
 coset $G_0/U_0$. 

But, on the other hand,  it turns out that  

\be
 \label{domenano}
\pi_0 (R_{YM})=0,
\ee  as it was argued in \cite{disc} with the aid of the manifest exact sequence consideration for the fibration 
\be \label {fibr}  U_0\otimes {\bf Z} \longmapsto SU(2) \longmapsto  SU(2)/(U_0\otimes {\bf Z}).
\ee   We will not write here this  exact sequence and recommend our readers the original paper \cite{disc} for this purpose.


\medskip Also it is important to know the second homotopy group $\pi_2 (R_{\rm YM})$. But this is easy to derive from the "primordial" natural isomorphism  (in this context, we forget for a moment about the "factorization"  (\ref{fact1}) of the gauge group $U(1)$; this is possible due to the natural isomorphism  (\ref{fact1})).
\be \label{contin} R_{\rm YM}\equiv SU(2)/U(1)\simeq S^2.     \ee 
But $\pi_2 S^2={\bf Z}$. Thus 
\be \label{pi2R}
\pi_2 ( R_{\rm YM})={\bf Z}.
\ee
Having learned  the homotopical groups for the vacuum manifold $R_{\rm YM}$, namely $\pi_0 ( R_{\rm YM})$, $\pi_1 ( R_{\rm YM})$  and $\pi_2 ( R_{\rm YM})$, we can trace the following "homotopical" relations:
Furthermore, it   becomes evident from the stated above  that
\be \label {thr}
\pi_1(R_{YM})= {\bf Z}= \pi_0 H\equiv \pi_0 (U_0\otimes {\bf Z})\ee
and
\be \label {pointd}
\pi_2(R_{YM})= {\bf Z}= \pi_1 H;\ee
\be \label{domno} \pi_2(R_{YM})=0. \ee
As it is well known (see e.g. the monograph \cite{Al.S.}), the first relations implies just the existence of thread topological defects (what is it we explained above). The second of these relations implies the existence of {\it point topological defects} \cite{Al.S.} \footnote{We will justify the relation $ \pi_1(U_0\otimes {\bf Z})={\bf Z}$ during the discussion in the present study.}. From the thermodynamics point of view, the existence of point topological defects is equivalent to violating the thermodynamical equilibrium in a one (singular) point of the degeneration space in the model discussed. 

More precisely, let us consider the set $U_1\in R^{k+1}$ ($k\ge 2$). The Minkowski space, inherent in the Minkowskian YAng-Mills-Higgs model with vacuum BPS monopoles quantized by Dirac,  is isomorphic to
$R^4$. There exist always two spheres $S_1^k$ and $S_2^k$ such that $U_1$ is located between these spheres.  Let $R$ be the investigated degeneration space and $F$ be the map $U_1\to R$ coinciding with the maps $f_1: S_1^k\to R$ and $f_2: S_1^k\to R$ on the  spheres $S_1^k$ and $S_2^k$, respectively. 

Moreover, let us consider the coordinate region $U$ (for instance, in the Minkowski space) over which the local thermodynamic  equilibrium is set. Then the map  $f:U\to R$ can be constructed. We are now interested in the case $k=2$; now we attempt to argue that namely this case controls the appearance of point topological defects inside the degeneration space. 

Let (in the  case $k=2$) $S^2\in U$; in this case we shall denote as $f$ also the restriction of the map $f:U\to R$ onto the sphere $ S^2$. If the map $f:S^2\to R$ is not homotopical to zero ($\pi_2 (R)\neq 0$), this map cannot be continue  onto the map $D^3\to R$, with $D^3$ being the ball restricted by the sphere $ S^2$. The said just means that  there is a {\it point defect} inside  the sphere $ S^2$\rm. The existence of such a defect comes to gaps (break points) in  definite order parameters characterized the studied local thermodynamic  equilibrium\rm. \par \medskip
More exactly, as $T<T_c$, where $T_c$ is the Curie point in which the second-order phase transition (accompanied in gauge theories by spontaneous breakdowns of  initial symmetries) occurs, the vacuum manifold (degeneration space) $R$ becomes (topologically) nontrivial and a local thermodynamic  equilibrium is set. But in   topologically nontrivial theories (gauge theories are patterns of them) always there  exist topological defects, involving  violating  local thermodynamic  equilibriums and  gaps in  order parameters.

\bigskip Point topological defects, the (geometrical and thermodynamical) origin of which was just clarified,  take the shape of {\it hedgehog topological defects} in abelian and non-abelian gauge models. Let us consider the mechanism of formation the hedgehog topological defects with the standard example of the Minkowskian YM theory involving the Higgs ($\phi$) and YM (gauge) modes. The action of such model has the look   \be \label{YM L}
S=-\frac {1}{4 g^2} \int d^4x F_{\mu \nu}^b F_b^{\mu \nu }+ \frac {1}{2} \int d^4x (D_\mu\phi,D^\mu\phi
) -\frac {\lambda}{4} \int d^4x \left[(\phi^b)^2- \frac{m^2}{\lambda}\right]^2,
\ee
with 
$$D_\mu\phi=\partial^\mu\phi+g[A^{\mu },\phi]$$ 
being the covariant derivative an $g$ being the YM coupling constant.\par
We suppose that   initial data of all the  fields (YM as well as Higgs ones) are given to within stationary
gauge transformations; the  manifold of these transformations has  a nontrivial structure of   three-dimensional paths in the group space of
the (initial) non-Abelian $SU(2)$ gauge group: 
\be
\label{3-path} 
\pi_3 (SU(2)) =\bf Z,   \ee 
with  $\bf Z$  being the group of integers: $n=0,\pm 1,.\pm 2,\dots$ \par 
In the case of  the $SU(2)$ (YMH)  gauge theory, YM fields $A^{\mu a}$ and Higgs ones, $\phi^a$, take
their values in the adjointttttt representation of the Lie algebra  of the  $SU(2)$ group.\par 
To get the converging action integral corresponding to  finite values of energy\rm, one should claim for the Higgs field $\phi ({\bf r})$ to be finite as ${\bf r} \to \infty$ in 
the  BPS limit \rm  $\lambda\to 0$ (as $m\to 0$) \cite{BPS,Al.S.}. 

The said  implies that $\phi^a$ should approach the minimum of the  potential $V\equiv \frac {\lambda}{4} (\frac {m^2}{\lambda}-\phi ^2)^2$: 
\be \label{Higs.as}
\phi^{a \infty} ({\bf n})\in M_0, \quad {\bf n} =\frac {\bf r}{r}, 
\ee
where $M_0$ is the  manifold of the minimum of the potential $V$ ({\it the vacuum
manifold}, in the classical sense): 
\be 
\label{min}
 M_0=\{\vert \vec\phi\vert= a;\quad a^2=m^2/ \lambda\} 
\ee 
as ${\bf r} \to \infty$. 

Thus $M_0$ consists of the points of the sphere $S^2$ in the three-dimensional $SU(2)$ group space. \par 
Indeed, the presence of Higgs  (would be)Goldstone \rm modes implies that the initial $SU(2)$ gauge symmetry inherent in the Minkowskian YM model is then spontaneously violated down to its $U(1)$ subgroup (via the Higgs mechanism  of the  $SU(2)$ gauge symmetry breakdown: cf. p. p. 243- 244 in  \cite{Cheng}). \par 

Since the vacuum manifold $M$ is invariant with respect to  the gauge
transformations of $U(1) \equiv H$ (in notation (\ref{RYM})) \footnote{Just this fact invariance of $M$ with respect to  the gauge
transformations $H$ may serve as the definition of vacuum manifolds.

A good explanation of this fact was given in the monograph \cite {Ryder}, in \S 8.1.}, it can be subdivided into the $H$-orbits of its points: $$M=\sum \limits_i H M_i$$
 ($M_i\neq M_j$ as $i\neq j$).  

So long as $H\subset  SU(2)\equiv G$,  the vacuum manifold $M$ is given as  the set of those transformations of  $G$  that do not  belong to $H$\rm. It follows from the invariance of $M$ with respect to $H$ that $ H M_i=M_i$ for all $i$: in other words, that $H$  is the \it stationary subgroup \rm of the point $M_i\in M$.  Thus (see \S 10.4 in \cite {Ryder}) 
 \be 
\label{sxema} 
M=G/H= \sum \limits_i H M_i, \ee 
and
\be \label{sxema'} 
G=H \oplus G/H\ee
It is just (\ref{sxema1})) drawn in more detail.

\bigskip In particular, 
 $$ SU(2)/U(1)=M_0\simeq S^2$$ as follows from (\ref{min}). Such a sketch is correct for each spontaneous breakdown of initial  symmetries in gauge theories. \par  \bigskip
 Since $\vert \vec \phi \vert \neq 0$ on $M_0\simeq S^2$, this manifold is topologically equivalent to ${\bf R}^3\setminus \{0\}$.
 On the other hand, the alone two-sphere $ S^2$  is topologically equivalent to  ${\bf R}^3\setminus \{0\}$ (see e.g. \S T1 in \cite{Al.S.}). \par
 One can show herewith 
(see \S T1 in \cite{Al.S.})  that
 \be  
\label{top4}
  \pi_2  ({\bf R}^3\setminus \{0\})= \pi_2  S^2=n, \quad n\in {\bf Z}. 
\ee 
Utilizing  latter Eq., one can associate an integer to a point topological defect.  This integer characterizes the topological type of the investigated defect.\par \medskip
The vacuum manifold $M_0=R\simeq S^2$ is connected and \it one-connected \rm ($\pi_1 S^2=0$). In such cases the homotopical group $\pi_2 (R)$ is Abelian (see \S T12 in \cite{Al.S.}).

equiv\bigskip One of most interesting point topological defects are those located at the origin of coordinates.  
(Minkowskian) Higgs theories with monopoles: for instance, \cite{Linde, H-mon,Polyakov, Wu}, are sources of such defects, point hedgehog topological defects. Point hedgehog topological defects come actually to the divergence of "magnetic" tensions and YM potentials at the origin of coordinates \cite{rem1}.  
E.g. the theory of t'Hooft-Polyakov monopoles \cite{Linde,Ryder,Cheng, H-mon,Polyakov} results YM potentials 
\be 
\label{Polyakov p}
A_i^a= \epsilon _{iab} \frac{r^b}{gr^2}. \ee
 Thus they diverge as $1/r$ at the origin of coordinates. \par
Higgs fields  are specified in this case by Eq.
\be 
\label{hedg} 
\phi^a \sim \frac {x^a}{r}f(r,a)
\ee
as $ {\bf r} \to \infty$;  $f(r,a)$ is a continuous function that  does
not change the natural topology \be
\label{top3}
\pi_2 S^2= \pi_3 (SU(2))=\pi_1(U(1))=\pi_1~S^1=\bf Z, 
\ee 
To reconcile the action (\ref{YM L}) with the anzatz (\ref{hedg}), it is necessary to set  $F \equiv m/2\sqrt{\lambda}$ instead of $f(a,r)$ in Eq. (\ref{hedg}).

\medskip t'Hooft-Polyakov monopoles \cite{H-mon,Polyakov} arise at the spontaneous 
breakdown of the initial $G\equiv SO(3)\simeq SU(2)$ gauge symmetry in the Georgi-Glashow model up to its $H\equiv U(1)$ subgroup.The natural isomorphism  $ SO(3)\simeq SU(2)$ involves then  \cite{rem1} $$R=G/H=S^2$$
 for the degeneration space $R$ and the topological chain 
(\ref{top3})  determining the existence of point hedgehog topological defects in the t'Hooft-Polyakov model.

It turns out that Higgs fields   converge at the origin of coordinates: they are directly proportional to the unit vector $\bf n$. \par \medskip

Let us set, following \cite{Al.S., Ryder, Cheng, H-mon, Polyakov}, 
\be  \label{Polyakov tens} 
F^{\mu\nu}= \frac{1}{a} \phi^a F_a^ {\mu\nu} - \frac{1}{ga^3} ~\epsilon _{abc} \phi^a (D^\mu \phi^b) (D^\nu \phi^c)  \ee for the YM "magnetic" strength $F^{\mu\nu}$. \par
 If we  (following \S 10.4 in \cite{Ryder}) introduce the vector \be  \label{Am}
A^\mu= \frac{1}{a} \phi^a A^\mu _a, 
\ee 
we can compute directly   \be 
\label{Polyakov tens1}
F^{\mu\nu}=\partial^\mu A^\nu -\partial^\nu A^\mu -  \frac{1}{ga^3} ~\epsilon _{abc} \phi^a (\partial^\mu \phi^b) (\partial^\nu \phi^c).   
\ee  Since $ A^\mu$ disappears at the spatial infinity, the "magnetic" tensor $ F^{\mu\nu}$ depends only on the Higgs field $\phi$ in this region. And moreover, in the gauge \cite{Al.S., Ryder} of the $\phi^2=\phi^3=0$ type the  tensor $ F^{\mu\nu}$ is reduced to the ordinary Maxwell tensor.

There are no "electric" fields in the  t'Hooft-Polyakov model  
\cite{Cheng, Ryder,H-mon, Polyakov}: \be \label{Veyl} F^{0i}=0,\ee
 at considering only stationary solutions, monopoles; also \be  
\label{radial pole}
 F^{ij}= - \frac{1}{gr^3}\epsilon^{ijk} r_k. 
\ee
  This corresponds to the \it radial \rm "magnetic" field \be
  \label{radial pole1}
 B_k=\frac{r_k}{gr^3}=- F^{ij}\epsilon_{ijk}.
\ee 
 Thus also the  radial  "magnetic" field $B_k$ diverges as $1/r^2$ at the origin of coordinates \cite{rem1}.

With  this example of t'Hooft-Polyakov monopoles we have demonstrated  that the order parameter in the  t'Hooft-Polyakov model \cite{Ryder, Cheng,H-mon, Polyakov} the radial  "magnetic" field $B_k$, possesses  the asymptote $r=0$\rm.  \par \medskip
In the Higgs sector of the t'Hooft-Polyakov model \cite{Ryder,Cheng, H-mon, Polyakov}
one can observe the following picture. The spatial asymptotic of Polyakov hedgehogs  \cite{Polyakov, Ryder} 
\be  \label{Higgs pole} \phi^a= F\frac{r^a}{r}\quad (r\to \infty); \quad F^2=m^2/4\lambda;
\ee
involves \cite{Ryder} \be  \label{Higgs der} D_i\phi^a\to 0
\ee
at the spatial infinity. \par

In turn, the latter fact implies the disappearance of the second, YMH, coupling item in the general Minkowskian YMH action functional (\ref{YM L}) for the concrete case of the t'Hooft-Polyakov model involving the "continuous" $\sim S^2$ vacuum geometry (\ref {min}), (\ref{contin}) \cite{rem1}.

Thus  there are no "electric"  fields in the t'Hooft-Polyakov  theory \cite{Ryder,Cheng, H-mon, Polyakov} quantized in the "heuristic" Faddeev-Popov (FP)  wise \cite{Fadd1,FP1}, and the role of Higgs fields, Polyakov hedgehogs (\ref{Higgs pole}), in the general Minkowskian  action (\ref{YM L}) is reduced somewhat due to the disappearance of the YMH coupling \footnote{The FP "heuristic" quantization \cite{Fadd1,FP1} of the t'Hooft-Polyakov model \cite{Ryder,Cheng, H-mon, Polyakov} can be reduced \cite{rem1} to fixing the temporal (Weyl) gauge $A_0=0$ via the Dirac delta-function $\delta (A_0)$ entering the appropriate FP path integral. 

This just results $F_{0i}^a= 0$ at assuming stationary t'Hooft-Polyakov monopole solutions in the Minkowskian Higgs model \cite{Ryder,Cheng, H-mon, Polyakov} 

Fixing the temporal gauge $A_0=0$ in FP path integrals is the actual way to remove "electric" fields $F_{0i}^a= 0$ also in another Minkowskian models with monopoles: for instance, in the Wu-Yang model \cite{Wu} or in the BPS \cite{BPS, Al.S., Gold} one.}.

 \bigskip Another good patterns of hedgehog point topological defects give Wu-Yang monopoles \cite{LP1,
LP2,Pervush2,Wu, David2}, with the "magnetic" (vacuum) tension  given by Eq. \be  \label{sb}
B^{i a}(\Phi_i)= \frac {x^ax^i}{gr^4};
\ee
it  also diverges as $1/r^2$ at the origin of coordinates.
vacuum 

\bigskip YMH vacuum BPS monopole solutions \cite{BPS} utilized in the studies about the Dirac fundamental quantization (Gauss-shell Hamiltonian resolving) \cite{Dir,Pervush2}  the YMH model is a specific generalization of the Wu-Yang solutions \cite{Wu}. The BPS monopole solutions \cite{BPS} are obtained from the standard  YMH action functional (\ref{YM L}) in the  {\it Bogomol'nyi-Prasad-Sommerfeld} (BPS)  limit \cite{rem1}
\be 
\label{lim} 
\lambda\to 0,~~~~~~m\to 0:~~~~~~~~~~ ~~~~~\frac{1}{\epsilon}\equiv\frac{gm}{\sqrt{\lambda}}\not =0. 
\ee 
\medskip

The crucial feature of the BPS model \cite{BPS} is that  the (vacuum) YM and Higgs solutions satisfy the {\it Bogomol'nyi equation} \cite{BPS,LP1, LP2, Al.S., Gold}
\be
\label{Bog} 
{\bf B} =\pm D \Phi,
\ee
derived (see \S$\Phi$11 in \cite{Al.S.}) at evaluating the  {\it Bogomol'nyi bound} \cite{LP1,LP2, rem1}
\be 
\label{Emin}
E_{\rm min}= 4\pi {\bf m }\frac {a}{g},~~~~~~~~~~~~\,\, ~~~~~a=\frac{m}{\sqrt{\lambda}};
\ee 
with $\bf m$ denoting the magnetic charge, of the YMH field configuration energy.

\medskip As it is easy to see, the { Bogomol'nyi equation} (\ref{Bog}) is nothing but the {\it potentiality condition} for the vacuum of the "classical" Minkowskian YM- Higgs model with BPS monopoles. It generalized the  potentiality condition in vector analysis:
\be \label{potcon} {\rm rot}  ~{\rm grad} ~{ \Phi}=0 \ee
for a scalar field $\Phi$ (to within a constant). Thus any  potential field may be represented as ${\rm grad} ~{ \Phi}$ (to within a constant). \par
It is a good prompt for us. 
In the Minkowskian YMH theory involving BPS monopole solutions (for instance, in the "classical" Minkowskian Higgs model \cite{BPS,Al.S.,Gold} with  BPS monopoles), there exists always such a scalar field. It is just the Higgs  scalar $\Phi$ represented as the Higgs BPS monopole in the vacuum sector of that theory. \par
hen it is easy to guess  that the Bogomol'nyi equation (\ref{Bog}),  having the look (\ref{potcon}), an be treated  as  the  potentiality condition for the Minkowskian YMH vacuum involving vacuum BPS  monopole solutions. It is so due to the Bianchi identity $D B=0$ which can be represented as $$\epsilon ^{ijk}\nabla _i F_{jk}^b =0$$ (at neglecting the items in $DB$ directly proportional to $g$ and $g^2$).\par
\medskip
Indeed, there can be drawn a highly transparent parallel between the  Minkowskian YMH vacuum involving vacuum BPS  monopole solutions and a liquid helium II specimen described in the Bogolubov-Landau model \cite {N.N.}. And it will be one of the directions in our investigations in the present study.  In particular, in Section 1 we shall discuss in what the general potentiality condition (\ref{potcon}) turns into. We shall show that this general potentiality condition implies the potential superfluid motion inside a liquid helium II specimen. \par

 \bigskip  An important property of the YM (vacuum) BPS monopole solutions is that they approach the Wu-Yang monopole solutions  \cite{Wu}  at the spatial infinity, while at the origin of coordinates the solutions are different from each other due to the manifest presence of the YM BPS ansatz.  

This fusion of solutions at the spatial infinity, being applied to the Dirac fundamental (Gauss-shell) quantization \cite {Dir} of the YMH BPS monopole model, leads to very interesting results.

Following \cite{Pervush2}, it is possible to resolve the Gauss-law constraint \be \label{Gr.eq}
D^2(\Phi ^{(0)})^{ab}({\bf x})G_b^c ({\bf x},{\bf y})=
\delta^{ac} \delta^3({\bf x}-{\bf y}), \ee
for the Green function $G ({\bf x},{\bf y})$ (here $a,b\dots$ are the $SU(2)$ group indices) in the Wu-Yang (BPS) monopole background. It can be argued that  only the spatial infinity  limit ${\bf x}\to \infty$ is essential at the calculations about the Eq. (\ref{Gr.eq}). In this limit the both YM monopole solutions merge as stated above. 

In the presence of the Wu-Yang monopole background \cite {Wu} one can represent the (contravariant) derivative squared $ D^2(\Phi ^{(0)})^{ab}({\bf x})
$ as \cite{Pervush2}
\be
\label{Gr.eqnmon}
D^2(\Phi ^{(0)})^{ab}({\bf x})
= \delta^{ab}\Delta -\frac {n^a n^b+\delta^{ab}}{r^2}+2(\frac {n^a}{r}\partial ^b-\frac {n^b}{r}\partial ^a), \ee
with  $$n_a(x)=x_a/r; \quad r=\vert {\bf x}\vert.$$ 
Let us now decompose the Green function $G^{ab}$ of the Gauss  equation (\ref{Gr.eq}) by the complete set of  orthogonal vectors in the color space:
\be \label{complete set}
G^{ab}({\bf x},{\bf y})= n^a(x) n^b(y)V_0(z)+ \sum \sb {\alpha=1,2} 
e^a_ \alpha (x)e^{b\alpha}(y)V_1(z);\quad z=\vert {\bf x}-{\bf y }\vert.
\ee
Substituting the latter expression into  Eq.  (\ref{Gr.eq}), one gets the Euler equation \be \label{Euler} \frac {d^2}{dz} V_n+ \frac {2}{z}\frac {d}{dz}V_n- \frac {n}{z^2}V_n =0; \quad n=0,1.
\ee
The general solution to latter  Eq. is \be
\label{Vn}
V_n (\vert {\bf x}-{\bf y} \vert)=d_n\vert {\bf x}-{\bf y} \vert ^{l^n_1}+c_n\vert {\bf x} -{\bf y} \vert^{l^n_2} \quad (n=0,1), \ee
with $d_n,~c_n$ being constants while $l^n_1,~l^n_2$ may be found as the roots of the equation 
$$(l^{n})^2+l^n=n,$$ i.e.
\be \label{korni}
l^n_1= -\frac {1+\sqrt{1+4n}}{2};~~~~~\quad l^n_2=\frac {-1+\sqrt{1+4n}}{2}.
\ee 
It is easy to see that for $n = 0$, at setting $d_0=-1/4\pi$, one gets
 the \it Coulomb-type potential\rm: \be \label{Coulomb}
l^0_1= -\frac {1+\sqrt{1}}{2}=-1 ;\quad l^0_2=\frac {-1+\sqrt{1}}{2}=0, \ee
\be
\label{Cp}
V_0 (\vert {\bf x}- {\bf y} \vert) =  -(1/ 4\pi) \vert {\bf x}- {\bf y} \vert ^{-1} + c_0;
\ee
while for $n = 1$, one comes to
the so-called "\it golden section\rm" potential with \be
\label{ris}
l^1_1= -\frac {1+\sqrt{5}}{2}\approx -1.618;\quad l^1_2=\frac {-1+\sqrt{5}}{2}\approx 0.618;
\ee 
\be
\label{ris1}
V_1 (\vert {\bf x}-{\bf y} \vert)= -d_1\vert {\bf x}-{\bf y} \vert ^{-1.618}+c_1\vert {\bf x}-{\bf y} \vert^{0.618}.
\ee
The latter potential (unlike the Coulomb-type one, (\ref{Cp})) implies always
the rearrangement of the naive perturbations series and  spontaneous chiral symmetry breakdown.

More precisely, in QCD this leads to the constituent mass of the gluonic  in the Feynman diagrams: this mass changes
the asymptotic freedom  formula \cite{Gr} in the region  of  low transferred  momenta in  such a wise that the
coupling constant $\alpha _{\rm QCD}(q^2\sim 0)$ becomes finite in this region.

\bigskip Finally, in our discussion about BPS monopole solutions, it is worth noting the following peculiarity of such (vacuum) solutions: BPS ansatzes, involving hyperbolic functions, (in YM as well as in Higgs sectors of the model in question) play the role \cite{ff} of specific electric form-factors.

For instance, in the paper  \cite{ff} (repeating the arguments \cite{Cheng}: see Section 15 ibid), it was argued that the total momentum \be\label{tmom} {\bf J}={\bf L}+{\bf T}\ee
of a particle in a magnetic monopole background (with $\bf L$ being its spatial angular momentum, including its spin, and $\bf T$ being the generator of the internal, for instance the gauge $U(1)$, symmetry) can be brought to the look
\be\label{JBPStotal}
{\bf J}={\bf r}\times m{\dot{\bf r}}+f_1^{BPS}({\bf n}\cdot{\bf T}){\bf n}+{\bf T} (1-f_1^{BPS}).
\ee
Here $\bf n$ is the unit vector in the radius $\bf r$ direction. $f_1^{BPS}$ is the YM BPS ansatz \cite{BPS, Al.S.}.

At the particular choice $\bf n$ to be the $z$-direction, ${\bf n}\cdot{\bf T}=T_3$. Then $({\bf n}\cdot {\bf T}){\bf n}=T_3\bf n$ and the second item in (\ref{JBPStotal}) can be represented as \be \label{form} f_1^{BPS} T_3{\bf n}, \ee
that implies, as can be argued \cite{ff}, the replacement $e\leftrightarrow f_1^{BPS}e$ in Eq. (\ref{JBPStotal}). Indeed, there is   quite a long line of reasoning leading to this conclusion (see \cite{Cheng, ff}) which we omit in our study.  What matters is that a "spreading" of the (elementary) electric charge $e$ occurs at that replacement, i.e. just that $f_1^{BPS}$ {\it plays a role of an electric form-factor}.

\bigskip Ending this discussion about topological defects, note   that {\it Dirac monopole solutions} \cite {Ryder, Cheng, Dirac, abel} arising in the $U(1)$ gauge theory also a kind of point hedgehog topological defects since the only singularity at the origin of coordinates is a "physical" one. Concerning the "Dirac string"  singularity \cite{Ryder, Cheng} (i.e. a line along the negative $z$-axis joining the origin of coordinates to infinity), it is a purely gauge artifact.

\medskip We recommend the recent paper \cite{abel} (Section 3 ibid) for the study the question of Dirac monopoles in the $U(1)$ gauge model and the Dirac fundamental quantization of that model.

\bigskip The article is organized as follows. Section 2 is devoted to the two-component phenomenological liquid helium theory. We show that this is indeed a system with {\it repulsion} forces. For contrast, we (in Subsection 2.2) analyze the superconductivity theory, where {\it attraction} forces  (more exactly, the {\it positive} coupling constant $g$) provides the minimum of the appropriate energy $E$  \cite{Landau52}.

Also the the most important goal of Section  2 will be constructing the Bogolubov helium-4 Hamiltonian, possessing the manifest $U(1)$ {\it gauge} symmetry. Section 2.3 is devoted namely this theme. 

In Section  2.4 we analyze the phenomenology of a superfluid helium specimen contained in a
turning (say, around its axis $z$) vessel, reminding readers of the arguments stated due to this in the monograph \cite{Landau52}. as a result, {\it global} vortices appear in such specimen violating superfluidity along rectilinear lines of their location. Parallel, we investigate the case of a  helium specimen at rest where {\it local} rectilinear vortices of the same topological nature arise \cite{Halatnikov}.

The ground section of our study is Section 3, the topological theory of ${\rm He}^4$. Our proposal in this context is following. We suppose that the $U(1)\cong S^1$ gauge symmetry inherent  the ${\rm He}^4$ model is violated in the $$U(1)\to\tilde U\equiv \bigcup  _{n\in {\bf Z}} U_1^n $$  wise, involving the disjoint union $\bigcup  _{n} U_1^n$ of of the topological
sectors  $U_1^n$. Such way  violating the $U(1)$ gauge symmetry provides simultaneously the superfuidity phenomena corresponding to the zero topological sector $U_1^0$, while in nonzero topological sector $U_1^n$  ($n\neq 0$) inside $\tilde U$ we observe (rectilinear) vortices \cite{Halatnikov}.  It is just the sign of the {\it first-order} phase transition taking place in the model!

\medskip. As a discussion upon the ground part,  we analyze briefly the perspectives of superconductivity in the framework of
discrete  $\tilde U$ ˜U vacuum geometry developed in the present study. It is very tempting, due to the
striking similarity between the superfluidity and superconductivity phenomena to utilize
the same framework for describing superconductivity. However, there are differences
between the both models, which we shall discuss. In particular, we  elucidate the role of the external magnetic field $H$. It turns out that the second-order phase transition takes place in superconductors inside the interval \cite{Landau52} $H\in [H_{c1};H_{c2}]$ ($H_{c1}$,$H_{c2}$ are, respectively, the lower and upper critical fields). Moreover, in this interval, one can observe germs of the normal ($n$) phase inside the superconductor ($s$) phase. This situation refers to as the mixed state in Type II
superconductors. At $H\leq H_{c_1}$, a specimen is purely in the s-state, while it is purely in the n-state at $H\geq H_{c_2}$.

\medskip In type I superconductors,  there is {\it only one}  critical field, $H_c$. The first-order phase condition takes place in this case. We confirm our conclusions about the two types of superconductors with the aid of thermodynamical calculations.

\bigskip Finally, in Appendix, we discuss and ground the isomorphism $$ \pi_0 (H)\simeq \pi_1 (R)$$ (here $H$ is the residual symmetry group and $R$ is the vacuum manifold in the model in question). Just this isomorphism provides the existence of thread topological defects in this model.

\section {The two-component phenomenological liquid helium theory. Superfluidity and vortices.}
\subsection{Introductory remarks.}
At one time Lev Davidovich Landau \cite{Landau} asserted that point of view that a superfluid liquid is a system with strong local coherences. 

This  served as a  base for the phenomenological two-component liquid helium theory \linebreak \cite{Landau}  \footnote {We should like to remind our readers that the two-component liquid helium theory served as an explanation for the superfluidity phenomenon in the liquid helium II, discovered experimentally in 1935-1941 y.y. by Piotr  Leonidovich Kapitza \cite{Kapitza}}, in which  coexist (at a  very low   temperature 
$T\to 0$)  two independent kinds of motion.   
The first one is  the superfluid  motion, the velocity of which does not exceed a critical value $v_0$: $v_0= {\rm ~min}~(\epsilon/p)$, for the ratio of the  energy $\epsilon$ and momentum $p$ for quantum excitations
spectrum in the liquid helium II. \par  

On the other hand, at values of $\epsilon/p$ exceeding this critical value  the dissipation occurs of the liquid helium energy via arising excitation quanta (with momenta $\bf p $ directed antiparallel to the velocity vector $\bf v$  \cite{Levich1}). Such dissipation of the liquid helium energy becomes advantageous \cite{Halatnikov,Levich1} just at 
$$ \epsilon+ {\bf p~v}<0 \Longrightarrow \epsilon -p~v<0.$$
Note herewith that quantum excitations also 
are  treated as collective motions of particles, helium molecules. \par
In the phenomenological two-components theory \cite{Landau} the most important role is played by collective quanta of short and long wavelengths.
Quanta of long wavelengths with $\epsilon \sim cp$ ($c$ being the sound velocity),  are  elastic (of extension and squeeze) longitudinal waves, referred to as \it phonons \rm in physical literature, while ones of short wavelengths, \it rotons\rm, contribute to
 the motion of excitation quanta with the kinetic energy $\sim p_0 ^2$ (and $p_0/\hbar =1.9 \times 10^8~{\rm cm}^{-1}$ \cite{Landau52}), as if they are usual
particles (see \S \S ~77, 78 of part 3 in \cite{Levich1}) \footnote{Graphically, the curve $\epsilon (p)$ for a liquid helium II attains its minima at the values of momentum $p\to 0$ (it is its global minimum) and $p=p_0$ (it is its local minimum) \cite{Landau52, Levich1}. \par Just this explains that  above two kinds of collective quanta in a liquid helium II specimen contribute effectively to its phenomenology and thermodynamics \cite{Halatnikov}.}.

The superfluidity comes to the motion of the quantum
liquid without  any friction: thus the viscosity of the superfluid component in helium II is equal to zero.  
This   gives an example of  {\it potential liquids}. \par
\medskip The second component in a helium II belongs to the excitations spectrum
of quasiparticles   with above  short  and  long wavelengths quanta.  The number of intermediate quanta is enough small in comparison
with the other two kinds and one usually neglects it.
\par
At velocities exceeding $v_0$ the dissipation of energy occurs due to an interaction between the liquid and the walls of the vessel where this liquid
is contained. \par

\bigskip Estimating and definitive revealing  the physical sense of $v_0$
 is possible only in the quantum theory. \par 
Such a theory was created by Nikolaj Nikolaevich Bogolubov  and his co-authors in the space of time between 1947 and 1958 y. y.  \cite{N.N.}. With the example of the model Hamilton operator \cite{N.N.,Levich,Nels,Landau52}
\be \label{Hamel}
\hat H= -\sum \limits_{a=1} ^N \frac{\hbar^2}{2m} \Delta_a +\frac{1}{2}  U (\vert {\bf r}_a -{\bf r}_b \vert), \ee
(with $U (\vert {\bf r}_a -{\bf r}_b \vert)$
being the  interaction energy between particles $a$ and $b$ and  $N$ being the complete number of  particles in the considered system)
brought in the diagonal form \be \label{diagonal'} \hat H=\hat H_0+ \sum \limits_{ {\bf p} \neq 0} \epsilon ({\bf p})
\hat \xi ^+ _{{\bf p}}\hat \xi  _{{\bf p}} \ee
via the {\it Bogolubov transformations}  \cite{N.N.,Levich,Landau52,Smir} \be
\label{Bogolubov1} \hat b^+ _{{\bf p}} =u_{{\bf p}} \hat \xi ^+_{{\bf p}} + v_{{\bf p}}\hat \xi  _{{\bf -p}}, \ee
with 
$$ u_{{\bf p}}^2- v_{{\bf p}}^2=1 $$
and the creation (respectively, annihilation) operators \be \label{bop1}
\hat b^+ _{{\bf p}} = \frac{\hat  a _0 \hat a^+ _{{\bf p}}}{\sqrt{n_0}},
 \quad \hat b _{{\bf p}} = \frac{\hat a^+ _0 \hat a _{{\bf p}}}{\sqrt{n_0}},  \ee
expressed through the "initial" creation (annihilation) operators entering actually the Bogolubov model Hamilton operator (\ref{Hamel}): respectively, $\hat a^+ _{{\bf p}}$ and $\hat a_ {{\bf p}}$ (including their values at  zero momenta ${\bf p}=0$ and corresponding to the number $n_0$ of helium atoms possessing these zero momenta;  these values are denoted by the index 0 \cite{Levich}), there 
was shown that  
the superfluidity is possible only in the liquid helium II treated as {\it an non-ideal Bose gas}, in which the repulsion forces between particles dominate over attraction forces\rm \footnote{The sign $-$ between the c-numbers squared $ u_{{\bf p}}^2$ and  $v_{{\bf p}}^2$ in (\ref{Bogolubov1}) corresponds to the Bose statistics. Note that in the superconductivity case, where the Bogolubov transformations are also applicable but over the Fermi statistics for the creation (annihilation) operators, this sign should change onto the $+$ one; see e.g. \S 39 in \cite{Landau52}.}. 

A crucial step in  constructing  the quantum superfluidity theory  \cite{N.N.} was  extracting  the c-number condensate 
\be \label {c-number condensate} \frac{1}{2} \frac{N^2}{V} \nu (0),\ee
with \cite{Levich} $$\nu ({\bf p}) =\int U(\vert {\bf q}\vert) e^{-i{\bf pq}} d{\bf q}$$
being the Fourier integral of the potential (repulsion) energy $ U(\vert {\bf q}\vert)$ ($ {\bf q}\equiv {\bf r}_1-{\bf r}_2$). 
\par
As a result, one  came to the following evaluation of $v_0$: \be
\label{kritich}
v_0=(\frac{\epsilon ({\bf p})}{\vert {\bf p}\vert}) _{{\bf p}\to 0} =  ( \sqrt{\frac{n_0\nu ({\bf p})}{Vm} + \frac{{\bf p}^2}{4m^2}})_{{\bf p}\to 0} =
 \sqrt{\frac{n_0\nu (0)}{Vm}}.
\ee
The limit ${\bf p}\to 0$ ensures a finite value of
$v_0$, confirmed  experimentally. \par Thus only long-wavelengths quanta are
responsible for the superfluidity in a helium, as Eq. (\ref{kritich})
shows. \par 
\bigskip For our further investigations there will be very useful and important the following alternative expression for the critical velocity $v_0$ of the superfluid motion in a liquid helium, obtained  in the quantum theory \cite{Landau52}.
This formula has the look \be \label{alternativ} {\bf  v}_0 =\frac{\hbar}{m} \nabla \Phi(t,{\bf r}),
\ee
where $m$ is the mass of a helium atom and $\Phi(t,{\bf r})$
is the phase of the helium Bose condensate wave function $\Xi (t,{\bf r})\in C$.
The latter one may serve as a complex order parameter in the Bogolubov-Landau  model of the liquid helium \cite{N.N., Landau}  and will play an essential role in our investigations about the topological properties of liquid helium in the present study; therefore there will be useful to cite here also its explicit look \cite{Landau52}: 
\be 
\label{Xi1}
 \Xi (t,{\bf r})= \sqrt {n_0(t,{\bf r})}~ e^{i\Phi(t,{\bf r})}, 
\ee 
with $ n_0(t,{\bf r})$  being the number of particles in the helium Bose condensate.

Analysing  the Bogolubov-Landau  model \cite{N.N., Landau} of the liquid helium II, we distinctly understand that cooperative
degrees of freedom may appears only in non-ideal Bose gases. \par 

\subsection {Superconductor as a system with attraction forces. Bogolubov transformations.} In this study we want to demonstrate that the superfluidity phenomena are always connected with the {\it negative} coupling constant between helium atoms, i.e. with a {\it repulsion} forces.

But to understand why this is so, let us  make the digression to the superconductivity case and show that this phenomenon is associated with the {\it positive} coupling constant between electrons constituting Cooper pairs.  We need this for a comparison  with  superfluidity, which will be very instructive! In our explanation, we follow \S 39 in \cite{Landau52}.

Let's begin with the fact that the Fermi statistics and the Pauli exclusion principle inherent originally in the superconductivity theory provide the existence of the {\it Fermi surface} in the momenta space which separates occupied from unoccupied electron states at zero temperature (actually, in the limit $T\to 0$). This also generates the definition of the  {\it Fermi energy}
(at absolute zero temperature $T=0$) usually referring to the energy difference between the highest and lowest occupied single-particle 
states in a quantum system of {\it non-interacting} fermions at absolute zero temperature. However, there is a more realistic concept of {\it Fermi level}. Unlike the Fermi energy, the Fermi level is defined  (in the same wise as the Fermi energy) at an arbitrary different from zero temperature. Besides that, the Fermi energy indeed an energy difference (usually corresponding to a {\it kinetic} energy), whereas the Fermi level is a total energy level including kinetic energy and {\it potential} (Coulomb) energy. More exactly, the Fermi energy can only be defined for non-interacting fermions, whereas the Fermi level remains well defined even in complex interacting systems, at thermodynamic equilibrium.

  A notion,  closely related to the Fermi level, is the notion of {\it Fermi surface} in the reciprocal (momentum) space in the $T\to 0$ limit. This surface  separates occupied from unoccupied electron states.  Such definition in a manner hybrid of the above notions "Fermi energy" and "Fermi level" thanks to the limit $T\to 0$ which still leaves room for (Coulomb) interactions between electrons \footnote{On the author opinion, the notion of Fermi surface can be extrapolated on an arbitrary $T\neq 0$.}.

\medskip In the above terminology, at the given momentum $\bf p$ of an electron, let us consider the expression 
 
\be \label {np}  \eta _p= p^2/2m -\mu,
\ee
involving the mass $m$ of the electron and the chemical potential $\mu$.   Since $\mu\approx p_F^2/2m$ (with $p_F$ being the Fermi momentum corresponding to the Fermi level), we have near the Fermi surface 
\be \label {np1}  \eta _F= v_F(p-p_F),
\ee
where $v_F=p_F/2m$. 

\medskip  In terms of the  creation/annihilation operators $\hat a^+_{\bf p,\alpha}$, $\hat a_{\bf p,\alpha}$, respectively, with $\alpha$ being the spinor index, the model Hamiltonian for the superconductivity theory acquires the look \cite{Landau52}
\be \label{hsc} \hat H=\sum _{{\bf p},\alpha} \frac {p^2}{2m}\hat a^+_{\bf p,\alpha}\hat a_{\bf p,\alpha} -\frac g V \sum _{{\bf p},{\bf p}'} (\hat a^+_{\bf p ',+}\hat a_{-\bf p',-}- \hat a^+_{-\bf p ,-} \hat a_{-\bf p,+}),
\ee
with $V$ being the volume occupied by the system of electrons and $g=4\pi \hbar ^2 \vert a\vert/m$ being the "coupling constant" (while the scattering length $a<0$). It can be argued that only the items with $\bf p _1=-\bf p _2 \equiv \bf p$ and $\bf p' _1=-\bf p' _2 \equiv \bf p '$ must be actually retained in the Hamiltonian (\ref{hsc}). This is associated with the predominant role of the interactions between pairs the particles with opposite momenta and spins. 
Furthermore, the suffixes $+$ and $-$ refer to the two values of the spin component. 

Let us now introduce the new Hamiltonian $\hat H'=\hat H-\mu \hat N$, where
$$\hat N= \hat a^+_{\bf p,\alpha} \hat a_{\bf p,\alpha} $$
is the particle number operator; the chemical potential $\mu$ is then determined by the condition that the mean value $\bar N$ is equal to the given number of particles in the system. 

\medskip Subtracting now $\mu \hat N$ from (\ref{hsc}), we can write $\hat H'$ as
\be \label{hprim}  \hat H'=\sum _{{\bf p},\alpha} \eta_p \hat a^+_{\bf p,\alpha}\hat a_{\bf p,\alpha}-\frac g V \sum _{{\bf p},{\bf p}'} (\hat a^+_{\bf p ',+}\hat a_{-\bf p',-}- a^+_{-\bf p ,-} \hat a_{-\bf p,+}).
\ee

\medskip Now, by analogy with the bosonic Bogolubov transforms  (\ref{Bogolubov1}), let us introduce the new set of creation/annihilation operators $\hat b_{\bf p \pm} ^+$ and $\hat b_{\bf p \pm }$, respectively related to the "old" set of operators $\hat a^+$ and $\hat a$ by the linear Bogolubov transforms of the 
\bea \label{bogf1} \hat b_{\bf p -}=u_{\bf p} \hat a_{\bf p -}+ v_{\bf p} \hat a_{-\bf p, +};  \nonumber\\ 
\hat b_{\bf p +}=u_{\bf p} \hat a_{\bf p + }- v_{\bf p} \hat a_{-\bf p, -}.
\eea
type. These operators must obey the similar Fermi anti-commutation relations as the "old" operators  $\hat a^+$ and $\hat a$: 
\be \label{acr}b_{\bf p \alpha}b_{\bf p \alpha}^+ +b_{\bf p \alpha}	^+b_{\bf p \alpha}=1.
\ee
For this to be so, the transformation coefficients must, in turn, satisfy
 \be \label{acr1} u_{\bf p}^2+v_{\bf p}^2=1
\ee
(cf.  Eq. (\ref{Bogolubov1}) in the superfluidity case obeyed the Bose statistics; because of this, the opposite sign arises therein, as discussed above). 

\medskip Now we can rewrite the Hamiltonian (\ref{hprim}) in terms of the operators $\hat b^+$ and $\hat b$ and the coefficients $ u_{\bf p}$, $ v_{\bf p}$:
\bea \label{hprim1} \hat H'=2 \sum_{\bf p}\eta_{\bf p} v_{\bf p}^2+\sum_{\bf p}\eta_{\bf p} ( u_{\bf p}^2-v_{\bf p}^2)(\hat b_{\bf p +}^+\hat b_{\bf p +}+\hat b_{\bf p -}^+\hat b_{\bf p -})\nonumber\\ +2\sum _{\bf p}\eta_{\bf p} u_{\bf p} v_{\bf p} (\hat b_{\bf p +}^+\hat b_{-\bf p, -}+\hat b_{-\bf p, -}\hat b_{\bf p +}) -\frac g V \sum_ {\bf p, \bf p'} \hat B_{\bf p'}^+ \hat B_{\bf p}, \nonumber\\ \hat B_{\bf p}=u_{\bf p}^2\hat b_{-\bf p, -}\hat b_{\bf p +}-v_{\bf p}^2\hat b_{\bf p +}^+\hat b_{-\bf p, -}^+ +v_{\bf p}u_{\bf p} (\hat b_{-\bf p,-}\hat b_{-\bf p,-}^+ -\hat b_{\bf p+}^+\hat b_{\bf p +}).
\eea
Herewith the coefficients $u_{\bf p}$ and $v_{\bf p}$ are now chosen from the condition that the energy $E$ of the system be a minimum for a given entropy. The entropy is given by the combinatorial expression \cite{Landau52}
\be \label{ent} S=-\sum_{\bf p, \alpha}[n_{\bf p, \alpha}\ln n_{\bf p, \alpha}+(1-n_{\bf p, \alpha})\ln (1-n_{\bf p, \alpha})].
\ee
The condition stated thus is equivalent to minimizing the energy for given quasi-particle occupation numbers $ n_{\bf p, \alpha}$.

In the Hamiltonian (\ref{hprim1}) the diagonal matrix elements are zero except for terms containing the products
$$\hat b_{\bf p, \alpha}^+\hat b_{\bf p, \alpha} =n_{\bf p, \alpha}; \quad \hat b_{\bf p, \alpha}\hat b_{\bf p, \alpha}^+ =1-n_{\bf p, \alpha}.$$ 
Hence 
\be \label{energy} E=2 \sum_{\bf p}\eta_{\bf p} v_{\bf p}^2+\sum_{\bf p}\eta_{\bf p} ( u_{\bf p}^2-v_{\bf p}^2)(n_{\bf p +}+n_{\bf p +})- \frac g V [\sum_{\bf p}u_{\bf p}v_{\bf p} (1-n_{\bf p +}n_{\bf p -})]^2. 
\ee
Varying this expression with respects to the parameters $u_{\bf p}$ and using the relation (\ref{acr1}) we find as the condition for a minimum 
$$ \frac {\delta E}{\delta u_{\bf p}} =- \frac 2{v_{\bf p}} (1-n_{\bf p +}-n_{\bf p -}) [2\eta_{\bf p}u_{\bf p}v_{\bf p}-\frac g V ( u_{\bf p}^2-v_{\bf p}^2)\sum_{\bf p '}u_{\bf p '} v_{\bf p '}(1-n_{\bf p ' +}-n_{\bf p ' -})]=0.$$ 
Hence
\be \label{cond.for.min}2\eta_{\bf p}u_{\bf p}v_{\bf p} =\Delta (u_{\bf p}^2-v_{\bf p}^2),
\ee
where $\Delta$ denotes the sum
\be \label{delta1} \Delta =\frac g V \sum_{\bf p }u_{\bf p } v_{\bf p }(1-n_{\bf p ' +}-n_{\bf p ' -}).
\ee
From (\ref{cond.for.min}), (\ref{delta1}) we can express  $u_{\bf p }$ and $v_{\bf p }$ in terms of $\eta_{\bf p}$ and $\Delta$:
\bea \label{uv}  \aligned u_{{\bf p}}^2 \\ v_{{\bf p}}^2 \endaligned  \} =\frac 1 2 (1\pm \frac{\eta_{\bf p}}{\sqrt{(\Delta ^2+\eta_{\bf p}^2)}}).
\eea
Substituting these values in (\ref{delta1}), we obtain an equation for $\Delta$:
\be \label{delta} \frac{g}{2V}\sum_{\bf p }\frac{(1-n_{\bf p  +}-n_{\bf p  -})}{\sqrt{(\Delta ^2+\eta_{\bf p}^2)}}=1.
\ee
In equilibrium, the quasi-particle occupation numbers are independent of the spin direction and are given by the Fermi distribution formula (with zero chemical potential):
\be \label{eql} n_{{\bf p}+}=n_{{\bf p}-}\equiv n_{\bf p}=[e^{\epsilon/T}+1]^{-1}.
\ee
Changing also from summation to integration over the $\bf p$-space, we can write this equation in the form 
\be \label{intform} \frac 1 2 g \int \frac{1-2n_{\bf p}}{\sqrt{\Delta^2+n_{\bf p}^2}} \frac{d^3p}{(2\pi \hbar)^3}=1.
\ee
Let us now analyze the relations derived above. We shall see that $\Delta$ plays a basic role in the theory of spectra of the type under consideration. Let us now calculate its value 
$\Delta_0$ at $T=0$. 

When  $T=0$, there are no quasi-particles, so $n_{\bf p}=0$ and Eq.  (\ref{intform}) becomes
\be \label{intform1} \frac{g}{2(2\pi \hbar)^3} \int \frac{4\pi p^2 dp}{\sqrt{(\Delta^2_0+n_{\bf p}^2)}}=1.
\ee
We may note immediately that this equation {\it could not have a solution for} $\Delta_0$ if $g<0$, i.e. {\it in the case of repulsions}, since the both sides would then have opposite signs. 

\medskip Thus we proved, after rather lengthy arguments,  that only {\it attraction} forces inside a superconducting system provide  solutions to the minimum condition for this system.

\bigskip Now let us analyze the case of superfluidity. It will be helpful here for us the \S 5 (Chapter 6) of the monograph \cite{Levich}. 

We will start our research by writing the superfluidity Hamiltonian. This will be a cornerstone of the all present study. 

\medskip It is easy to see \cite{Levich} that this has the general shape 
\be \label{haths} \hat H=-\sum _{\alpha=1}^N \frac {\hbar^2}{2m} \Delta_\alpha + \frac 1 2 \sum U (\vert {\bf r}_\alpha - {\bf r}_\beta\vert),
\ee
where $N$ is the complete number of particles in the system and $ U (\vert {\bf r}_\alpha - {\bf r}_\beta\vert)$ is the interaction energy between the particles $\alpha$ and $\beta$. In the dilute gas approximation, the only pair interactions can be taken into account. 

It is comfortable, for the further consideration,  to regard the gas as that placed into the cube with the edge $L$. Then the Hamiltonian of a free particle
$$ \hat H_\alpha= -\frac {\hbar ^2}{2m} \Delta_\alpha$$
will possess a discrete spectrum. The momentum of the free particle will have the projections running through a discrete series of values

$$ p_x=\frac {2\pi \hbar}{L}n_x, \quad p_y=\frac {2\pi \hbar}{L}n_y, \quad p_z=\frac {2\pi \hbar}{L}n_z,$$
where $n_x$, $n_y$, $n_z$ are arbitrary integers including zero. The eigenfunctions of the operator $\hat H_\alpha$ normalized onto the volume $V=L^3$ have the look 
$$ \psi ({\bf r})=\frac 1{\sqrt V}e^{\frac i \hbar {\bf p}{\bf r}}.$$
Let us apply now the second quantization method to our system (\ref{haths}) of interacting particles. 

It's done like this (see \S 99 Charpter 5 in \cite{Levich}).  Let $E_k$ be the energy of the particle in the $k$ state.  The authors \cite{Levich} postulate the following shape of the quantum Hamilton operator for the case of {\it non-interacting} particles located in the given background field:
\be \label{haths1}\hat H = \sum _{i=1}^N \hat H_i =\sum _{i=1}^N (\hat T_i +U(\xi))=\sum _{i=1}^N (-\frac{\hbar ^2}{2m} \Delta_i+U(\xi)),
 \ee
where $U(\xi)$ is the potential energy of the particle $i$ in the background field and $\hat T_i$ is the operator of its kinetic energy. It is easy to see that Eq. (\ref{haths1}) is applicable to the case of the pairwise interaction in the dilute gas approximation. 

If now $\hat a_k$, $\hat a_k^+$ are the Bose annihilation/creation operators for the given $k^{\rm th}$ particle, it is easy to represent the Hamilton operator (\ref{haths1}) in the shape
\be  \label{haths2}\hat H= \sum _k E_k \hat a_k^+\hat a_k,
\ee
that is, obviously, equivalent to the
\be  \label{haths21} \hat H = \sum _{k,l} (\hat H_i)_{lk}\hat a_k^+\hat a_k.
\ee
If now we choose as $\psi_k$ the wave functions of the operator $\hat T_k$ corresponding to the eigenvalues $\epsilon_k$, then Eq. (\ref{haths21}) is rewritten in the form 
\be  \label{haths22} \hat H =\sum _k \epsilon_k \hat a_k^+\hat a_k +\sum _{k,l} \hat a_l^+\hat a_k \int \psi^*_l U(\xi)\psi_k d\xi.
\ee
In the case of a system of particles with a pairwise interaction, the operator of the  interaction energy has the look
\be\label{W} \frac 1 2 \sum _{i\neq l} W(\xi_i, \xi_l).\ee
Let us now introduce the operator
$$ \hat L_2=\sum _{i,j=1}^N \hat L(\xi_i, \xi_j),$$
which, in the second quantization representation, is expressed by the formula
\be \label{l2} \Hat L_2=\sum_{k,p,l,m} (lm \vert \hat L (\xi, \xi')\vert kp) \hat a_l^+\hat a_m^+\hat a_k\hat a_p,
\ee
with matrix elements
\be \label{me} (l,m\vert \hat L (\xi, \xi')\vert k,p)=\int \psi_l^* (\xi)\psi_m^* (\xi ')\hat L (\xi, \xi')\psi_k (\xi)\psi_p (\xi ') d\xi d\xi'.
\ee
The operator (\ref{W}) is a particular case of the general formula (\ref{me}). This allows us to write down the Hamilton operator $\hat H$ as
\be  \label{haths23} \hat H =\sum _k E_k \hat a_k^+\hat a_k +\frac 1 2 \sum _{k,p,l,m} (lm \vert W (\xi, \xi')\vert kp) \hat a_l^+\hat a_m^+\hat a_k\hat a_p.
\ee

\subsection{The gauge $U(1)$ symmetry of the He $^4$ Hamiltonian.}  Returning again to the non-ideal gas case (\ref{haths}), we apply Eq. (\ref{haths21}) to this case. This results
\be  \label{hath22} \hat H= \sum E({\bf p}_k)\hat a _{{\bf p}_k}^+\hat a _{{\bf p}_k}+ \frac 1 2 \sum_{i,k,l,m} (lm \vert U \vert ki)\hat a _{{\bf p}_l}^+\hat a _{{\bf p}_m}^+\hat a _{{\bf p}_k
}\hat a _{{\bf p}_l} =\hat H_0+ \hat H',
\ee
with the sum in (\ref{hath22}) over the all possible discrete values of  momenta (positive al well as negative). The energy  $E({\bf p}_k)$ is the energy of the free particle
$$ E({\bf p}_k)= \frac {{\bf p}_k^2}{2m}.$$
The interaction energy $U$ entering the matrix element of the Hamiltonian depends, by virtue of what has been said, only on the coordinates of the two particles. 

\medskip In order to calculate the matrix element, one can utilize the explicit look of the  free particle wave function 
\bea \label{wf}   (lm \vert U \vert ik)= \nonumber \\  \frac 1 {V^2} \int _V e^{-\frac i \hbar {(\bf p}_l {\bf r} _1 +{\bf p}_m \bf r _2)} U(\vert {\bf r} _1-{\bf r} _2\vert)e^{\frac i \hbar {(\bf p}_l \bf r _1 +{\bf p}_m \bf r _2)} dV_1 dV_2.\eea
Introducing the new variables ${\bf q}= {\bf r} _1-{\bf r} _2$ and ${\bf R}=( {\bf r} _1+{\bf r} _2)/2$ and integrating with accounting that 
\bea \label{integr} \int _V e^{\frac i \hbar {\bf p}{\bf r}} =\{ \aligned V \quad {\rm at} \quad {p=0}\\ 0 \quad {\rm at} \quad {p\neq 0}\endaligned
\eea
we get
\bea \label{wf1} (lm \vert U \vert ik)=\left \{ \aligned \frac 1 V \nu ({\bf p}_l-{\bf p}_i) \quad {\rm at} \quad {\bf p}_l+{\bf p}_m={\bf p}_i+{\bf p}_ k\\  0 \quad  {\rm at} \quad {\bf p}_l+{\bf p}_m\neq {\bf p}_i+{\bf p}_ k   \endaligned \right \},
\eea
where 
$$  \nu ({\bf p})=\int U(\vert {\bf q}\vert) e^{-i{\bf p}{\bf q}} d{\bf q}. $$
Integrating over all the angles can be done directly since $U$ depends only on the absolute value of the vector $\bf q$.  This results
\bea \label {nu1} \nu ({\bf p})=\int U(\vert {\bf q}\vert) q^2dq \int e^{ipq\cos \theta} \sin \theta \int d\varphi = \nonumber \\ 4\pi \int U(\vert {\bf q}\vert) \frac {\sin (pq)}{pq} q^2dq.
\eea
We can see that the function $\nu ({\bf p})$ is a real one, while $\nu ({\bf p})=\nu (-{\bf p})$. Then the operator (\ref{hath22}) can be recast to the shape 
\be \label{hat221} \hat H 
= \sum _k \frac {p_k^2}{2m} \hat a _{{\bf p}_k}^+\hat a _{{\bf p}_k} +\frac 1 2 \sum \frac 1 V \nu ({\bf p}_l-{\bf p}_i)\hat a _{{\bf p}l} \hat a _{{\bf p}_m}  a _{{\bf p}j}  a _{{\bf p}k} .
\ee
In the second item of (\ref{hat221}) the sum is taken only over those values of the momenta $\bf p_i$, $\bf p_k$, $\bf p_l$, $\bf p_m$ for which the momentum conservation relation 
$$ \bf p_l+ \bf p_m =\bf p_i+ \bf p_k$$
is satisfied. 

The further task comes to specifying the eigenvalues of the Hamiltonian (\ref{hat221}), i.e. to diagonalizing the energy matrix.   At investigating the lower energy states, the perturbation theory is not applicable.   It is clearly visible from the fact that, in  the lower energy states, i.e. at the small momenta  values, the kinetic energy of the system goes to zero. Conversely, the interaction energy has a finite value and is large in comparison with  the kinetic energy. That's why an especial approximate method was developed by N. N. Bogoliubov for the investigation of weakly excited states. 

Let us consider firstly a system with any interaction  absent. In the ground energy state of such a system, the momenta of all the particles equal to zero:
\bea \label{grund} n_{\bf p} =0 \quad {\rm at} \quad p\neq 0; \nonumber \\ n_{\bf p} =N \quad {\rm at} \quad p=0.,
\eea
where $n_{\bf p}$ the number of particles with the momentum $\bf p$. It is naturally to think that the majority of particles will possess the momenta equal to zero even at a weak interaction turned on.  in view of the above, let us adopt that
\bea \label{ex} \left \{ \aligned \hat a_0^+ \hat a_0=n_0 \approx N \\  \hat a_{\bf p}^+ \hat a_{\bf p} =n_{\bf p}  \ll N. \endaligned \right \}
\eea
The operators $\hat a_0^+$ and  $\hat a_0$ satisfy the standard commutation relation 
$$ \hat a_0\hat a_0^+ -\hat a_0^+\hat a_0 =1.$$
Since $\hat a_0^+ \hat a_0=n_0$ is large in comparison with 1, we will neglect further  the noncommutativity of these operators, i.e. we will replace the operators $\hat a_0$ and  $\hat a_0^+$ with c-numbers. 

The assumptions (\ref{grund}) and (\ref{ex}) simplify greatly the expression for the interaction  Hamiltonian. Namely, in the sum over all the momenta, it should be left only the large terms in which the multipliers $\hat a_0^+$ and  $\hat a_0$ enter pairwise or four times. There are
$$ \hat a_0^+ \hat a_0^+\hat a_0 \hat a_0, \quad \hat a_0^+ \hat a_0^+\hat a_{\bf p} \hat a_{\bf p}, \quad \hat a_0^+\hat a_{\bf p}^+\hat a_0\hat a_{\bf p}^+$$ 
etc. 

Vice versa, the items of the type $\hat a_0^+\hat a_{{\bf p}m}\hat a_{{\bf p}_l}\hat a_{{\bf p}k}$  where ${\bf p}_m \neq 0$, ${\bf p}_l  \neq 0$ and ${\bf p}_k  \neq 0$, can be omitted. 

Thus we find for $\hat H'$ 
\bea \label{243} \hat H'= \sum \sum_{{\bf p}l+{\bf p}m={\bf p}i+{\bf p}k} \nu ({\bf p}_l -{\bf p}_i) \hat a_{{\bf p}l}^+\hat a_{{\bf p}m}^+ \hat a_{{\bf p}i}  \hat a_{{\bf p}k} \cong \nonumber \\ 
\sum _{{\bf p}\neq 0} \nu ({\bf p}) \{  \hat a_{{\bf p}}\hat a_0^+\hat a_{{\bf p}}\hat a_0 +\hat a_0^+\hat a_{{\bf p}}^+\hat a_{{\bf p}}\hat a_{{\bf p}} +\hat a _{-{\bf p}}^+ \hat a _{-{\bf p}}^+ \hat a_0 \hat a_0\} + \nonumber \\  \nu (0) \hat a_0^+ \hat a_0^+  \hat a_0 \hat a_0 \cong \nu (0) N^2 + \sum _{\bf p \neq 0} \nu ({\bf p}) \{2\hat a_{{\bf p}}^+\hat a_{{\bf p}}n_0+\hat a_{{\bf p}}^+\hat a_{{\bf p}}^+n_0+\hat a _{-{\bf p}}\hat a _{-{\bf p}} n_0\}.
\eea
Since the complete number $N$ of the particles in the system is 
$$ N=n_0+\sum_{{\bf p}\neq 0}\hat a_{{\bf p}}^+\hat a_{{\bf p}},$$
where the second item is small in comparison with the first one, we have, in the same approximation, 
\bea \label{244} \hat H'=\sum \nu ({\bf p}_l -{\bf p}_i) \hat a_{{\bf p}l}^+\hat a_{{\bf p}m}^+ \hat a_{{\bf p}i}  \hat a_{{\bf p}k} \cong \nonumber \\ \nu (0) N^2 +2N  \sum _{\bf p \neq 0} \nu ({\bf p}) (\hat a_{{\bf p}}^+\hat a_{{\bf p}}+\hat a_{-{\bf p}}^+\hat a_{-{\bf p}})+ \nonumber \\N  \sum _{\bf p \neq 0} \nu ({\bf p}) (\hat a_{{\bf p}}^+\hat a_{-{\bf p}}++\hat a_{-{\bf p}}^+\hat a_{{\bf p}}).
 \eea
The omitted items have the order $\sqrt N$. The complete Hamiltonian $\hat H$, up to the value of the $N$ order, acquires the look
\bea \label{245} \hat H =\sum _{\bf p \neq 0} \{ \frac {p^2}{2m}+\frac {n_0} V \nu ({\bf p})\} \hat a_{{\bf p}}^+\hat a_{{\bf p}} + \frac 1 2 \frac {N^2}V \nu (0)+ \nonumber \\ \frac N {2V} \sum _{{\bf p}\neq 0} \nu ({\bf p})\hat a_{{\bf p}}^+ \hat a_{-{\bf p}}^+ + \frac N {2V} \sum _{{\bf p}\neq 0} \nu ({\bf p}) \hat a_{{\bf p}} \hat a_{-{\bf p}}.
\eea
Let us introduce the new operators $\hat b_{\bf p}^+$ and $\hat b_{\bf p}$:
\be \label{bb} \hat b_{\bf p}=\frac {\hat a_0^+\hat a_{{\bf p}}}{\sqrt {n_0}}, \quad \hat b_{\bf p}=\frac {\hat a_0\hat a_{{\bf p}}^+}{\sqrt {n_0}}.
\ee
The operators $\hat b_{\bf p}^+$ and $\hat b_{\bf p}$ thus defined satisfy the same commutation relations that the operators $\hat a_{\bf p}^+$ and $\hat a_{\bf p}$ since we consider $\hat a_0^+$ and $\hat a_0$ as a c-numbers. Besides that, as it is easy to see,
\be \label{bb1}\hat b_{\bf p}^+\hat b_{\bf p}=\hat a_{\bf p}^+\hat a_{\bf p}=n_{\bf p}
\ee
since $\hat a_0^+ \hat a_0=n_0$. With the aid of the operators $\hat b_{\bf p}^+$ and $\hat b_{\bf p}$, the Hamiltonian (\ref{245}) can be rewritten as
\bea \label{248} \hat H =\sum _{\bf p \neq 0}  \frac {p^2}{2m}\hat b_{\bf p}^+\hat b_{\bf p}+ + \frac 1 2 \frac {N^2}V \nu (0)+ \nonumber \\ \frac{n_0}{2V} \sum _{\bf p \neq 0} \nu ({\bf p}) [\hat b_{\bf p}^+\hat b_{-{\bf p}}+\hat b_{\bf p}\hat b_{-{\bf p}}+2\hat b_{\bf p}^+\hat b_{\bf p}]. 
\eea
In order to bring the  Hamiltonian (\ref{248}) to the diagonal look, let us commit a linear transformation to the new operators $\hat \varsigma_{\bf p}$, $\hat \varsigma_{\bf p}^+$:
\bea \label{sig}   \hat b_{\bf p}=u_{\bf p}\hat \varsigma_{\bf p} +v_{\bf p}\hat \varsigma_{-{\bf p}}^+, \nonumber \\  \hat b_{\bf p}^+ = u_{\bf p}\hat \varsigma_{\bf p}^+ +v_{\bf p}\hat \varsigma_{-{\bf p}},
\eea
with $$ u^2_{\bf p}- v^2_{\bf p}=1;\quad u_{\bf p} =u_{-\bf p}; \quad v_{\bf p}=v_{-\bf p}.$$

If $u_{\bf p}$ and $v_{\bf p}$ are real functions of the momentum  $\bf p$, the the new Bose operators $\hat \varsigma_{\bf p}$ and $\hat \varsigma_{\bf p}^+$ satisfy the standard commutation relations. Substituting (\ref{sig}) in (\ref{248}), and demanding that the coefficients at the operators of the $\hat \varsigma_{\bf p}^+\hat \varsigma_{\bf p}$ and $\hat \varsigma_{\bf p}\hat \varsigma_{\bf p}$ types turn to zero, we find the functions $u_{\bf p}$ and $v_{\bf p}$. A simple but somewhat long calculation results
\bea \label{250} u_{\bf p}=\frac 1 {\sqrt{1-A^2_{\bf p}}}, \quad v_{\bf p}=\frac{A^2_{\bf p}}{\sqrt{1-A^2_{\bf p}}}, \nonumber \\ A_{\bf p}= \frac V {n_0\nu({\bf p})} \{ \epsilon ({\bf p})-\frac {p^2}{2m}-\frac{n_0}V\nu({\bf p})\},\nonumber \\ \epsilon ({\bf p})= \sqrt {\frac{n_0}V\frac{p^2\nu({\bf p})}{m}+ \frac{p^4}{4 m^2}}.
\eea
Wherein only the diagonal items persist in the operator $\hat H$, and it brought to the form 
\be \label{251} \hat H = \hat H_0 +\sum _{{\bf p}\neq 0} \epsilon ({\bf p}) \hat \varsigma_{\bf p}^+ \hat \varsigma_{\bf p}.
\ee
The eigenvalues of the Hamilton operator (\ref{251}) are
\be \label {ev} E=E_0+\sum _{{\bf p}\neq 0}\epsilon ({\bf p}) n' ({\bf p}),
\ee
where
\be \label {e0} E_0=\frac {N^2}{2V}\nu(0)+\sum _{{\bf p}\neq 0} \frac 1 2 [\epsilon ({\bf p})-\frac{p^2}{2m}-\frac{n_0}{V}\nu({\bf p})],
\ee
where $n'({\bf p})$ are arbitrary integers. 

The complete momentum $\bf P$ of the system of particles also can be found:
$${\bf P}=\sum _{{\bf p}\neq 0}{\bf p}  \hat a_{\bf p}^+ \hat a_{\bf p}=\sum _{{\bf p}\neq 0} {\bf p} \hat b_{\bf p}^+ \hat b_{\bf p}.$$
If, in the last equality, go over to the operators $ \hat \varsigma_{\bf p}^+$ and $ \hat \varsigma_{\bf p}$ from the operators $ \hat b_{\bf p}^+$  and $ \hat b_{\bf p}$, we get
\be \label{P} {\bf P}=\sum _{\bf p} {\bf p}\hat \varsigma_{\bf p}^+\hat \varsigma_{\bf p}=\sum _{\bf p}{\bf p} n'_ {\bf p}
\ee
Eqs. (\ref{e0}) and (\ref{P}) permit a simple corpuscular interpretation. We see that the energy of the system is presented as the sum of two items. The first item $E_0$ represents the energy of the ground (lower) state. The second item can be interpreted as the summary energy of quasiparticles representing the collective excitations of the system while $ n'_ {\bf p}$ is the number of elementary excitations in the state with the momentum $\bf p$. The energy of each elementary excitation is $\epsilon ({\bf p})$, (\ref{250}). 

At small momenta, the  excitation energy can be represented in the shape 
 \be \label{ep}\epsilon ({\bf p})=\sqrt{\frac{\nu(0)}{mV_0}} \vert {\bf p} \vert .
\ee
Here $V_0=V/n_0\approx V/N$ is the volume per particle. It follows from the formula (\ref{ep}) that the inequality  
\be \label{rep} \nu (0)=\int U(q) d {\bf q} >0
\ee
must be satisfied which corresponds {\it to the  predominance of repulsive forces}. In the reverse situation, when $\nu (0)<0$, the energy becomes imaginary that corresponds to the unstable state of system. 

\bigskip Thus we proved just now that repulsive forces are dominated in a non-ideal Bose gas. There is still the proof (at which conditions) whether the superfluidity phenomena can occur in such a gas when the inequality (\ref{rep}) is satisfied. 

Suppose that the additional velocity $\bf v$ is given to the all totality of particles with respect to some stationary reference frame: for example, to the walls of the vessel or tube containing this non-ideal Bose gas. Then one can think that all the results just obtained will be fair also in the reference frame moving with the  velocity $\bf v$  with respect to some stationary one. If the energy equal $E$ in the moving reference frame while $E^v$ in the stationary one, then they are related as
\be \label{ev1}E^v = 
E+ \frac {N m {\bf v}^2}2+ {\bf vP},
\ee
where $\bf P$ is the complete momentum of the Bose gas in the moving reference frame. Utilizing the expressions (\ref{ep}) and (\ref{P}), we get
\be \label{ev2}E^v =E_0+\frac {N m {\bf v}^2}2+\sum_{\bf p} n_p'\{  \epsilon(p)+{\bf vp}\}. 
\ee
In order to slow down the totality of particles, it is necessary that the excitations appear with the momenta directed against the velocity $\bf v$. The increment of the energy when an one such excitation appears is
\be \label{deltae}
\Delta \epsilon =\epsilon ({\bf p})- \vert {\bf v}\vert  \vert {\bf p} \vert .
\ee
If $\Delta \epsilon >0$, the appearance of excitations is energetically disadvantageous. This means that the system will move infinitely long with the velocity $\bf v$ and without excitations emerging. Any braking in the system is absent and it possesses a {\it superfluidity}. Let us now formulate the superfluidity condition. For the value $\Delta \epsilon$ to be positive, it is necessary that 
$$ \frac {\epsilon ({\bf p})}{\vert {\bf p} \vert}>\vert {\bf v}\vert $$
at arbitrary $\bf p$. Let us denote as ${\bf v}_0$ the minimal value of the ratio $\frac {\epsilon ({\bf p})}{\vert {\bf p} \vert}$. Then at moving with the velocity ${\bf v}<{\bf v}_0$ the superfluidity takes place. In other words, a superfluidity is possible in the discussed system if the inequality
\be \label{krit} \vert {\bf v}_0\vert ={\rm min} (\frac {\epsilon ({\bf p})}{\vert {\bf p} \vert}) >0
\ee
is satisfied. It is just the same critical velocity for superfluidity we have discussed superficially at beginning this section. 
	
\medskip Thus two important results us demonstrated in this section for non-ideal Bose gases- First, (\ref{rep}), that repulsive forces predominate in such system. The second one, the possibility of superfluidity phenomena at the velocities  not exceeding $\vert {\bf v}_0\vert $ by their absolute values.

\medskip Due to Eq. (\ref{250}) for $\epsilon ({\bf p})$, 

$$ {\bf v}_0=(\frac {\epsilon ({\bf p})}{\vert {\bf p} \vert})_{{\bf p}\to 0} =(\sqrt{\frac{n_0\nu ({\bf p})}{Vm}+\frac {{\bf p}^2}{4m^2}})_{{\bf p}\to 0}= \sqrt {\frac {n_0\nu(0)}{mV}}. $$
If $\nu(0)>0$, ${\bf v}_0$ is real and positive.  Thus in the case when repulsion forces prevail in the given non-ideal gas there exist a   real and positive value ${\bf v}_0$ and this Bose gas possess a manifest superfluidity. 

If any interaction between particles is absent (the ideal Bose gas case), then $U(q)=0$, $\nu ({\bf p})=0$ and the excitations energy is given by the formula
\be \label{ideal} \epsilon ({\bf p}) =\frac {{\bf p}^2}{2m}.
\ee
Herewith ${\bf v}_0=0$ and thus an ideal Bose gas {\it does not possess a superfluidity}. 

\medskip Thus we see that superfluidity appears in a non-ideal Bose gas and is absent in an ideal Bose gas. A  superfluidity is not related with a specifics of a Bose system. For her existence, a special formula of the energy spectrum of collective excitations is required.  Namely, according to  (\ref{krit}), it is necessary for this that the ratio of the minimal  excitation energy {\it as a whole} to the momentum of this excitation has a finite value. 

Vice verse, the spectrum of an ideal Bose gas at small excitations has the look (\ref{ideal}) and does not satisfy (\ref{krit}). 

An excitations spectrum of interacting Bose particles satisfies  (\ref{krit}) only  if repulsion forces are available between particles. Attraction forces do not lead to the superfluidity phenomena in such a non-ideal Bose system. It is necessary however to stress that the following property of Bose particles was used at obtaining this result. We thought that the majority of interacting particles are in the state with the zero momentum, i.e. that they form {\it a condensate}  in the momentum space. This serves as a necessary condition for the superfluidity emerging in a Bose gas. However, the just specified condition is not a sufficient one since a condensate is formed also in an ideal gas however  a superfluidity is not observed. The difference between the condensates in an  ideal gas and non-ideal one is seen from the following reasoning. Let an ideal Bose gas to move with some velocity $\bf v$ relative to the walls of the vessel. If one of the particles stops, as a result of an interaction with the wall, then the rest mass will continue its motion with less kinetic energy. The consecutive repetition of this process will slow down the gas eventually. 

The case is different for a gas which particles are interconnected by the repulsion forces. The stop of individual particles  {\it is impossible} here. The interaction with the wall should slow down or, which is the same, to excite system as a whole. At the motion velocity ${\bf v}>{\bf v}_0$ it turns out to be impossible.

\bigskip  Our previous analysis of the non-ideal Bose gas Hamiltonian (applied to the case of ${\rm He} ^4$ system rested in some reference frame) allows us to rewrite this operator in terms of creation/annihilation operators depending manifestly on the radius-vector $\bf r$ and time $t$.  It is a "classical" look for such operator widely used in the modern literature about ${\rm He} ^4$. In the present study we refer onto the paper \cite{Nels} at writing down such Hamiltonian which we will call {\it the model ${\rm He} ^4$ at rest Hamiltonian} in the present study:
\begin{eqnarray} \widehat{\cal H} &=& \int d^3r~\hat a^+ ({\bf r},t)\left({-\hbar^2\over 2m}\;\nabla^2\right)
\hat a({\bf r},t)\nonumber\\
&\quad +&{1\over 2} \int d^3r\int d^3r'\hat a^+ ({\bf r},t)~\hat a^+({\bf r}',t)
V(|{\bf r}-{\bf r}|)~
\hat a({\bf r}',t)~\hat a({\bf r},t)
\label{eq:three1}
\end{eqnarray} 
(with $ V(|{\bf r}-{\bf r}|)$ being the  bosonic pair potential  in the helium).  
This operator actually can be treated as the (inverse) Fourier transformation of the operator (\ref{hat221}). 

\medskip It turns out that the bosonic creation/annihilation operators ($\hat a^+({\bf r},t)$ and $\hat a({\bf r},t)$, respectively) can be expressed in terms of the phase $\Phi (t,{\bf r})$  of the Bose condensate wave function \footnote{Indeed, these operators  are, perhaps, complex functions of another variable: say, ${\tilde \Phi} (t,{\bf r})$. But, without loss of generality, one can  think  that all the enumerated operators $\hat a^+({\bf r},t)$ and $\hat a({\bf r},t)$ possess the same phase $\Phi (t,{\bf r})$ that coincides with the one for the Bose condensate wave function. }. 

In order to find a real connection between the creation/annihilation operators $\hat a^+({\bf r},t)$ and $\hat a({\bf r},t)$ and the the phase $\Phi (t,{\bf r})$  of the Bose condensate wave function, we have to notice two facts. Firstly (as it was mentioned above)  the complete number $N$ of the particles in the system is \cite{Levich}
$$ N=n_0+\sum_{{\bf p}\neq 0}\hat a_{{\bf p}}^+\hat a_{{\bf p}},$$ with $n_0$ being the number of "particles" inside the Bose condensate (the latter fact in a good agreement with the general interpretation of a vacuum as a surface where the energy of a system takes its minimum).

Secondly, following \cite{Landau52}, let us write down the {\it operator} $\hat \Psi ({\bf r},t)$ for  the non-ideal Bose gas in the Heisenberg picture. It is   
\be \label {heisenberg} \hat \Psi ({\bf r},t)= \frac 1{\sqrt V} \sum _{\bf p} \hat a _{\bf p} {\rm exp} \{ \frac i \hbar {\bf p}{\bf r}-\frac i \hbar \frac{p^2}{2m} t\}. 
\ee
As it was explained above, since $\hat a_0^+ \hat a_0=n_0\gg 1$, we can neglect further  the noncommutativity of these operators, i.e. we  replace the operators $\hat a_0$ and  $\hat a_0^+$ with c-numbers. Applying this property to the above Heisenberg operator (\ref {heisenberg}), at $\vert p\vert =0$ we consider the appropriate part of this operator, which is precisely 
\be \label{ksju} \hat \Xi =\hat a_0/\sqrt V.
\ee
In the general case of an arbitrary Bose liquid, since the condensate contains a macroscopically large number of particles, changing this number by 1 does not essentially affects the state of the system. One can speak that the result of adding/removing  one particle (mode) in the condensate is to convert a state of a system of $N$ particles into the  "same" state of a system of $N\pm 1$ particles. In particular, the ground state remains the ground state. Let $ \hat \Xi$ and $ \hat \Xi^+$ denote the part of the $\psi$ operatorsthat changes the number of particles in the condensate by 1; then, by definition,
$$\hat \Xi \vert m, N+1> =\Xi  \vert m, N>, $$ 
$$\hat \Xi^+ \vert m, N>=\Xi^* \vert m, N+1>,$$
where the symbols $ \vert m, N>$ and $ \vert m, N+1>$ denote two "like" states differing only as regards the number of particles in the system, and $\Xi$ is the complex number. These statements are rigorously valid in the $N\to\infty$ limit. Hence the definition of $\Xi$ can be written as
\bea \label{limiting} {\rm lim}
_  {N\to\infty} <m,N \vert \hat \Xi \vert m, N+1 >=\Xi, \nonumber \\
{\rm lim}
_  {N\to\infty} <m,N+1 \vert \hat \Xi^+ \vert m, N >=\Xi^*;
\eea
the limit is taken for a given finite value of the liquid  density  $N/V$.

The quantity $\Xi$ defined via (\ref{limiting}) is a function of coordinates and time representing the particle wave function in the condensate state. It is normalized by the condition $\vert \Xi\vert ^2=n_0$, and can, therefore, be expressed as 
\be \label {Xin} \Xi(t,{\bf r})=\sqrt {n_0(t,{\bf r})}e^{-i\Phi(t,{\bf r})}
\ee

(in fact now we showed the origin of Eq. (\ref{Xi1}) above).

The current density calculated from the wave function (\ref{Xin}) is  (see \S 19 in the monograph \cite{Landau3})
\be \label {cur} {\bf j}_{\rm cond} = \frac {i\hbar}{2m} (\Xi \nabla \Xi^*-\Xi^*\nabla \Xi)=
\frac \hbar m n_0\nabla \Phi,
\ee
where $m$ is the mass of the liquid particle. This has the significance of the macroscopic  current density of condensate particles, and may be equated to $n_0{\bf v}_s$, where ${\bf v}_s$ is the  macroscopic velocity of this motion. It is easy to see that the critical velocity ${\bf v}_0$, given via (\ref{krit}), is the particular (but extremely important!) example of such macroscopic velocity. 

From a comparison expressions (\ref{cur}) and (\ref{Xin}), we find 
\be \label{vs}{\bf v}_s =(\hbar/m)\nabla \Phi.
\ee
Respectively, 
\be \label{v00}{\bf v}_0 =(\hbar/m)\nabla \Phi \approx {\bf v}_s.
\ee

\medskip Our reasoning about the operators $\hat \Xi$ and $\hat \Psi$ allows us to write 
\be \label {fixi} \hat \Psi= \hat \Xi +\hat \Psi'; \quad \hat \Psi^+= \hat \Xi ^+ +\hat \Psi'^+.
\ee
Then the "excitation" part $\hat \Psi'$ converts the state $\vert m,N>$ into states orthogonal to it, i.e.  
\bea \label{ortog} {\lim}_{N\to  \infty} <m,N\vert \hat \Psi'\vert m, N+1>=0,\nonumber \\{\lim}_{N\to  \infty} <m,N+1\vert \hat \Psi'^+\vert m, N>=0 \eea
In the limit $N\to \infty$, the difference between the states $\vert m, N>$ and $\vert m, N+1>$ disappears entirely, and, in this sense, $\Xi$ becomes the mean value of the operator $\hat \Psi$ for that state. 

\medskip This pretty long discourse proves in fact that the phase $\Phi$ entering the value $\Xi(t,{\bf r})$, (\ref{Xin}), enters also the  operator $\hat \Psi (t,{\bf r})$, (\ref{heisenberg}), for the non-ideal Bose gas in the Heisenberg structure. 

\bigskip Returning again to the "model" He$^4$ Hamiltonian (\ref{eq:three1}), we, relying onto the above investigations about the phase   $ \Phi (t,{\bf r})$, can conclude that this Hamiltonian $\widehat{\cal H} $ is manifestly invariant with respect to $U(1)$ {\it gauge} transformations 
\be
\label{preobr.fazi} 
\Phi (t,{\bf r})\to \Phi (t,{\bf r}) +  n\lambda ({\bf r}), \quad n\in {\bf Z}. \ee  
At these transformations the wave function (\ref {Xi1})  is multiplied by  stationary factors  
\be 
\label{preobr.fazi1}
v_n({\bf r})= \exp (i~n \lambda ({\bf r})).  
\ee 
 quantity 

\medskip As a next in turn consequence of our above phase   $\hat \Phi (t,{\bf r})$ of the (complex) quantity $ \Xi (t,{\bf r})$ (the latter one can be interpreted as an eigenvalue of the Heisenberg operator $\hat \Xi$, (\ref{ksju}), for the Bose condensate) is the obvious possibility to interpret the vacuum expectation value $<\vert \Xi \vert^2(t,{\bf r})>$ as the {\it order parameter} in the Landau-Bogolubov  He$^4$ model. This is in a good agreement with the above remark that $\Xi$ is actually the mean value of the operator $\hat \Psi$. 

So $< \vert \Xi\vert ^2 (t,{\bf r})>=0$ for the helium I modification \cite{Levich1} \footnote{The helium I is the "high-temperature" helium modification, turning into the solid state at low temperatures.
However at pressures lower than 30 atm., the second-order phase transition occurs in the Curie point $T_c\to 0$ with appearing  the liquid helium II modification, possessing manifest superfluid properties. \par The transparent evidence in favour of such second-order phase transition occurring is the discontinuity \cite{Levich1} in the plot of the helium heat capacity in the Curie point $T_c\to 0$.\par For instance \cite{Nels}, in the case of superfluid helium films, the experimental data result
$$ C_p(T)\equiv T(\frac{\partial S}{\partial T})\approx c_1+c_2\exp[c_3/\vert T-T_c\vert^{1/2}]    $$
for the helium heat capacity $ C_p(T)$,
with $S$ being the appropriate entropy; $c_i$ ($i=1,2,3$) are constants.}, corresponding to the initial $U(1)$ gauge symmetry of the helium Hamiltonian $\widehat{\cal H}$, while $<\vert\Xi\vert^2 (t,{\bf r})>\neq 0$ upon this symmetry is violated. The way of this violation in the model presented in this study will be us discussed below. 

\medskip Alternatively, the critical velocity ${\bf v}_0$ of superfluid motions in a helium II specimen, given via (\ref{alternativ}) (and (\ref{v00}))  also can serve as an order parameter in the helium theory. Really, $\vert {\bf v}_0\vert =0$ in the symmetrical phase, while $\vert {\bf v}_0\vert \neq 0$ when the $U(1)$ gauge symmetry of the Bogolubov Hamiltonian $\widehat{\cal H}$ \cite{Nels}, (\ref {eq:three1}), is violated. 

\bigskip  The  $U(1)$ gauge matrices (\ref {preobr.fazi1}),  bear a trace of  nontrivial topologies since they depend on integers $n$. There is some some (and at the same time profound!) analogy between these and so called {\it topological multipliers} involved \cite{LP1,LP2, rem2,Pervush2,fund,David2} at the Dirac fundamental quantization of the Minkowsian YMH model with the vacuum BPS monopole background. 

The essential part of the mentioned model  \cite{LP1,LP2, rem2,Pervush2,fund,David2} is writing down the Gauss law constrained Hamiltonian of the YMH model (that means expressing the temporal components $A_0$ of gauge fields $A$ in terms of their spatial components in the model Hamiltonian). Additionally the definite "transverse" gauge is imposed onto these spatial components\cite{rem2,Pervush2}:  
\be
\label{Gauss}
\frac {\delta W}{\delta A^a_0}=0 \Longleftrightarrow [D^2(A)]^{ac}A_{0c}= D^{ac}_i(A)\partial_0 A_{c}^i
\ee
($a=1,2,3$ are the $SU(2)$ group indices  and $i=1,2,3$ are the spatial indices)
with the covariant (Coulomb) gauge \be
\label{Aparallel}
A^{a\parallel}\sim [D^{ac}_i(\Phi ^{(0)})A_c^{i ~(0)}]=0\vert _{t=0}
\ee
(involving the YM BPS monopole background $\Phi ^{(0)}$ in the zero topological sector of the Minkowskian Higgs model with vacuum BPS monopoles). $D$ is the covariant derivative involving YM gauge modes.

In the cited literature  \cite{LP1,LP2, rem2,Pervush2,fund,David2},   the YM Gauss law constraint (\ref{Gauss}) is solved in terms of nonzero stationary initial data:\be
\label{init}
\partial _0 A_i ^c =0 \Longrightarrow  A_i ^c(t,{\bf x})=\Phi_i^{c(0)}({\bf x}):
\ee
the YM vacuum BPS monopole solutions in the Minkowskian Higgs model (in the zero topological sector of that model) is one of examples
  resolving (\ref{init}) the YM Gauss law constraint (\ref{Gauss}).
	
	The said allows to rewrite the Gauss law constraint (\ref{Gauss}) as \be
\label{Gauss2}
\partial _0 [D^{ac}_i(\Phi ^{(0)})A_c^{i~ (0)}]=0
\ee 
upon the removal temporal YM components $A_0^a$ from the left-hand side of this equation. Also, with taking account the Coulomb gauge (\ref{Aparallel}), the YM Gauss law constraint
 (\ref{Gauss2}) acquires the alternative look \cite {LP1} \be \label{Aparallel1} \partial _t A^{a\parallel}[A_c^{i(0)}(t,{\bf x})] =0\ee
over the set of (topologically trivial) vacuum YM BPS monopole solutions (\ref{init}).

As an important consequence of Eq. (\ref{Aparallel}), it is possible to work with those (topologically trivial) vacuum YM BPS solutions which are transverse in the sense  (\ref{Aparallel}) \footnote{Indeed, Eq. (\ref{Aparallel}), as that applied to the classical YM BPS ansatz \cite{BPS} is difficult to solve  because of the covariant derivative presented there.}. It is a novelty in comparison with the "classical" BPS monopole theory \cite{BPS}. 

In these our calculations we used the manifest commutativity of the operators $D$ and $\partial_0$. 

\bigskip  Indeed, the YM vacuum possesses always a nontrivial topological structure regardless the space in which the model "lives". In particular, initial data of any classical solution in the Minkowski space-time are also topologically
degenerated. Thus it is worth to investigate topologically degenerated vacuum solutions in the Minkowski
space-time in the class of functions corresponding to finite energy densities.

\bigskip

The topological degeneration of initial data means the existence of such a vacuum is stipulated by the fact that the homotopies group of all the three-dimensional
paths (loops) in the $SU(2)$ group manifold is nontrivial:
\be
\label{top11}
\pi_3 (SU(2)) =\bf Z.
\ee
We should always take account of this ”initial” topology at the analysis of the YM vacuum (independently
on the space where we study it).

But the degeneration of initial data means (in our case, for the Minkowskian BPS monopole vacuum)  that not only the classical vacua, but also all the fields: in particular, excitations ({\it multipoles}) ${\bar A}$ over such a vacuum \footnote{Following \cite{Pervush2}, a  multipole is considered as a weak perturbative part with asymptotics at the spatial infinity $$\bar A_i(t_0,\vec{x})|_{\rm asymptotics} =
 O(\frac{1}{r^{1+l}})~~~~(l > 1)~. $$} :\be  \label{multipoli}
 A_i^{ (0)}=\Phi_i^{ (0)}+ {\bar A}_i^{ (0)}, \ee 
in the transverse gauge \rm (\ref{Gauss2}) are degenerated\rm: 
\be 
\label{degeneration1}
{\hat A}^{(n)}_i= v^{(n)}({\bf x}) ({\hat A}_i^{ (0)}+ \partial _i)v^{(n)}({\bf x})^{-1},\quad v^{(n)}({\bf x})=
\exp [n\hat\Phi _0({\bf x})]. 
\ee Herewith, by analogy with (\ref{multipoli}), we can write down for a nonzero topological sector of the considered Minkowskian non-Abelian theory \cite {LP1,LP2}: 
\be
 \label{s12}
 \hat A_i^{ (n)}(t,{\bf x}) =\hat\Phi_i^{ (n)}({\bf x}) + {\hat{\bar A}}_i^{ (n)} (t,{\bf x}).
\ee We can write down explicitly  gauge transformations leading to this topological degeneration \cite {LP1,LP2}: 
\bea
\label{degeneration}
 \hat A_k^D = v^{(n)}({\bf x})T \exp \left\{\int  \limits_{t_0}^t d {\bar t}\hat A _0(\bar t,  {\bf x})\right\}\left({\hat A}_k^{(0)}+\partial_k\right ) \left[v^{(n)}({\bf x}) T \exp \left\{\int  \limits_{t_0}^t d {\bar t} \hat A _0(\bar t,{\bf x})\right\}\right]^{-1}, 
\eea
 where  the symbol $T$ stands for the time ordering of the matrices under the exponent sign. 

These YM fields got the name {\it topological Dirac variables } in the series of papers \cite{LP1,LP2, rem2,Pervush2,fund,David2}.

In the initial time
instant $t_0$ the topological degeneration of initial data comes thus to  "large" stationary matrices $v^{(n)}({\bf x})$ \footnote{The terminology "large" originates from the papers \cite {Jack}. Following  \cite {Jack}, we also shall refer to the topologically trivial matrices $v^{(0)}({\bf x})$ as to the "small" ones.}, depending on  topological numbers $n\neq 0$ and called the factors of the Gribov topological degeneration \cite{Gribov}  or simply the \it Gribov factors\rm.  \par 

\medskip Let us now analyze briefly the some properties of these multipliers. 

In the majority of  our formulas in this part,  YM fields $\hat A$ are present.

These  fields have indeed the following look in the Planckian $\hbar$, $c$ units \cite{Pervush3}: \be \label{hatka}
\hat A_\mu = g \frac {A_\mu^a\tau _a}{2i \hbar c}.  
\ee
The account of $\hbar$ and $c$ in latter Eq. is closely connected with the actual value $g/(\hbar c)$ of the strong coupling constant.

Further, following the papers \cite{LP1,rem2,Pervush1}, we introduce the matrices \be \label{dress}
U(t,{\bf x})= v({\bf x})T \exp \{\int  \limits_{t_0}^t 
[\frac {1}{D^2(\Phi^{\rm BPS })} \partial_0 D_k (\Phi^{\rm BPS }){\hat A}^k]~d\bar t ~\}. \ee

It is the manifest result extracting the temporal component of the YM field, $A_0^a$, from the Gauss law constraint (\ref{Gauss}). 

Following the work  \cite{LP1}, let us denote as $U^D[A]$ the exponential expression in (\ref{dress}); this expression may be rewritten \cite{LP1,Pervush1} as
\be
\label{UD}
U^D[A]= \exp \{\frac {1}{D^2(\Phi^{BPS })} D_k (\Phi^{BPS }) {\hat A}^k\}
\ee
over the stationary BPS monopole background 
and with the normalization $t-t_0=1$. 

We shall refer to the matrices $U^D[A]$ as  to the \it Dirac \rm "\it dressing\rm" of non-Abelian \linebreak fields (following \cite{LP1,Pervush2}). 

Meanwhile, temporal component of topological Dirac variables $\hat A^D$ would be removed (following Dirac \cite{Dir}): these grounds come to the nondynamical status of temporal components of YM fields. 

Thus \cite{David2} \be
\label{udalenie}
U(t,{\bf x}) (A_0^{(0)}+\partial_0) U^{-1}(t,{\bf x})=0.
\ee
Eq. (\ref {udalenie}) can serve for specifying {\it Dirac } matrices $ U(t,{\bf x})$.

\medskip
In order for Dirac variables (\ref{degeneration}) to be gauge invariant, it is necessary \cite{David2,Azimov} that exponential multipliers $ U(t,{\bf x})$, (\ref{dress}), entering (\ref{degeneration}), \be
\label{gauge}
{\hat A}^u_i=u(t; {\bf x})({\hat A}_i+\partial_i)u^{-1}(t,{\bf x}).\ee
The said may be written down as the transformations law for $ U(t,{\bf x})$:
\be
\label{z-n dlja U}
U(t,{\bf x})\to U_u (t,{\bf x}) =  u^{-1}(t,{\bf x}) U(t,{\bf x}).
\ee 
If the transformations law (\ref{z-n dlja U}) for matrices $ U(t,{\bf x})$ takes place, it is easy to demonstrate that (topological) Dirac variables (\ref{degeneration}) are indeed gauge invariant performing the following computations proposed in Ref. \cite{Azimov}.

Denoting matrices $ U(t,{\bf x})$ as $v[A]$, one write \cite{Azimov} (perhaps within nonlinearities of higher orders of smallness by the strong coupling constant $g$, due to (\ref{hatka})) \be \label{proverca}
 \hat A_i^D[A^{u}]= v[A]~ u^{-1} u(\hat A_i+   \partial_i) u^{-1} u ~v[A] ^{-1} = \hat A_i^D \ee
(the fact that matrices $v[A]$ and $u$ are commute with each other was utilized at the computations (\ref{proverca}); for the $U(1)\subset SU(2)$ embedding taking place in the Minkowskian Higgs theory in question, this is obviously and is not associated with additional difficulties).

\medskip The fact that the  functionals (\ref{degeneration})  are {\it transverse YM fields} follows from the possibility to approach them by transverse Dirac variables inherent in 
four-dimensional  constraint-shell QED \cite{Pervush2,Azimov,Nguen2} in the approximation \cite{Pervush2} \be 
\label{rjad} A^0_c= \frac {1}{\Delta}\partial_0 \partial_iA_{c}^i +\dots
\ee 
of the {\it linearized} YM theory (i.e. in the lowest order of the perturbation theory by the YM coupling constant $g$).

\bigskip  The next important property of (topological) Dirac variables (\ref{degeneration})  are indeed {\it manifestly Lorentz covariant}.  The arguments in favour of this fact were provided in the papers \cite{Pervush2, math}. A short historical retrospective would be in order here. Relativistic properties
of Dirac variables in gauge theories were investigated in the papers \cite{Heisenberg} (with the reference
to the unpublished note by von Neumann), and then this job was continued by I. V.
Polubarinov in his review \cite{Polubarinov}. \par 

These investigations displayed that there exist such relativistic transformations of Dirac variables that maintain transverse gauges of  fields. 
More precisely,  Dirac variables ${\hat A}^{(0)D}$ observed in a rest  reference frame $\eta_\mu^0=(1,0,0,0)$ in the Minkowski space-time  (thus $\partial_i{\hat A}^{(0)D}_i=0$), in a moving reference frame $$ \eta ^{\prime}  = {\eta^0 }+\delta^0_L{\eta_0} $$
are also transverse, but now regarding the new reference frame $\eta ^{\prime}$ \cite{Pervush2, David3}, \be \label{trp} \partial_\mu{\hat A}^{D\prime}_\mu=0.
\ee
In particular, $A_0(\eta^0) = A_0(\eta^{0\prime}) =0$, i.e.  the Dirac removal (\ref{udalenie}) \cite{David3} of temporal components of gauge fields is transferred from the rest to the moving reference frame. 
In this consideration \cite{Pervush2,Heisenberg,Polubarinov},  $\delta^0_L $  are ordinary total Lorentz transformations of coordinates, involving  appropriate transformations of fields (bosonic and fermionic).  
When one transforms fields entering the  gauge theory into Dirac variables in a rest reference frame $\eta_0$ and then goes over to a moving reference frame $\eta^\prime$, Dirac variables $\hat A^D$, $\psi^D$, $\phi^D$   are suffered relativistic transformations consisting of two therms. The first item is the response of Dirac variables onto ordinary total Lorentz transformations
of coordinates (Lorentz busts)
$$x'_k=  x_k+ \epsilon_ k t, \quad t'= t+ \epsilon_ k x_k, \quad \vert \epsilon_ k \vert \ll 1. $$ The second therm corresponds to "gauge" Lorentz transformations $\Lambda(x)$ of Dirac variables  $\hat A^D$, $\psi^D$, $\phi^D$  \cite{Pervush2, Nguen2}\footnote{It may be demonstrated \cite{Pervush2, Nguen2,Azimov,Arsen} that the  transformations turning gauge fields $A$ into Dirac variables $\hat A^D$, imply the $\psi^D= v^T({\bf x},t)\psi$ transformations for fermionic fields $\psi$ and $\phi^D= v^T({\bf x},t)\phi$ transformations for spin 0 fields (in particular, for the Higgs vacuum BPS monopole modes).}: $$  \Lambda(x)\sim \epsilon_ k \dot A^D_k (x)\Delta^{-1},    $$ 
with 
$$\frac{1}{\Delta}f(x)=-\frac{1}{4\pi}\int d^3y\frac{f(y)}{|\bf{x}-{\bf y}|}
$$
for any continuous function $ f(x)$. 

Thus any relativistic transformation for  Dirac variables may be represented as the sum of two enumerated therms. 
For instance \cite{Pervush2}, \begin{eqnarray}
\label{ltf} 
 A_{k}^{D} [ A_{i} &+& \delta_{L}^{0} A ] - A_{k}^{D} [ A ]  =  \delta_{L}^{0} A_{k}^{D} + \partial_{k} \Lambda,  \end{eqnarray} 
 \begin{eqnarray}
\label{ltf1}
 \psi^{D} [ A &+& \delta_{L}^{0} A , \psi + \delta_{L}^{0} \psi ] -  \psi^{D} [ A, \psi ] = \delta_{L}^{0} \psi^{D} + i e \Lambda  (x^{\prime}) \psi^{D}.  \end{eqnarray}  

\bigskip In (\ref{degeneration1}), $\hat\Phi _0({\bf x})$ is the so-called {\it Gribov phase}. To understand the nature of this Gribov phase, note the following. As it was demonstrated in Ref. \cite{LP2,LP1,rem2,Pervush2}, the Coulomb constraint-shell gauge keeps its look in each topological class of the Minkowskian Higgs model with vacuum
BPS monopole solutions quantized by Dirac if the Gribov phase $\hat\Phi _0({\bf x})$ entering Eqs. (\ref {degeneration1}), (\ref {degeneration}) for topological Dirac variables in that model, satisfies \it the equation of the Gribov ambiguity \rm  (or simply the \it Gribov equation\rm)  
\be
\label{Gribov.eq} [D^2 _i(\Phi _k^{(0)})]^{ab}\Phi_{(0)b} =0.
\ee 
 The origin of latter Eq. is in the standard definition of a "magnetic"  field, 
\be \label {magnet} B_i^a= \epsilon_{ijk} (\partial^j A^{ak} +\frac {g}{2}\epsilon ^{abc}A_{b}^j A_{c}^k).\ee
Really, the values $D_i A^{ia}$ (in particular, $D_i A^{i D}$ if topological Dirac variables $A^D$ are in question) have the same dimension  that a "magnetic" YM field $ B_i^a$, given via (\ref{magnet}).

Then it is easy to see that the Gribov ambiguity equation (\ref{Gribov.eq}) is the consequence of the Bogomol'nyi equation (\ref{Bog}), implicating (topologically trivial) Higgs vacuum BPS monopole modes $\Phi_{(0)}$. 

 Speaking about the connection between the Bogomol'nyi and Gribov ambiguity equations, note that this connection may be given via the Bianchi identity
\be \label{Bianchi}\epsilon ^{ijk}\nabla _i F_{jk}^b =0,\ee that is equivalent to 
$$ D B=0$$
in terms of the (vacuum) "magnetic"  field $\bf B$, (\ref{magnet}).

\bigskip Its turns out, there exists a beautiful analogy between the superfluidity as  a kind of {\it potential} motion inside a He$^4$ specimen and the Minkowskian BPS monopole YMH vacuum quantized in the Dirac Gauss-shell scheme. 

Let us recall some information from hydrodynamics. As it is well known (see e.g. \S 10 in \cite{gidrodinamika})   the well-known  continuity equation
$$ \frac{\partial \rho }{\partial t}+ {\rm div} \rho {\bf v}=0$$
  (with $\rho $ being the density of the considered liquid) is simplified in a radical way as $\rho = {\rm const}$ (i.e. when the density remains constant along the whole volume besetting by the liquid during the whole time of motion) The  condition $\rho = {\rm const}$ means  that the liquid is {\it incompressible} \cite{gidrodinamika}.. \par 
In  this case the continuity equation acquires the  simplest look 
\be
 \label{idealn}
 {\rm div}~ {\bf v}=0.
\ee

or
$$   \Delta \phi=0      $$
if ${\bf v}={\rm grad}~\phi$ for a scalar field $\phi$, i.e. if the considered liquid is {\it potential}.

Thereafter, it should be recalled \cite{rem1} that the vacuum "magnetic" field $\bf B$ plays the same role that the (critical) superfluiduty velocity ${\bf v}_0$ of the superfluid component in a liquid helium I (given via (\ref{v00})). Then one comes to the Gribov ambiguity equation (\ref {Gribov.eq}) instead Eq. $\Delta \phi=0 $ upon replacing
$$ {\bf v}_0\Leftrightarrow {\bf B}; \quad   \phi \Leftrightarrow  \Phi.              $$

\bigskip Note that  the (topological) Dirac variables  can be constructed also in another gauge models. 

So, for instance, such variables were built  for the "toy" model, the {\it two-dimensional Schwinger model} \cite{Schwinger1, Gogilidze} (the one space dimension plus the time) involving the gauge as well  field $A$. The remarkable feature of the two-dimensional Schwinger model \cite{Schwinger1, Gogilidze}  is fixing the weyl gauge $A_0=0$ for the gauge field $A$- It is interesting to note that this is dual (in a definite sense) to the gauge imposed onto the field $A$ in QED (1+1) by Coleman, Jackiw and Susskind  \cite{Col} where the gauge $A_1=0$ was chosen. 

\medskip The  two-dimensional Schwinger model \cite{Schwinger1, Gogilidze} involves the exact $U(1)\simeq S^1$ symmetry since no massive field provided therein. 

By analogy with Eq. (\ref{fact1}), we factorize the initial $U(1)$ gauge symmetry of the model \cite{Schwinger1, Gogilidze} as \be
\label{QED11} 
H_1 \equiv U(1)\simeq U_0\otimes {\bf Z}.
\ee
Really, the discrete group space of $H_1$ may be identified with the circle $S^1$ cutting by its topological sectors. Vice verse, one may glue together back the circle $S^1\simeq U(1)$ from these "peaces". Just this justifies the isomorphism (\ref{QED11}).

If $\Lambda (x)$ fis a gauge transformation of the gauge field $A$ such that
\be
\label{stationar} 
A'_1(x,t)= \exp (i \Lambda (x)) (A_1(x,t) + i \frac{\partial_1}{e}) \exp (-i \Lambda (x)), 
\ee 
with the spatial asymptotic 
\be \label{spa} \lim _{\vert x\vert \to \pm \infty} \Lambda(x)=0,\ee one can construct the {\it topologically nontrivial Weyl base element} \cite{Fock,Weyl,London1} \be \label{bas.el} 
P_1\equiv  \exp (i \Lambda (x)) 
\ee 

This implies, in a natural way, a nontrivial topology of the configuration space $\{A_1(x,t)\}$. This can be seen from the spatial  asymptotic  (\ref{spa}) for $\Lambda (x)$.  In this case we get \be
\label{ntcon} 
 \lim _{\vert x\vert \to \pm \infty} \exp (i \Lambda^{(n)} (x)) \equiv \lim _{\vert x\vert \to \pm \infty} P_1^{(n)}(x) =1, 
\ee 
or 
\be
\label{ntcon1}
 \Lambda^{(n)} (\infty)-  \Lambda^{(n)} (-\infty) =2\pi n; \quad n \in {\bf Z}.  
\ee Herewith it is convenient to pick out the subsets of positive, $n_+$, and negative, $n_-$, numbers among integers $ n \in {\bf Z}$.

By analogy with the soliton theory \cite{Al.S.,Cheng,Coleman1,Raj}, we  assume that values of the  field $A_1$ at the spatial infinity  belong to the vacuum of the discussed model. 

The map (\ref{ntcon})  associates the straight line $\Bbb R$
(the "ends" of which are identified) to  the circle  $S^1\simeq U(1)$) for which
\be
\label{ctop} 
 \pi_1 S^1= \pi_1 (U(1))= {\bf Z}.
\ee Indeed,  we should rewrite  the formula (\ref{ntcon1}) as 
\be
\label{ctop1}   \Lambda^{(n_\pm)} (\infty)-  \Lambda^{(n_\pm)} (-\infty) =2\pi n_\pm\hbar \ee
 in the Planckian  units \footnote{Weyl base elements $P_1$ take, indeed, the look \cite{Ilieva2} $$\exp (i \frac{ \Lambda (x)}{ \hbar})$$ 
in the Planckian  units.
\par 
In this case one achieves that the exponential Weyl base elements $P_1$ are really dimensionless by assuming for $ \Lambda (x)$ to have the dimension of action \cite{Pervush4}.

Moreover, such look of Weyl  base elements $P_1$, together with Eq. (\ref{ctop1}) for specifying the function $\Lambda^{(n)} (x)$,  promote   achieving the asymptotic (\ref{ntcon}) for these elements at the spatial infinity.
\par 
Herewith occurs the actual cancellation of the Planck constant $\hbar $  in expressions (\ref {ntcon}).
}. Thus  the above  nontrivial topology disappears in the classical limit \rm 
$\hbar\to 0$. \par  To satisfy the spatial asymptotic (\ref {ctop1}), it is expediently to assume \be\label{ctop11}
\Lambda ^{(n_\pm)}(x)= \hbar ~2\pi n_\pm \frac {x} {R}   
\ee
(involving the distance $R$, which we rush to infinity) at arbitrary values of the spatial coordinate $x$ \cite{Pervush4}.

The Weyl base elements  $P_1$, (\ref {bas.el}), involving functions $\Lambda^{(n)}(x)$ \cite{Pervush4}, (\ref {ctop11}), with the spatial asymptotic (\ref {ctop1}) form the vacuum manifold $R_1$ which is {\it isomorphic}, as it is easy to see, to the  gauge group $H_1$. 

\medskip  Note that the $U(1)$ gauge symmetry generated by the map (\ref{ntcon}) is indeed {\it exact} since there are no mass items in the Schwinger model \cite{Schwinger1, Gogilidze}.  Thus the isomorphism (\ref{QED11}) is rather formal, the circle $S^1$ remains {\it unbroken}. In the present study about the superfluid helium we'll see  that in order to justify the superfluidity phenomena as well as the vortices \cite{Halatnikov} inside a He$^4$ specimen, it is necessary that the circle $S^1
  \simeq  U(1)$ {\it decays into topological domains}, "numbering" by the set $\bf Z$ of integers, in such a wise that {\it domain walls} arise between them.

\bigskip We omit here the "technology"  how to construct the Dirac variables in the Schwinger model \cite{Schwinger1, Gogilidze} and send our reader to the paper \cite{Gogilidze} for studying this issue.  Note only that all the calculations in \cite{Gogilidze} are performed in the Weyl gauge $A_0=0$. Herewith it's natural to wonder what  will look like the calculations about the Dirac quantization of the Schwinger model in the dual gauge    $A_1=0$ by Coleman and other \cite{Col}. It is an interesting challenge for future research.


\bigskip Let's say a few more words about Gauss-shell 4-dimensional QED (including the electric  four-potential $A_\mu$ and the electron-positron fermionic field $\psi$). A profound analysis of this theory was performed in the papers \cite{Pervush3,Nguen2} (unfortunately, today  somewhat forgotten). In the  study \cite{abel}, this author just reminded to his readers (in Section 2 therein)  the main points of Gauss-shell 4-dimensional QCD, the emergence  there Dirac variables (which are {\it topologically trivial} this time) as a result of setting to zero the temporal component $A_0$ of the four-potential \footnote{This is again a nondynamical degree of freedom; for the same reasons  that were indicated in Introduction \cite{Gitman}.}, which can be expressed through three other (spatial) components via the Gauss law constraint.   

As it was demonstrated in \cite{abel} (this result is in a certain sense additional to those demonstrated in \cite{Pervush3}), the mentioned Dirac variables $A^{D}_i$ ($i=1,2$) can be expressed in the shape of the {\it retarding potential} (see e.g., \S62 in \cite{A.I.}) in the presence of "transverse"' fermionic currents $j^{D}$ (the $i=1,2$ components of which have the standard look \cite{A.I.} in terms of the fermionic Dirac variables $\psi^D$, $\bar \psi^D$).

The also interesting and important result is that \cite{abel,Nguen2} the temporal component of the "Dirac" current $j^D$ is related to the temporal component $ A_0^T $ of the "Dirac" time-like
three-vector  $A^D=(A_0^T, {\bf A}_i^D)$ ($i=1,2$), which, in turn, is the solution to the Poisson equation 
\be \label{dub} A_0^T=\frac{1}{\Delta} j_0^D(x).\ee

\medskip The important point of the investigations \cite{abel} is also the topic "Relativistic covariance of Dirac variables in Gauss-shell 4-dimensional QCD" (See Section 2.2. therein).

\bigskip The probably most important, in the paper \cite{abel} is Section 3 where the attempt (although far from finished) to construct the Dirac topological variables for the complete $U(1)$ gauge model. As it is well known, the nontrivial topology arises therein due to the nonzero   fundamental group
\be \label {fundament}
\pi_1 (U(1))= \pi_1 S^1= {\bf Z},
\ee
in turn specified by the radius $\vert {\bf x}\vert <\infty$ of the circle $S^1$.

It is almost obvious that QED corresponds to the trivial (unit) element of the $\pi_1 (U(1))$ group. This can be seen at least from the natural n of Maxwell equations with introducing the tensor $\tilde F_{\mu \nu}$ dual to the Maxwell stress tensor $ F_{\mu \nu}$ and the {\it magnetic current} $k_\mu$ dual to the electric current $j_\mu$. 

This dualization generates, in a natural way \cite{Ryder,Cheng}, the {\it Dirac quantization condition} \cite{Dirac}
\be \label{dq}
q_eq_m=\frac 1 2 n \hbar c; \quad n\in {\bf Z};
\ee
for the  electric, $q_e$, and magnetic, $q_m$, charges, respectively, and simultaneously the {\it Dirac monopole configuration} that is, indeed, the gauge potential $A_i$ ($i=1,2,3$) singular along the negative $z$-axis (in the Cartesian coordinates). Herewith integers $n$
play the role of topological numbers related to the topological chain (\ref{fundament}).

At $n=0$ and $q_m=0$, Eq.  (\ref{dq}), i.e. the Dirac quantization condition,   is  satisfied at {\it arbitrary} electric charge $q_e$. Thus we come to QED actually! Q.E.D.  

\bigskip. I would like to list the basic results of Ref. \cite{abel} regarding the $U(1)$ gauge theory, including constructing Dirac (topological) variables therein. 

1.  It was argued  that Dirac (topological) variables include the flux  $\Phi(r,\theta)$  (in the spherical coordinates) across the part of the sphere specefied by some values of $r$ and $\theta$ \cite{Ryder,Cheng}. It is remarkably that indeed the flux  $\Phi(r,\theta)$ depends on topological numbers $n$- In this is the principal distinction of topological multipliers in the $U(1)$ gauge model \cite{abel} from those in the  YMH gauge model with vacuum BPS monopole solutions \cite{LP1,LP2,
Pervush2, David2, David3} where such multipliers entering the Dirac topological variables include the phase $\hat \Phi_0(x)$, which is  \cite {LP1,LP2, Pervush2}
\be
 \label{phasis}
{\hat \Phi}_0(r)= -i\pi \frac {\tau ^a x_a}{r}f_{01}^{BPS}(r), \quad 
f_{01}^{BPS}(r)=[\frac{1}{\tanh (r/\epsilon)}-\frac{\epsilon}{r} ],
 \ee  i.e. it {\it does not depend} on the topological numbers $n$. 

2. In order to improve the  ultraviolet behavior of the model in question (i.e. to avoid redundant divergences in appropriate Feynman diagramms and in the total model Hamiltonian) it was proposed in \cite{abel} the specific look for the topological Dirac variables depending now explicitly on the gauge potentials $A_\mu^{(n)}$ ($\mu=0,1,2,3$ in the Minkowski space) and the topological numbers $n\in \bf Z$. In other words, in each topological class {\it their independent topological Dirac variables}. In this is the difference of the topological Dirac variables in the $U(1)$ gauge model from those in the  YMH gauge model with vacuum BPS monopole modes \cite{LP1,LP2,
Pervush2, David2, David3} where all the topological Dirac variables with various $n\in \bf Z$ depend explicitly on   the {\it topologically trivial} gauge potentials $A_\mu^{(0)}$ ($\mu=0,1,2,3$) and the Gribov topological multipliers $v^{(n)}({\bf x})$. 

3. The proof of the transversality property for the topological Dirac variables in the model \cite{abel} is an enough not easy thing and involves some cumbersome mathematics affecting
the Dirac monopole configuration \cite{Ryder, Cheng} and the "Gribov phase" $\Phi (r,\theta)$.  Also it  would be interesting to discuss the Lorentz-Poincare transformation properties of these Dirac variables, as that was done for the Dirac variables in  Gauss-shell 4-dimensional QED. But this topic remained beyond the paper \cite{abel}. 

4. The question about the temporal components   $A_0^{Dm(n)}$ (where the index $D$ stands for the Dirac variables while the index $m$ for the Dirac monopole configuration) of the topological Dirac variables is a refined  and important one simultaneously. So ruling out  the temporal components $A_0^{Dm(n)}$ with $n\neq 0$ implies the absence of electrical charges $q_e$ creating actually these modes. But then $q_m = n/q_e\to \infty$. This involves the infinite magnetic tension $\bf B$ at definite conditions\footnote{For instance, for the Dirac monopole configuration  \cite{Ryder, Cheng} the magnetic tension $\bf B$ has the radial component 

$$ {\bf B}_r\sim \frac {q_m} {r^2} \frac{1-\cos \theta}{\sin \theta}.$$  Such a magnetic tension $\bf B$ can diverge at the spatial infinity if and only if the magnetic charge $q_m$ treated as the cyclic surface integral  $$ 4\pi q_m=\oint_S {\bf B}\cdot d{\bf S}$$ i.e. the area bounded by an integration contour, increases faster than $r^2$.}. Thus this is undesirable!  Therefore, the author of \cite{abel} {\it recommends} to retain the temporal components of the Dirac variables  $A_\mu^{Dm(n)}$ with $n\neq 0$.  Spatial components of these potentials (can be treated as physical fields due to their gauge invariance) are the Dirac monopole (vacuum) configurations. Such configurations  are the patterns of {\it dyonic} solutions, possessing  both the electric and magnetic charges. 

4$'$. A something interesting occurs in the zero topological sector of the Abelian gauge model \cite{abel}.  The Dirac quantization condition (\ref{dq}) applied to the topological  number $n=0$ at $q_m=0$ results {\it arbitrary} electric charges $q_e$ \footnote{The purely electric case $q_m=0$ is electrostatic at arbitrary $q_e\neq 0$, that generates an electrostatic potential $\varphi$ due to the Poisson equation $\triangle \varphi =q_e$. In this case (see \S 46, \S 47 in \cite{Landau2}), no flat electromagnetic waves (photons) are possible. These can arise only if $q_e=0$ when the additional condition ${\rm div}{\bf A}=0$ is imposed.}, while if  $q_m=0$, we come to the {\it purely magnetic} theory. Thus  the topologically trivial "temporal" Dirac variables are $A_0^{Dm(0)}=0$ for this purely magnetic configuration. Simultaneously, $A_i^{Dm(0)}\neq0$ ($i=1,2,3$). These are specific topologically trivial Dirac monopole solutions. 

5. As a  "by-product" of the investigations \cite{abel} (namely the point 4$'$ above), it was point out that purely magnetic topologically trivial (gauge) configurations are possible, with zero temporal components of gauge potentials. These were named {\it magnons} in the paper  \cite{abel}.  The remarkable feature of such configurations is that they {\it can not interact} with photons  (due to the absence of electric charges) and {\it thus are invisible}. And this is just the pattern of {\it dark matter}. 

\bigskip  After our large excursion in the topic "topological Dirac variables in different gauge models", let us return again to the He$^4$ model Hamiltonian (\ref{eq:three1}). While there is no talk about manifest constructing topological Dirac variables in the superfluid  He$^4$ model; it is a very difficult task. On the author opinion, it's easier to do in the various superconductivity models involving manifestly gauge potentials. But the topological multipliers (\ref{preobr.fazi1}) remind those in other gauge theories. 

\medskip The important property of these topological multipliers is the following one. Let us consider the set of {\it permissible superfluid velocities} $\vert {\bf v}_s \vert \leq \vert {\bf v}_0 \vert $ of a He$^4$ specimen.  At  acting by a "large" exponential multiplier $v_n({\bf r})$, (\ref{preobr.fazi1}), onto such a permissible superfluid velocity $\vert {\bf v}_s \vert$,  this exponential multiplier transforms this velocity into another
permissible superfluid velocity   ${\bf v}'_s$:  
\be 
\label{preobr.skorosti}  {\bf v}'_s= {\bf v}_s + \frac{\hbar}{m} n \nabla \lambda ({\bf r}); \quad n\in {\bf Z}.   \ee
This gauge transformation describes correctly a "large" gauge orbit.  Thus "large" transformations (\ref{preobr.fazi1}) act as authomorphisms in this "superfluidity sector".

Indeed, one can associate the vector three-dimensional space  $\bf V$ of permissible superfluid velocities ${\bf v}_s\leq {\bf v}_0$ ($\vert {\bf v}_s \vert \neq 0$) to the above mentioned "superfluidity sector". As we shall argue further, studying the phenomenology of the superfluid helium \cite {Landau52}, this "superfluidity sector" includes the superfluid phenomena in the topologically trivial sector $n=0$ as well as thread vortices  \cite{Halatnikov} in the topologically nontrivial sectors $n\neq 0$.   This implies, in turn, the Gribov topological degeneration of trivial topologies in a helium (that are
responsible for the superfluidity therein).

\medskip It is possible (but this will be the subject of study in one of the future works)  to write down explicitly external $U(1)$ 1-forms corresponding to the Bogolubov-Landau superfluid helium theory.  Note that this constructing external $U(1)$ 1-forms allows to consider the the holonomies group ${\cal H}$ such that elements $b_\gamma \in {\cal H}$ belonging to nontrivial topologies $n\neq 0$ correspond to thread vortices  \cite{Halatnikov} while  the  $n=0$ (unit) holonomy element corresponds to the superfluid motion.  

\bigskip Above, the complex-valued helium Bose condensate, (\ref{Xi1}) 
wave function $\Xi (t,{\bf r})$ was introduced.   
It is easy to see that vov of this  wave function, $<\vert \Xi\vert ^2 (t,{\bf r})>$, can serve as the  quite naturally as the
order parameter in the Landau-Bogolubov helium theory described by the Hamiltonian (\ref{eq:three1}).  In particular, for the helium I modification  \cite{Levich1}  \footnote{The helium I is the "high-temperature" helium modification, turning into the solid state at low temperatures.
However at pressures lower than 30 atm., the second-order phase transition occurs in the Curie point $T_c\to 0$ with appearing  the liquid helium II modification, possessing manifest superfluid properties. \par
The transparent evidence in favour of such second-order phase transition occurring is the discontinuity \cite{Levich1} in the plot of the helium heat capacity in the Curie point $T_c\to 0$.\par 
For instance \cite{Nels},  in the case of superfluid helium films, the experimental data result
$$ C_p(T)\equiv T(\frac{\partial S}{\partial T})\approx c_1+c_2\exp[c_3/\vert T-T_c\vert^{1/2}]    $$
for the helium heat capacity $ C_p(T)$,
with $S$ being the appropriate entropy; $c_i$ ($i=1,2,3$) are constants.},  corresponding to the initial $U(1)$ gauge symmetry of the helium Hamiltonian  $\widehat{\cal H}$,  while $<\vert\Xi\vert^2 (t,{\bf r})>\neq 0$ upon this symmetry is violated (in the "discrete" wise, as it will be discussed below) corresponding to the helium II modification. 

\medskip Alternatively, the critical velocity ${\bf v}_0$  of superfluid motions in a helium II specimen  also can serve as an order parameter in the helium theory.
Really, $\vert {\bf v}_0\vert =0$ in the symmetrical phase, while $\vert {\bf v}_0\vert \neq 0$ when the $U(1)$ gauge symmetry of the Bogolubov Hamiltonian $\widehat{\cal H}$  is violated. 

It is worth to remember, in this context, the connection between $\vert {\bf v}_0\vert$ and $<\vert\Xi\vert (t,{\bf r})>$ which is given actually via  (\ref{cur}).

\bigskip  Thus now it's time, in the our study, to get started studying the phenomenology of the superfluid helium specimen contained in a turning vessel. We'll follow up the appearance of {\it global} {\it rigid} rectilinear vortex threads \cite{Landau52} inside such  superfluid helium specimen around which rotary motions of the liquid He occur and the potential nature of a superfluid liquid is broken down. It will be argued that these rigid rotations form the group $O(2)$.  Besides that we will show that the {\it spontaneous} break down of superfluidity takes place with appearance of {\it local} vortices \cite{Halatnikov} possessing circulation velocities ${\bf v}^{(n)}$ ($n\in {\bf Z}$) in each topological sector of the He$^4$ model. Note that these  circulation velocities are closely connected with the  holonomies elements $b_\gamma \in {\cal H}$ discussed above. This presence of global as well as local vortices in the model discussed will serve as a basis for us for constructing the helium-four topological theory. It is just the principal goal of the present study. 

\subsection{The phenomenology of of the superfluid helium specimen contained in a turning vessel. Global and local vortices.}
Let us imagine, following \cite{Landau52},  that an {\it ordinary} liquid   is contained in a cylindrical vessel turning around its axis $z$. 
 Friction forces between such liquid and the walls of the vessel lead, finally, to its rotation  as a whole  together with the vessel. 

Vice versa, in a  {\it superfluid} liquid only the normal part of the liquid, corresponding to the excitations spectrum, is carried away with the turning vessel, while the superfluid component remains at  rest according to the potentiality condition of the  (\ref{potcon}) type.

 At a rotation of a liquid as a whole  its velocity is  ${\bf  v}=[\vec \Omega {\bf  n}]$,  with  $\vec \Omega$ being the angular velocity of the rotation, while the unit vector ${\bf  n}$ is directed outward, perpendicular to the surface of the liquid.  Herewith $${\rm rot}~{\bf  v}= 2 ~\vec \Omega \neq 0.$$   
	However at large values of $\bf v$, the state when the superfluid component is at  rest is not advantageous from the thermodynamic  point of view. 
  The condition for the thermodynamic equilibrium means for a turning system that the value 
	\be 
\label{vrasch}
 E_{\rm rot}=E-{\bf  M}\vec \Omega,
\ee  
where $E$ and $\bf  M$ are, respectively, the energy and rotary momentum  of the  system with respect to the chosen rest reference frame, attains its minimum. \par
The item  $-{\bf  M}\vec \Omega$  in  latter Eq. implies (at large values of  $\vec \Omega$)  that a state with ${\bf  M}\vec \Omega>0$ is  more advantageous than an one with ${\bf  M}=0$. As the rotation velocity of the vessel increases, a rotary motion of the superfluid component arises finally.   \par 
The seeming contradiction between this assertion and the potential nature of a superfluid liquid is removed if one will think that  the potential nature of a superfluid liquid is broken down along especial lines in the liquid called {\it  the vortex threads} \cite{Landau52}  \footnote{This can be demonstrated mathematically. Really, if the critical velocity ${\bf v}_0$ of the superfluid motion in a helium II is different from zero in the given laboratory reference frame, then at going over to the rotating reference frame,
$$ {\bf v}_0 \to {\bf v}_0+ [\vec \Omega {\bf  n}].$$ The curl from  r.h.s. of the latter Eq., $2\vec \Omega$, is different from zero, while the ${\rm rot}~{\bf v}_0 =0$.  Thus one can distinguish the areas of superfluidity and vortices in a  helium II specimen considered in the rotating reference frame  on the basis of this simple mathematics.}.

 Around such lines   rotary motions of the liquid occur, one calls them \it  potential rotations \rm \cite{Landau52},  while ${\rm rot}~{\bf  v}_0=0$ in the remaining  part of the spatial volume. 
 The latter assertion was made in 1949 y. by 
Onsager and later, in 1955 y. was developed further by Feynman and Hall \cite{Fey}.  Vortex threads in a liquid have usually thickness of atomic distances; from the macroscopic point of view, they are considered as infinitely thin.   \par 

heir existence does not contradict Eq. (\ref{alternativ})  for ${\bf v}_0$ since this expression envisages an enough slow change of ${\bf  v}_0$ in the spatial volume (with the behaviour $O(1/r^{n})$, $n>0$, at the spatial infinity, as we shall make sure below \cite{Landau52}), while in an infinitely small neighborhood of a  vortex thread ${\bf  v}_0$  it may change anyhow quickly.\par 
 As a rule, the presence of vortex threads in a liquid  yields their macroscopic and herewith enough large contribution in the helium Hamiltonian; then the state  of the liquid involving vortex threads cannot be treated as a quantum perturbation. \par \medskip
Let us now consider vortex threads  from the kinematic  point of view. They can be treated   as  singularity lines in the distribution of the critical motion velocity ${\bf  v}_0$ for a liquid helium II specimen,  provoked by  potential rotations of this liquid together with the cylindrical vessel where it is contained.

Each vortex thread can be characterized by the definite value (we shall denote it as $2\pi \kappa$) of the circulation of the velocity vector ${\bf  v}$ along a closed way around the given vortex thread: 
\be
 \label{circulation}
\oint {\bf  v} d~ {\bf  l} = 2\pi \kappa.
\ee
  This  value does not depend on a choice of the integration way.
	
	According to the Stokes formula, this integral is equal  to the flux of ${\rm rot}~{\bf  v}$ through the surface restricted by the closed contour in the integral (\ref{circulation}).  As ${\rm rot}~{\bf  v}=0$, the integral (\ref{circulation}) vanishes.

The vector $\bf v$ is always directed along tangential lines of horizontal circles with centres lain on the axis $z$ (see e.g. Fig. 8 in \cite{Al.S.}).\par

Vortex threads cannot be interrupted inside the liquid: they can be either closed or finished at the boundary of the considered liquid (if this  liquid is infinite, then  vortex threads are also infinitely long) \footnote{More exactly, at the boundary, the fluid velocity must match the no-slip condition. This means that the fluid particles at the boundary have zero velocity relative to the boundary.  Consequently, the vortex thread experiences a shearing effect near the boundary. The fluid particles near the wall slow down, affecting the vortex thread’s behavior.

As the vortex thread approaches the boundary, it faces two possibilities:

1. Closure: The vortex thread can loop back on itself, forming a closed path. This closure occurs when the fluid motion near the boundary opposes the vortex’s rotation.

2. Dissipation: Alternatively, the vortex thread can dissipate. This happens when the shearing effect near the boundary disrupts the coherent motion, causing the vortex to unravel.

In either case, the vortex thread does not abruptly vanish; it either closes upon itself or gradually dissipates.

}.\par

The condition (\ref{circulation})   allows  to find the distribution of the rotary velocity.
  In the  simplest case of rectilinear vortex threads in an infinitely stretched liquid, lines of force are circles, and the rotary velocity  vector $\bf v$ is directed  along their tangential lines.  These circles lie in planes perpendicular to the given vortex thread (see e.g. Fig. 5 in \cite{Al.S.}). \medskip
	
	The theorem about the average value applied  to the integral (\ref{circulation})   results $2 \pi r v =2\pi r \vert {\bf v}\vert$ on its left hand; whence 
\be
\label{ots}
v= \frac{\kappa}{r}, \ee 
with $r$ being the distance from the vortex thread. 
\par
Note that the rotary velocity $\bf v$ decreases with  increasing  the distance from the rotation axis (vortex thread).  \par
For a vortex thread of an arbitrary shape,  the rotary velocity $\bf v$ can be expressed by means of Eq.
 \be 
\label{Bio}
 {\bf v}= \frac{\kappa}{2} \int \frac{[d{\bf L}\cdot {\bf R}]}{R^3}, 
\ee
  with $d{\bf L}$  being the element of the vortex thread and ${\bf R}$  being the radius-vector drawing from this element in the observing point for the rotary velocity $\bf v$.   This expression remind us the well-known \it Biot-Savart formula \rm for the magnetic field  created by a linear current. \par
	The formal agreement of the  both theories can be achieved at comparing the circulation  of the rotation velocity (\ref{circulation})  with the one of the magnetic field $\vec {\cal H}$ created by a linear current $J$: 
$$ \oint \vec {\cal H} d{\bf l}=4\pi J/c.$$ Then at the replacements  
\be 
\label{repl}
\vec {\cal H} \Longrightarrow {\bf v}; \quad J/c \Longrightarrow  \kappa /2
\ee
one comes to the  theory (\ref{Bio}), (\ref {ots}) \cite{Landau52}.\par
From Eq.  (\ref{Bio}) we see that the rotary velocity vector $\bf v$   is orthogonal to the element $d{\bf L}$ of the vortex thread, while the critical 
velocity vector ${\bf v}_0$ (characterized  potential motions  of the superfluid component in the helium II) is \it colinear \rm to the vector $d{\bf L}$,  
so long as the superfluid liquid moves along the walls of the (narrow) cylindrical vessel where it is contained, i.e. parallel to the rotation axis $z$.  Thus the vectors $\bf v$  and ${\bf v}_0$ are, indeed, \it mutually orthogonal\rm, while ${\rm rot}~{\bf v}=2\vec \Omega$ is colinear to ${\bf v}_0$. \par  
 One can think herewith that  
$$ d{\bf L}\sim d {\bf v}_0 ~dt.$$
This allows  to find the relation between the elements $d v$ and 
$ d v_0$.  From (\ref{Bio}) we deduce 
\be 
\label {diff.sootn}
 dv= \frac{\kappa}{2}~ dL\cdot R^{-2} \sin \phi \sim \frac{\kappa}{2}~ d {v}_0 ~dt \cdot R^{-2} \sin \phi,
\ee
where $\phi$ is the angle between the radius-vector $\bf R$ and  element $d{\bf L}$ of the vortex thread. 

 As $\phi=\pi/2$ ($\sin \phi =1$), the investigated element $d{\bf L}$ of the vortex thread intersects under the right angle the plane where the circle lies and to which the vector ${\bf v}$ is tangential. In this case $ v= 
v_0$ if $R=1$ and  
\be 
\label{trebov}  \frac{\kappa}{2}\int \limits _{t_1}^{t_2} dt =1.
 \ee  
Such normalization of the velocities is very convenient. It allows, without difficulties, to pass from  ${\bf v}$ to ${\bf v}_0$  in the calculations about vortex threads: e.g. when a scalar product: as, for instance, in  (\ref {circulation}),  is in question. \par\medskip

As we have seen above, these vectors are mutually orthogonal.  One  deals in this case  with two thermodynamic phases inside  the liquid helium II \cite{disc}. 

The first one corresponds to  potential motions of the superfluid component (with the velocity does not exceeding ${\bf v}_0$ \cite{Landau,Levich1})  parallel to the rotation axis $z$, i.e. to the wall of the cylindrical vessel where the specimen is contained.  
The second one corresponds to  potential rotations around a vortex thread (which is parallel to the axis $z$) with the rotary velocity ${\bf v}$\rm. \par  
The gap between the both vectors (these, as it is easy to see, can be treated as the order parameters characterized the above thermodynamic phases) testifies to the  first-order phase transition between the above phases inside  the liquid helium II at the temperature $T=0$\rm.  The calculations below \cite{Landau52} will confirm this fact. 
\par 
\medskip  At distances from the given vortex thread in the liquid helium II small in  comparison with its curvature radius, Eq. (\ref{Bio})  is reduced to Eq. (\ref{ots}). The both Eqs. (\ref{Bio}) and (\ref{ots})  point out the singular behaviour of ${v}=v_0$ at the origin of coordinates.  
This becomes possible only due to the ${\rm rot}~{\bf v}$ being different from zero,  i.e. when the potential nature of the superfluid liquid is violated and vortex threads appear\rm.  In fact,  the singular behavior of $\bf v$ at the origin of coordinates can be treated as the sufficient condition for this. \par

\medskip    
  The quantum nature of vortex threads in a superfluid liquid implies that the constant $\kappa$ takes only discrete set of its values. Really, if one utilizes Eq. (\ref{alternativ}) \cite{Landau52}, specifying the critical  superfluidity velocity  ${\bf v}_0$ through the phase $\Phi$ of the  Bose condensate wave function, one finds the circulation of ${\bf v}$ (equal to the circulation of ${\bf v}_0$ in the us assumed normalization $v=v_0$):
\be \label{oint}
\oint {\bf v}_0 d{\bf l}= \frac{\hbar}{m}\oint \nabla \Phi d{\bf l}= \frac{\hbar}{m}\Delta \Phi, 
\ee 
with $\Delta \Phi$ being the change  of the phase $\Phi$ at a complete round about the  integration contour. \par
Since the wave function (\ref {Xi1})  of the  Bose condensate is  completely specified, the change  of the phase $\Phi$ at a complete round of the  integration contour can take only integer multiples of $2\pi$. Therefore
\be \label{kappa}
\kappa = n \hbar/m, \quad n\in {\bf Z}. \ee
At substituting the latter expression for $\kappa$, by taking manifestly the account of topologies $n$, into the circulation integral (\ref{circulation}), one gets
  \be 
\label{zirk1}
 \oint  {\bf v} d{\bf l}= 2\pi n \frac {\hbar}{m }; \quad n\in {\bf Z}. 
\ee
We learn from  latter Eq. that  any potential rotation of the liquid helium II together with the cylindrical vessel where it is contained is a quantum effect directly proportional to the Planck constant \rm $\hbar$. 
 It is easy to see that   Eq. (\ref{zirk1})  implies  the topological dependence of the rotary velocity $\bf v$. \par 
This is an example which is very important for us. We shall give mathematical grounding  the expression (\ref{kappa}) later on.

Note however that topologies $n>1$ are suppressed in the theory of superfluid liquid (we shall  show this also below), while  circulations with $n=1$ ($\kappa =  \hbar/m$) are thermodynamically advantageous, i.e. stable. \par 
\medskip  Now we are able to calculate \cite{Landau52, disc}  the change  of the energy of the liquid in the presence of a vortex thread. 
 From the axial symmetry reasons it is obvious that 
\be 
\label{vklad}
 \Delta E=\int \frac{\rho_s v_0^2}{2} dV= \frac{\rho_s}{2} L \int v_0^2\cdot 2\pi r dr= L \rho_s \pi \kappa ^2 \int \frac{dr}{r} 
\ee  
(here $L$ is  the length of the vessel; $\rho_s$ is  the density of the superfluid component in the vessel). \par 
The integration by $dr$ would be carried out here in the interval between the radius $R$ of the vessel and an arbitrary value $r\sim a$ of the atomic distances order. 
At  values lesser than $a$, the macroscopic treatment of 
vortex threads loses its sense. \par 
Because of the logarithmic divergence of the integral  (\ref{vklad}), its value is few "sensitive"  to  $a$ at the down integration limit. \par 
Thus    
\be \label{vklad1}
\Delta E = L \pi \rho_s \frac{\hbar ^2}{m^2} \ln \frac{R}{a}
\ee 
as $n=1$: the latter expression has the so-called logarithmic exactness, i.e. one treats as large both the ratio $R/a$ and its logarithm . \par

One neglects, in the considered theory \cite{Landau52},  motions of   vortex threads, implying, as a rule,   changes in the density $\rho_s$ of the superfluid component. It is justly if and only if the main contribution in the energy integral  (\ref{vklad1}) occurs at large distances $r$, at which changes of the density $\rho_s$ found to be too small.  For the same reason one neglects also changes in the interior energy of the superfluid component in comparison with $\Delta E$. \par \medskip
The just calculated change (gap) in the liquid helium II energy, due to the appearance of  vortices in this liquid,  is the second evidence \cite{disc} \rm (along with the gap between the directions of ${\bf v}_0$ and ${\bf v}$; these indeed are mutually orthogonal, as we have discussed above)  in favor  of the 
first-order phase transition occurring in this case\rm.  \par\medskip
 The rotary momentum of the  liquid helium II turning  proves to be \cite{Landau52}
\be
\label{moment vr} M= \int \rho_s v_0 ~r dV= \rho_s \kappa  \int dV \sim L\pi R^2 \frac{\hbar}{m} \rho_s \quad {\rm for}~ n=1. 
\ee
\medskip
The appearance of   vortex threads becomes thermodynamically advantageously if 
 \be 
\label{nerav1}  \Delta E_{\rm rot}= \Delta E -M\Omega <0, \ee 
i.e. if    
\be
\label{nerav}
 \Omega >\Omega_{\rm cr}= \frac{\hbar}{mR^2}\ln \frac{R}{a}. 
\ee  
When $n>1$, vortex threads become thermodynamically unstable since one would, in this case, multiply the energy $\Delta E  \propto \kappa^2$ by the factor $n^2$ at the given $n$, while the rotary momentum $M \propto \kappa$ is  multiplied herewith only by $n$. Then $\Delta E_{\rm rot}$    increases rapidly.\par
At  further increasing  the rotary velocity $\vec \Omega$ of the cylindrical vessel,  when it exceeds the critical value (\ref{nerav}),  new vortex threads appear; as $\Omega \gg \Omega_{\rm cr}$, the number of vortex threads becomes very large. 

In this case their distribution in a cross section of the vessel 
goes  to the uniform one, and, as the  supreme case, implicates    rotations of the liquid helium II {\it as a whole}.  
 One can make sure of this by noting that the number of vortex threads increases directly proportional to $\Omega$, as we shall show below. \par 
Then the second item in (\ref{nerav1}),   $-M \Omega$, increases as $\Omega^2$, while the first item in (\ref{nerav1}) increases only as $\Omega$.  
Thus one can neglect it in comparison with the second one. \par 
By specifying the minimum of $\Delta E_{\rm rot}$, one specifies  simultaneously the maximum of  $M$. The latter one is attained just at   rotations of the liquid (helium II) as a whole.\par 
The number of vortex threads at a fixed (and herewith very large) value of $\Omega$ can be specified   easily at claiming \cite{Landau52} that the circulation of the velocity vector $\bf v$ corresponds to   rotations of the liquid as a whole at the large number of vortex threads. \par 
At  choosing (at such assuming) the integration way in (\ref {circulation})  to encircle an unit of area in the plane perpendicular to the rotation axis, 
\be 
\label{whole}
 \oint {\bf v} d {\bf l}= \nu \cdot 2\pi \kappa =2\pi \nu \frac{\hbar}{m}, \quad n=1.
\ee
with $\nu$ being the density of the vortex threads distribution in the cross section  of the vessel. \par 
On the other hand,  since
$${\rm rot}~{\bf v}= 2\vec \Omega, $$
the circulation  integral (\ref {circulation}) is  equal to $2\Omega$ for the unit of  area in the plane perpendicular to the rotation axis. \par 
Comparing the both expressions for the circulation  of ${\bf v}$, we find 
\be 
\label{plotn} 
\nu = m\Omega/ \pi \hbar \Leftrightarrow 1/m =\Omega/(\nu \pi \hbar). \ee 
The latter relation just proves the above assertion that $M\Omega$ involving $m^{-1}$ is  directly proportional to $\Omega^2$. 

Concerning  $\Delta E$, it is also directly proportional to $\Omega^2$ (as that involving $m^{-2}$) but contains besides that the multiplier $\hbar^2$, i.e. it is a value of the highest order of smallness.

\medskip {\it The appearance of vortex threads violates, in a definite sense, the superfluidity}.  
Elementary excitations, forming the normal component in  a helium II, now are distributed among vortex threads and impart  them (and therefore also the superfluid component) a deal of their momentum. 
In other words, this means  that effective dissipative friction forces appear between both the mentioned components\rm.
In general, vortex threads move with the flux  of the liquid. \par
At $T=0$, when the liquid helium II is completely superfluid, each of elements $ d {\bf l}$ moves  with the velocity ${\bf v}_s$ (tangential to lines of force). 
If  $T\neq 0$, friction forces between the  superfluid and normal components  (whose cause is the distribution of excitation quanta among vortex threads) prohibit moving off the superfluid component.\par  \bigskip  From the thermodynamic point of view, this us calculated increment $\Delta E$ of the liguid helium energy in the presence of a vortex thread, a manifest rotary effect, can be treated as the {\it latent heat}. It is just the sign of the first-order phase transition taking place in the liquid helium  II turning model \cite{Landau52}. The essence of this first-order phase transition is in coexisting two thermodynamic phases: of collective solid potential rotations and superfluid motions setting by the Bogolubov Hamiltonian $\widehat{\cal H}$.  We consider our theory in the limit $T\to 0$ for the absolute temperature of environment.  In this limit, all as if freezing. We will speak further that {\it frozen transition of the first order} occurs in the liquid helium  II turning model \cite{Landau52}.

As to alone superfluid motions inside a liquid helium II specimen, represent the {\it stable state} corresponding to the {\it second-order} phase transition occurred therein and coming to violating the $U(1)$ gauge symmetry of the helium Hamiltonian $\widehat{\cal H}$ \cite{Nels} (we discussed this above).

\bigskip  Now let us consider the case of a rested helium ($\rm He^4$) specimen. It is described just by the model Hamiltonian (\ref{eq:three1}). As we have emphasized repeatedly in the present study, 
the potentiality of  superfluid motions in a liquid helium is expressed by the condition \cite{ Landau,Levich1,gidrodinamika} $$ {\rm rot}~ {\bf v}_0=0,$$
with the critical velocity ${\bf v}_0$ of the potential  superfluid motion given via (\ref{alternativ}) \cite{Landau52}. \par
Due to the Stokes theorem, this condition can be written  down as the equality to zero of the circulation of ${\bf v}_0$ along an arbitrary contour $\Sigma$:
\be 
\label{zc}
\oint \limits_{\Sigma} {\bf v}_0 d{\bf l}=0. 
\ee 
At multiplying the latter relation left by the   mass $m$ of a helium atom, 
one gets, on the left hand side of  latter Eq., the value that is the subject of the quantization procedure in quantum mechanics. \par
In connection with this fact, in the monograph \cite{Halatnikov}  there was proposed to consider Eq. (\ref{zc}) {\it as a particular case} of the more general  quantum 
condition \rm   
\be \label{zirk}
m ~ \oint \limits_{\Sigma} {\bf v}^{(n)} d{\bf l}= 2\pi n \hbar; \quad n\in {\bf Z}.
\ee 
Additionally, linear velocities vectors $ {\bf v}^{(n)} $ can be chosen (for instance, by means of a Galilean transform) in such a way that $\vert{\bf v}^{(n)}\vert=\vert {\bf v}_0\vert $ independently on a concrete topological charge $n\in{\bf Z}$.
It is the same normalization  we have utilized above at the analysis of rectilinear threads in the liquid helium II turning model \cite{Landau52}.  

Thus we come again to Eq. (\ref{zirk1}) \cite{Landau52} for vortices, but this time in the helium considered in the \it rest reference frame\rm. We see thus that {\it the superfluid potential motion in the helium is only a particular case of the more general theory}. \par
\medskip Comparing Eq. (\ref{zirk}) with the Biot-Savart formula (\ref {Bio}) \cite{Landau52}  for the circulation velocity $\bf v$ (the both Eqs. describe the same phenomenology), we see  that the critical velocity $ {\bf v}_0$ of superfluid motions inside the given helium II specimen has its singularities in infinitesimal thin neighbourhoods of (rectilinear) threads in the helium (at rest) where the superfluid properties of the helium disappear. Thus a helium Bose condensate is indeed \it singular\rm, and  this singularity is the cause of thread topological defects taking place even in the case when one considers the helium in a rest reference frame.
\par
\bigskip
The concrete nature of singularities in a helium Bose condensate located in a rest reference frame was investigated in Ref. \cite{Nels}.\par
The principal result was got in Ref. \cite{Nels}  concerning singularities in a helium Bose condensate (it was reproduced also in \cite{disc})  is that the order parameter of the helium theory (its role is played by $< \vert \Xi \vert ^2 (t,{\bf r})>$, the vacuum expectation value of the helium Bose condensate wave function $ \Xi$, (\ref {Xi1})) suffers  logarithmic discontinuities in the points ${\bf r}={\bf r}_\alpha$ (on the plane $z=0$) where (rectilinear) quantum  vortices are located in a liquid helium II specimen at rest. Such  discontinuities in plots of order parameters is the sign of  
first-order phase transitions occurring in physical theories \cite{disc}.  In particular, this is true for the case \cite{Halatnikov} of quantum  vortices  in a liquid helium II specimen at rest. \par

To prove the latter statement, the correlation (Green) function
$$  G(r)\equiv<\Xi(t,{\bf r}),\Xi^*(t,{0})>, $$
different from zero for the liquid helium II modification, was considered in \cite{Nels}.\par 
It can be recast to the look \cite{Nels}
\begin{eqnarray}
G(r)&=&\langle\Xi(t,{\bf r}),\Xi^*(t,0)\rangle\nonumber \\
&=& \Xi_0^2\exp\left[-{1\over 2}\langle[\Phi(t,{\bf r})-
\Phi(t,0)]^2\rangle\right],
\label{eq:seventeen}
\end{eqnarray}
where the last line is a general property of Gaussian fluctuations. 

 To a first approximation, one can neglect fluctuations in the amplitude of $\Xi$, setting it to be a $\Xi_0$.
In this case the vacuum expectation value $< \vert \Xi \vert ^2 (t,{\bf r})>$ of the helium Bose condensate wave function $ \Xi$ will equal to $G(0)$, that is the constant value $(\Xi_0)^2$ to a first approximation. \par
\medskip
On the other hand,  the above discussed expression (\ref{zirk}),  \cite{Halatnikov},  allowing to express  topological indices $n\in {\bf Z}$ via circular velocities ${\bf v}^{(n)}$, is mathematically equivalent to  \cite{Nels}  \begin{equation}
\frac 1 {2\pi }\oint_{\Gamma}\nabla \Phi ~ d{\bf l}=
\int_\Omega d^2r\;n_v({\bf r}),
\label{omeg}
\end{equation}
with the vortex "charge density" given as \begin{equation}
n_v({\bf r})=\sum_{\alpha=1}^N\;
n_\alpha\delta({\bf r}-{\bf r}_\alpha)
\label{eq:thirtyone}
\end{equation}
for a collection $N$ of vortices
located at positions $\{ {\bf r}_\alpha \}$ with integer charges $\{n_\alpha\}$.  Herewith the contour $\Gamma$ enclosing many "elementary" vortices such that
 \be
\oint\;\Phi\cdot d {\bf l}=
2\pi s_\alpha\;.
\label{twentynine}
\ee  with $s_\alpha=\pm 1$.  Then we can replace the sum of "elementary" vortices by an one vortex with the fixed topology $N\in \bf Z$. 

Without loss of generality, it can be set $z=0$  in all further calculations about $ \nabla \Phi$ \cite{Nels}.  In (\ref{omeg}), Eq. (\ref{alternativ}) \cite{Landau52, rem1}, with appropriate replacing $ {\bf v}_0\to{\bf v}^{(n)}$, was utilized.

Applying  then the Stokes formula to   (\ref{omeg}),  one find upon integrating:
\begin{eqnarray}
\epsilon_{ij}\partial_i\partial_j\Phi (t,{\bf r})&=&
\partial_x\partial_y\Phi-\partial_y\partial_x\Phi\nonumber \\
&\approx& n_v({\bf r}).
\label{eq:thirtytwo}
\end{eqnarray}
To cast this Eq. in a
more familiar form, one can introduce \cite{Nels} the value $\tilde \Phi(t,{\bf r})$ dual to $\Phi(t,{\bf r})$:
\begin{equation}
\partial_i\Phi(t,{\bf r})=\epsilon_{ij}\partial_j
\tilde\Phi(t,{\bf r}).
\label{eq:thirtythree}
\end{equation}
Then
\begin{equation}
\nabla^2\tilde \Phi(t,{\bf r})=n_v({\bf r}).
\label{eq:thirtyfour}
\end{equation}
In particular,
\be 
\label{trivili}
\nabla^2\tilde \Phi(t,{\bf r})=0
\ee 
in the case when vortices are absent, i.e. in the case of purely superfluid motions  inside a liquid helium II specimen.

Eq. (\ref{eq:thirtyfour}) is the particular case of the Poisson equation. Thus it permits its
\begin{equation}
\tilde\Phi(t,{\bf r})=\sum_{\alpha}\;n_\alpha G({\bf r},{\bf r}_\alpha)
\label{eq:thirtyfive}
\end{equation}
solution, where the Green function $ G({\bf r},{\bf r}_\alpha)$  (defined as on the left-hand side of Eq. (\ref {eq:seventeen})) satisfies
\begin{equation}
\nabla^2 G({\bf r},{\bf r}_\alpha)=\delta({\bf r}-{\bf r}_\alpha).
\label{eq:thirtysix}
\end{equation}
For $\vert{\bf r}-{\bf r}_\alpha \vert $ large enough and both the points far from any boundaries, the Green function $ G({\bf r},{\bf r}_\alpha)$  has the look (see \S11.8 in \cite{V.S.Vladimirov})\begin{equation}
G({\bf r},{\bf r}_j)\approx{1\over 2\pi}\ln\left(
{|{\bf r}-{\bf r}_j|\over
\xi_0}\right)+C
\label{eq:thirtyseven}
\end{equation}
on the  $z=0$ plane,
where $C$ is a constant  contributing to the vortex core energy.

The origin of the parameter $\xi_0$ entering Eq. (\ref{eq:thirtyseven}) is following \cite{Nels}.  It is the characteristic length related to the coefficients of the first two terms of Eq. \cite{Landau52}
\begin{equation}
{F\over T}=
\int\; d^2r\left[{1\over 2}\;A|\nabla~\Xi|^2+
{1\over 2}\;a|\Xi|^2+b|\Xi|^4+
\cdots\right]
\label{eq:eleven}
\end{equation}
for the helium free energy $F$.

In this case $\xi_0$ can be defined as \cite{Nels} 
\be \label{xio}
\xi_0=\sqrt{A/\vert a\vert },
\ee
with $a$ given as $a=a'(T-T_c)$.

As it follows from (\ref{eq:thirtyseven}),  at setting $C=0$, the Green function $ G({\bf r},{\bf r}_j)$ diverges logarithmically when  $\vert{\bf r}-{\bf r}_j \vert \to\infty$, while it  approaches zero when 
$$ \vert{\bf r}-{\bf r}_j \vert \to\xi_0.$$
Additionally, $ G({\bf r},{\bf r}_j)$ diverges logarithmically also at $\vert{\bf r}-{\bf r}_j \vert \to 0$. The latter property of $ G({\bf r},{\bf r}_j)$, as we shall se soon, is very important.

\medskip The just performed analysis of  $ G({\bf r},{\bf r}_j)$  allows to draw the series important conclusions about the phase $\tilde\Phi(t,{\bf r})$ of the helium wave function  $\Xi$ (taking over the complete collection $N$ of topologies: including the trivial one $s_\alpha=0$, corresponding to superfluid motions \cite{Landau,Halatnikov},  So, due to (\ref {eq:seventeen}), 
\be \label{srednee}
<\vert (\Xi) \vert ^{2}(t,{\bf r})> \equiv <\Xi (t,{\bf r}),\Xi^*(t,{\bf r})>\approx G(0)=1/N \sum\limits _\alpha < \vert \Xi^{2} \vert _\alpha (t,{\bf r}) >,\ee
serving the order parameter in the helium  theory, diverges logarithmically at each index $\alpha (n)$ ($n\in{\bf Z}$) in  the same (\ref{eq:thirtyseven}) \cite{Nels}  sense that the Green function $ G({\bf r}_\alpha,{\bf r}_\alpha)
\approx G(0)$, i.e. in the points $\{{\bf r}_\alpha \}$ where quantum  vortices are located in a liquid helium II specimen at rest \footnote{In (\ref {srednee}) the Green function $G(0)$ is different from that given via Eq. (\ref {eq:seventeen})  involving the constant amplitude $\Xi_0$. 
Now the amplitude of $\Xi (t,{\bf r})$ wouldn't be constant, and this implies that $G(0)$ has the nontrivial look $ G({\bf r}_\alpha,{\bf r}_\alpha)$ \cite{Nels}, (\ref {eq:thirtyseven}).}

This is associated immediately with (rectilinear) quantum vortices arising spontaneously \cite{Halatnikov} in a liquid helium II specimen at rest. 
Herewith  also the phase $\tilde \Phi(t,{\bf r})$  of the helium wave function  $\Xi$ possesses the same behaviour that the Green function $ G({\bf r},{\bf r}_j)$ (it is correctly due to (\ref{eq:thirtyfive})). \par
The said is true for nontrivial vortices topologies $n\neq 0$ due to  (\ref {eq:thirtyfive}). However for $n=0$, i.e. in the "superfluid case", $\tilde \Phi(t,{\bf r})$ becomes an uncertain value when $\vert{\bf r}-{\bf r}_j \vert \to\infty$ (or when $\vert{\bf r}-{\bf r}_j \vert \to 0$).\par 
\medskip

Summarizing up, now we can assert that in the presence of (rectilinear) quantum vortices \cite{Halatnikov} in a liquid helium II specimen, the order parameter
$<\vert \Xi\vert ^2(t,{\bf r})>$ in the helium model suffers discontinuities of the logarithmic nature in the points $\{{\bf r}_\alpha \}$ in which these (rectilinear) quantum vortices are located (the long distances logarithmic singularities of $ G(0)$ are  not associated, probably, with quantum vortices). \par 
The discovered discontinuity (of the logarithmic nature) in the plot of the vacuum expectation value $<\vert \Xi\vert ^2 (t,{\bf r})>$ for the helium Bose condensate wave function $ \Xi$, serving the order parameter in the helium theory, just testifies in favor of first-order phase transition occurring in a liquid helium II rested specimen with arising therein quantum vortices. And once again, as in the liquid helium turning model \cite{Landau52}, one can speak here about the frozen first-order phase transition in the $T\to 0$ limit!
\bigskip

In the monograph \cite{Halatnikov} (in \S 31)  there was demonstrated, by means of the variation  of the value  $E_{\rm rot}$  \footnote{Herewith one implies \cite{Halatnikov}   that the total energy $E$ of the helium takes account of \it local \rm vortices, arising as a result of  spontaneous violating  the superfluidity in a helium on the quantum level. \par
 In \cite{Halatnikov} there was shown that local  vortices contribute with the logarithmic exactness in the total energy $E$, by analogy with (\ref {vklad1}) \cite{Landau52} in the case of a global rotation of a liquid helium together with the cylindrical vessel where it is contained.}, given via Eq. (\ref{vrasch}) \cite{Landau52}, by the velocities $v$ and $v_0$,  that takes place the following equation of motion:
\be 
\label{variant}
 (v- \Omega r)\frac {1}{r} \frac {\partial}{\partial r} (v~ r) +  \frac {\kappa}{4\pi} \frac {\partial} {\partial r} \frac {1}{r} \frac {\partial} {\partial r} (v_0~r)=0. 
\ee 
In this Eq. $v=\vert {\bf v}\vert$ in the first item is the rigid rotation velocity  of the liquid helium II turning together with the cylindrical vessel where it is contained, that is determined by Eq. (\ref{circulation}),  while in the second item the  velocity $v_0$ of the "superfluid" motion, obeyed  (\ref{zirk}), is presented.  \par
Thus two contributions are present in Eq. (\ref{variant}). It results, in particular,
\be \label{solid} v= \Omega r. \ee
This solution corresponds \cite{Halatnikov} to   collective (rigid) {\it solid} rotations of the liquid helium II together with the cylindrical vessel where it is contained \rm  . 
 This result will be very important for us.

\section {The topological theory of ${\bf He}^4$.}
Now this is an opportune moment to draw a parallel  between the discussed "superfluidity-vortices" theory and the YMH model  \cite{ LP1,LP2,Pervush2, David2,David3}

Already in the paper \cite{Pervush1},  there was stressed the great role of alike collective (cooperative) "solid" rotations in non-Abelian gauge theories. 
In the  works  \cite{David2,David3, Pervush2,LP2,LP2} (see also \cite{Arsen}) the nature of such cooperative "solid" rotations in non-Abelian models was revealed with the example of the Minkowkian YMH theory quantized by Dirac, involving the physical vacuum with BPS monopole solutions and the topological degeneration of initial data. \par 
There was demonstrated  that in the Dirac fundamental quantization scheme \cite{Dir}, coming to the Gauss-shell reduction of the Minkowkian YMH model.  such cooperative rotations are manifested as vacuum "electric" monopoles (zero mode solutions) (\ref{zero}), induced, in turn, by Higgs vacuum BPS  monopoles modes $\Phi_{(0)}^c ({\bf x})$ belonging to the zero topological sector of this model.

In the recent study this author ground the existence  of  cooperative "solid" rotations in Minkowskian non-Abelian models quantized by Dirac  by profound topological causes: namely by the Gribov "discrete" factorization  (\ref{fact1})  \cite{disc,Pervush1}  of the  residual symmetry group. \par
In turn, this involves \cite{disc} thread topological defects  (accompanied by various rotary effects) in the Minkowskian non-Abelian gauge model \cite{LP1,LP2,Pervush2,David2,David3}  in which the Gribov "discrete" factorization (\ref{fact1}) of the residual symmetry is assumed in order to justify various rotary effects and the Dirac fundamental quantization \cite{Dir} of that gauge model. \par
On the other hand, there is an important distinction between the Minkowskian non-Abelian gauge theory \cite{LP1,LP2,Pervush2,David2,David3} and the model \cite{Landau52} of the liquid helium II turning together with the cylindrical  vessel where it is contained.
This distinction lies in the nature of collective rotations in the both enumerated models. \par
So in the model \cite{Landau52} of the liquid helium II turning together with the cylindrical  vessel where it is contained there are {\it rigid} $O(2)$ rotations, while in the Minkowskian non-Abelian theory  \cite{LP1,LP2,Pervush2,David2,David3} there are {\it local gauge  } rotations generated by the violated: say, $SU(2)$, initial gauge symmetry.  \par
The latter fact regarding the Minkowskian non-Abelian theory \cite{LP1,LP2,Pervush2,David2,David3}  is closely connected with the Josephson effect \cite{rem3,Pervush4} taking place in the appropriate physical vacuum: at the absolute zero of temperature $T=0$  it comes, generally speaking, to persistent circular motions without exterior sources. 

Unlike the case of the Minkowskian non-Abelian physical vacuum \cite{LP1,LP2,Pervush2,David2,David3}  quantized by Dirac, in the case \cite{Landau52}  of  a liquid helium II specimen turning together with the cylindrical  vessel where it is contained its rotary motions possess rather the "forced" nature induced by the rigid $O(2)$ symmetry group maintained at such rotations. \par
The case \cite{Halatnikov} of  a liquid helium II specimen at rest, involving {\it local} thread vortices, is a more suitable case for the comparison with the Minkowskian non-Abelian theory \cite{LP1,LP2,Pervush2,David2,David3}. 

We shall soon make sure  that rotary phenomena \cite{Halatnikov} taking place in a liquid helium II specimen at rest  are also specific displays of the general Josephson effect \cite{Pervush4}. And this makes related the  helium at rest model \cite{Halatnikov}  with the Minkowskian non-Abelian theory  \cite{LP1,LP2,Pervush2,David2,David3} or with QED$_{(1+1)}$ \cite{Gogilidze,Ilieva2}. 
\par 
\medskip  As we have already ascertained \cite{disc}, the Bogolubov helium Hamiltonian 
(\ref {eq:three1}) \cite{N.N.,Levich,Nels}  is invariant with respect to $U(1)$ gauge transformations (\ref{preobr.fazi}). Among these gauge transformations, one can pick out "small" and "large" ones in the terminology by Faddeev and Jackiw \cite{Fadd2}, i.e. those that correspond to the topologies $n=0$ and $n\neq 0$, respectively. \par
On the other hand, as we learn from Eq. (\ref{zirk}), superfluid potential motions in the helium, as a sign that the initial $U(1)$ gauge symmetry of the Bogolubov Hamiltonian (\ref {eq:three1}) is violated, are possible only as $n=0$ (or in the classical limit $\hbar\to 0$). \par 
Below we shall show, utilizing   the holonomies group arguments, that the conditions $n=0$ and $\hbar\to 0$ result the same
concerning  the superfluidity in a helium: namely they yield the trivial topology $n=0$ in a superfluid  helium\rm. This trivial topology corresponds to "small" Faddeev-Jackiw matrices\rm. \par
We have already supposed \cite{disc}  that  the initial $U(1)$ gauge symmetry of the Bogolubov Hamiltonian 
(\ref {eq:three1}) is violated in the (\ref{fact1})  fashion.  More exactly, 
\be
\label{subgr}
 U(1) \simeq U_0 \otimes {\bf Z}\longrightarrow\tilde U  \simeq {\bf Z}.
\ee
As a visual picture of the group space $\tilde U$, appearing therein,  one can imagine \cite{disc} the circle $S^1$ cutting up by its "topological sectors" separated from each other. \par 

\bigskip Eq. (\ref{subgr}) permits an interesting "geometrical"  interpretation. On the "geometrical" point of view, the just described way to break down the initial (assumed to be {\it continuous} \cite{disc} due to the freedom in the choice of the nature of the [initial group] geometry, us explained in Section 3), $ U(1)$ gauge symmetry of the helium Hamiltonian (\ref{eq:three1})  comes to separating from each other  all the homotopical classes of ways in the $ U(1)\simeq S^1$ group space,  that turns it in a  multi-connected  space. 
More exactly, the homotopical classes of the one-dimensional ways in the   $ U(1)$ group space (characterized by eigen topological numbers $n\in \bf Z$) becomes \it quite separable sets \rm in this case \cite{Engel}. This means  that always there exists such a   function $f: \tilde U\to I$ ($I\equiv [0,1]$) for two homotopical classes $A$ and $B$ in $ U(1)$ that $f(x)=0$ as $x\in A$, while $f(x)=1$ as $x\in B$. Here $x$ is a (one-dimensional) way. 
One speaks in this case that $f$ \it separates \rm the sets $A$ and $B$.\par 
Vice verse, the actual isomorphism 
$$\tilde U\simeq U(1),$$
we learn from (\ref{subgr}), comes to the back "agglutination" of the circle $S^1$ of its "topological sectors".
\par 
\medskip  Let's go into detail about this all. At first, how it happens the $U(1)\to \tilde U$ process. We suppose that topological sectors of the $U(1)$ group space (i.e. actually the homotopical classes $[u]^n$ of loops, (\ref{cll})) become separated, during this process, by domain walls. Herewith it can happen that such a quite separate set \cite{Engel} (defined above) is the mixture of different topologies. In this case, due to the abelian nature of $U(1)$, the topology of the domain is determined by the large in modulus topological number when all $n_i$ of the same sign.  If such topological numbers have different signs,  we merely operate with these as with integers. Thus one can assert that each separated domain possesses {\it the fixed topology}. 

Also, it can be given the alternative definition of $ \tilde U$ as {\it the disjoint union of the topological sectors} $U_1^n$  ($n\in {\bf Z}$), i.e. 
\be \label {disj}\tilde U=\bigcup _{n\in {\bf Z}} U_1^n.
\ee  Now we give the sketch of the proof that  $U(1)\to \tilde U$ is indeed {\it an isomorphism}. The process of gluing the sectors $U_1^n$  together  involves ensuring that the transition between adjacent sectors is smooth and continuous. This means that the boundary of each sector $U_1^n$ match up correctly with the boundary of the sectors $U_1^{n+1}$ and $U_1^{n}$.  During this gluing process, we need to check for any residual sets or gaps that might arise. These residual sets would indicate discontinuities or mismatches in the reconstruction of $U(1)$.  

To ensure that there are no residual sets, we would typically:

$\bullet$ Define a continuous map: construct a continuous map that takes each point in the disjoint union and maps it to a corresponding point on $U(1) \simeq S^1$. 

$\bullet$ Verify smooth transitions: ensure that the transition between  $U^1_n$ and $U^1_{n+1}$  is smooth, meaning there are no jumps or gaps.

$\bullet$ Topological consistency: check that the overall topology of the reconstructed space matches that of  $U(1)$.

How to construct a  smooth transition between $U_1^{n+1}$ and $U_1^{n}$? Following (\ref{cll}),  the loops from the sector $U_1^{n+1}$ can be gotten from those from the  sector  $U_1^{n}$ in the shape $$[u]^{n+1} =[e_{p0}]\circ \tau ^n \circ \tau =[u]^n \circ \tau.$$
 The map $\phi: \gamma_n \to \gamma_{n+1}$  (for the loops $\gamma_n \in [u]^n$ and $\gamma_{n+1} \in [u]^{n+1}$) is really an isomorphism: it is indeed  bijective mapping. To make sure of this  we recommend our readers Lecture 3 in the monograph \cite{Postn4}.  Another very important for us conclusion is that the fundamental group of one-dimensional loops $\pi_1 (U(1), p_0)\simeq \tau^n$ (following\cite{Postn4}, $p_0$ is a pre-selected point on the sphere $S^1\simeq U(1)$) acts as an automorphism within the disjoint union $\tilde U$. In fact, this is equivalent to say that $\tilde U$ is invariant relatively the (gauge) group $U(1)$, i.e, {\it can be treated as the degeneration space} in definite contexts: for instance, in the topological $\rm He ^4$ model, discussed in the present study. 

Herewith any  homotopical class $\tau$ (possessing the topological number $n=1$ and treated \cite{Postn4} as the generator of the cyclical group  $\pi_1 (U(1), p_0)$) can be treated naturally as the operator {\it raising} the topology: from $n$  to $n+1$. In particular, such operator transforms the topological sector $U_1^n$  into  $U_1^{n+1}$. Inversely,  $\tau^{-1}$  can be treated as operator {\it 
lowering} topology: from $n$  to $n-1$. And again an interesting question arises: how this is embodied in the  $\rm He ^4$ phenomenology?

\bigskip  To conclude this topic, note that the isomorphism (\ref{subgr}) (if it will be proven strictly!) permits its interesting interpretation as {\it the covering} of the space $U(1)\cong S^1$ by the  disjoint union $\tilde U$. Herewith the (gauge group) space $U(1)\cong S^1$ serves as the total space of this covering while $\tilde U$ (given via (\ref{disj}))  serves as its base. We recommend our readers the monograph \cite{Postn4} (see Lecture 2 in this  monograph) for a detailed study of this question. However, there is an important  difference between the definition of  covering given in \cite{Postn4} with that us used. We have the total space  $U(1)\cong S^1$ with $\pi_1 [U(1)] ={\bf Z}$. At the same time, the definition of  covering given in \cite{Postn4} provides that the total space of a covering must be {\it linearly connected}: $\pi_1 {\cal E}=0$ for the total space ${\cal E}$.  There is no contradiction here! Such "relaxation" in the definition is quite legal and permissible in differential topology. 

However, it is also possible, of course, to utilize in our case the straight line $\cal R$ which is linearly  connected ($\pi_1 {\cal R}=0$) as the total space ${\cal E}$. As it was shown in the discussed monograph \cite{Postn4} (see Example 1 in Lecture 2 therein) this line covers the circle $S^1=\{(x,y); x^2+y^2=1\}$ in the way 
$$ \pi (t)=(\cos 2\pi t; ~\sin  2\pi t).$$ By transitivity, we then can construct the covering ${\cal R} \to \tilde U$ which will satisfy the requirement (in the monograph \cite{Postn4}) for the total space $ {\cal E}$ to be linearly connected.

\bigskip Just the isomorphism (\ref{subgr})  implies the appearance of vortices in the helium  as well as  the presence  of the superfluid component there. 
 This is immediate manifesting  the Gribov topological degeneration in helium \cite{disc} as a purely quantum effect (as we shall make sure in this below). \par
In this case the residual  symmetry group for the liquid helium specimen turning together with the cylindrical vessel where it is contained \cite{Landau52} will be 
\be \label{Hgr}
H\equiv O(2) \otimes U_0 \otimes {\bf Z}.
\ee
Herewith the origin of the rigid $O(2)$ rotations becomes  evident from our discussion in Section 2.4. 

Prove now that
\be \label{pi0H}
\pi_0 (H)= {\bf Z}.
\ee 
The nature of $O(2)$ is such that it contains the subgroup of ({\it rigid}) $SO(2)\cong U(1)$ rotations isomorphic, in a natural wise, to the circle $S^1$ which is {\it connected}.  This means, and it will be extremely important for our further discussion, that these rigid $SO(2)$ rotations on the some {\it constant} angle $\alpha$ do not change the Bogolubov Hamiltonian $\hat {\cal H}$ , (\ref{eq:three1}). This is so since the operator $\hat {\cal H}$ contains the mutually Hermitian-conjugated creating/annihilation operators $\hat a^\dagger ({\bf r},t)$,  $\hat a ({\bf r},t)$, respectively.  So the global $SO(2)$ rotation, involving the exponent $\exp(i\alpha)$ do not alter the products of the $\hat a^\dagger \hat a$ type. 

 Besides that, there are also reflections: these are not connected to the identity, meaning 
$ O(2)$
 has two components: one connected to the identity (the rigid $SO(2)$ rotations), and one associated with the reflections. So,  $ \pi_0 O(2)={\mathbb Z}_2$.  On the other hand, we see that 
$$ O(2)\cong SO(2)\otimes {\mathbb Z}_2. $$  Herewith now we do not "discretize " $SO(2)\cong U(1)$  as we do this in the present study  for the local $U(1)$ symmetry. So we speak that $\pi_0 SO(2)=0$!

 To calculate the number of connection components, i.e. the group $\pi_0$ of the tensor product $H$, utilize the simple rule for the $\pi_n$ of the tensor product of (topological) groups. For instance, for $\pi_0$ and $n=3$, $$ \pi_0 (G_1\otimes G_2\otimes G_3)=\pi_0 (G_1) \otimes\pi_0 (G_2) \otimes \pi_0 (G_3)$$
(and so for each $\pi_n$ and any number of groups).

Since $\pi_0 U_0=0$ and $\pi_0 {\bf Z}={\bf Z}$,

\be \label{H0} \pi_0 H ={\bf Z} \otimes {\mathbb Z}_2.\ee

 Let us consider $ {\bf Z} \otimes {\mathbb Z}_2$ in relation to the ${\rm He}^4$ specimen contained in the turning cylinder. We should recall herewith that the topological numbers $n\in {\bf Z}$ can be derived from Eq. (\ref{zirk}) as 
\be \label{zirk11} n=\frac m{2\pi \hbar} \oint _\Sigma {\bf v}^{(n)}; \quad m/\hbar ~{\rm is ~fixed}. \ee

These $n$ can be treated naturally as winding numbers (Chern-Simons functionals) in the ${\rm He}^4$ topological model.

In this context, the action of the global $O(2)$ rotations (more concretely, its ${\mathbb Z}_2$ subgroup), i.e. the group ${\mathbb Z}_2$ onto the group $\bf Z$ comes to the  transformations $n\to \pm n$, in other words, {\it a vortex maps onto itself or into its ant-vortex}!  It is remarkable 
We still have an infinite number of topological sectors marked with integers, but just now there is a symmetry that “turns over” them. Thus (with the account of said above about the constant angle $\alpha$) the global $O(2)$ rotations do not change the picture we discuss in the present study.

As a result, we can rewrite (\ref{H0}) as \be \label{H01} \pi_0 H ={\bf Z} \otimes {\mathbb Z}_2\cong {\bf Z}.\ee

\medskip Let us check that the appropriate degeneration space $R_{\rm turn}=G/H$ is indeed topologically equivalent to 
 $${\bf R}^2\setminus \{0\}\simeq S^1$$
in the studied model \cite{Landau52} of a liquid helium II turning. \par 
Utilizing the associative property of the matrix product  $ U_0  \otimes {\bf Z}\otimes O(2)\simeq G$ for the initial symmetry group \footnote{Indeed, it is an intermediate symmetry upon the "discretization" of the initial $U(1)$ gauge symmetry of the Bogolubov Hamiltonian $\hat {\cal H}$.} , one can rewrite it as  
\be
\label{resultant1}
  U_0  \otimes [{\bf Z}\otimes O(2)]\equiv G. 
\ee 
The matrix expression standing in the square brackets  in latter Eq.  commutes with the group of "small" (Abelian) gauge transformations $U_0$; thus it can be treated as a "coefficient" behind $U_0$.\par 
In turn, this determines the specific application of Eqs. (\ref{sxema1}), (\ref{sxema}) \cite{Ryder} in the theory of vortices in a liquid helium II. One would merely replace the residual symmetry group $H'$ by ${\Bbb I}$, the  unit matrix (which always can be treated as a group), in Eq. (\ref{sxema})    and multiple it (e.g. left) by the  "coefficient" ${\bf Z}\otimes O(2)$. 
In this case Eq. (\ref{sxema})  acquires the look
\be
\label{sxema2} 
[{\bf Z}\otimes O(2)] \otimes U_0  =[{\bf Z}\otimes O(2)] \otimes ({\Bbb I}+ U_0/ {\Bbb I}). 
\ee
Whence  the degeneration space $R_{\rm turn}$ is
 \be
 \label{R1} R_{\rm turn}=
 [{\bf Z}\otimes O(2)] \otimes U_0 \cong  {\bf Z}\otimes SO(2)\otimes {\mathbb Z}_2\otimes U_0.\ee 
in the considered helium turning theory \cite{Landau52}. As we have proven just, one can speak that this degeneration space $R$ is isomorphic to $\tilde U$ ({\it at least, on the level of phenomenology}!) due to the ${\mathbb Z}_2$ transformations which map a gauge vortex with the topology $n$ oto itself or onto its antivortex with the topology $-n$.  In particular,  the trivial topological sector $n=0$ maps always onto itself; in other words, global $ O(2) $ rotations {\it do not change superfluid properties of helium}! And plus also the global rotations onto the angle $\alpha$, which leave the Hamiltonian $\hat{\cal H}$ invariant. Note in this context that the rigid $SO(2)\cong U(1)$ symmetry saves the number of (quasi)particles in the specimen. This, as it can be proven, does not change the thermodynamical properties of ${\rm He^4}$. 

\medskip Thus $R_{\rm turn}$ is topologically equivalent to 
\be \label{rturn}S^1\cong U(1)\cong U_0\otimes {\bf Z}\cong \tilde U.\ee   This is correctly modulo global $O(2)$ transformations, as we have just shown!

Therefore we can speak that the degeneration space $R_{\rm turn}$ inherits all the topological properties of $\tilde U$. Namely, 

\be \label{rturn-top} \pi_0 R_{\rm turn} =\pi_1 R_{\rm turn} ={\bf Z}.\ee

\medskip  As a digression, it is interesting to note that the {\it gauge} group  $ U_0$ acts free in the space of "large" ($n\neq 0$) matrices, forming there $ U_0$-orbits. Each such orbit can be expressed in terms of the holonomies group  $H_{\rm hol}$ \cite{LP1,Al.S.} formed by one-dimensional loops (belonging to the fundamental group $\pi_1 (U(1))$ and characterized by their topological numbers $n\in \bf Z$ due to the Pontryagin degree of a map theory arguments \cite{Postn3}.\par 
  Thus each matrix  from ${\bf Z}\otimes U_0$
corresponds to a "large" (or "small" if $n=0$) gauge transformation with the fixed topological number $n\in \bf Z$, and one can associate the appropriate homotopical class from $\pi_1 (U(1))=\bf Z$ to this $n$.\par

\bigskip  Now, from ({\ref{H01}), (\ref{rturn-top}) we read off the relation 
\be \label{verteksu} \pi_0 H= \pi_1 R_{\rm turn}={\bf Z},\ee 
which guarantees the existence of vortices, i.e. thread topological defects, inside $R_{\rm turn}$.

In this concrete case it implies the presence of (rectilinear) thread topological defects (vortices) in the quested model.
These rectilinear vortices contribute with  the item $\Delta E$ \cite{Landau52}, (\ref {vklad1}),  additional to the Bogolubov helium (diagonalized) Hamiltonian $\widehat{\cal H} $ \cite{Nels}. Just this increasing $\Delta E$ the helium energy (in comparison with  that given by $\widehat{\cal H} $ and referring to the helium specimen at rest) is the sign of the first order phase transition occurring. Recall herewith that the group $O(2)$ of global rotations does not change the termodynamical properties of liquid ${\rm He}^4$!

\medskip
Also, there are  walls between different topological domains inside the discrete liquid helium II turning degeneration space $R_{\rm turn}$, (\ref{R1}), because of Eq (\ref{rturn-top}).

On the other hand, there are no point topological defects inside $R_{\rm turn}$ since $\pi_2 S^1=0$. 

\bigskip  Finally, let us write down manifestly our results for the liquid helium II at rest. In fact we already know these results from our previous study. Denote the appropriate degeneration space as $R_r$. It is easy to see from the previous analysis that it coincides with the residual gauge symmetry group:
\be \label{Rr}R_r\cong\tilde U\cong U_0\otimes {\bf Z}.\ee

This again provides thread topological defects and domain walls inside this manifold due to standard reasoning  us discussed in this study. 

\bigskip Return again to Eq. (\ref{subgr}).  As we have stressed repeatedly in this study, the Bogolubov helium Hamiltonian 
(\ref {eq:three1}) \cite{N.N.,Levich,Nels}  is invariant with respect to $U(1)$ gauge transformations (\ref{preobr.fazi}). Among these gauge transformations, one can pick out "small" and "large" ones in the terminology by Faddeev and Jackiw \cite{Fadd2}, i.e. those that correspond to the topologies $n=0$ and $n\neq 0$, respectively. 
On the other hand, as we learn from Eq. (\ref{zirk}),  superfluid potential motions in the helium, as a sign that the initial $U(1)$ gauge symmetry of the Bogolubov Hamiltonian (\ref {eq:three1}) is violated, are possible only as $n=0$ (or in the semiclassical limit $\hbar\to 0$). \par
Let us show that the conditions $n=0$ and $\hbar\to 0$ result the same
concerning  the superfluidity in a helium: namely they yield the trivial topology $n=0$ in a superfluid  helium\rm. This trivial topology corresponds to "small" Faddeev-Jackiw matrices\rm. \par

On the other hand, it is easy to see that that "large" Faddeev-Jackiw exponential multipliers (\ref{preobr.fazi1}) corresponds to  {\it violating  the superfluidity in helium and appearing thread vortices} \cite{Halatnikov} therein\rm. \par
The Gribov  degeneration of the trivial topology $n=0$ in a superfluid  helium ($=$violating  the superfluidity in helium and appearing rectilinear vortices therein!) is thus {\it a purely quantum effect} directly proportional to the Planck constant $\hbar$, vanishing in the semiclassical limit $\hbar \to 0$.

In the resting helium case, this violating takes place \cite{Halatnikov} again in infinite thin neighborhoods of vortex  threads;  and  again  thread topological defects arise, as in the case \cite{Landau52} of a liquid helium  turning together with the cylindrical  vessel where it is contained. \par
Eq. (\ref{subgr}) confirms this fact: $$\pi_0 (\tilde U)\equiv \pi_1 (R_r)= {\bf Z}.$$
Herewith it can be argued (and we did this above repeating the said in \cite{disc}) that the degeneration space $R_r$ in the  helium at rest model \cite{Halatnikov} is  topologically equivalent to the circle $S^1$. \par

\bigskip It is interesting to compare our results about the liquid helium II  and those gotten in the monograph \cite{Volovik}. 

More exactly, in the  monograph \cite{Volovik}, the group of {\it rigid} space rotations $SO(3)_{\bf L}$ around the axes $x$, $y$, $z$ (involving the orbital momentum ${\bf L}$) for a ${\rm He}^4$ specimen (instead of $O(2)$ in our case) \footnote{The mentioned rigid $SO(3)_{\bf L}$ is the natural group of rigid space rotations in the 3-dimensional Euclidian space which does not change the Bogolubov Hamiltonian $\hat {\cal H}$, the number of (quasi)particles and the thermodynamical properties of liquid ${\rm He}^4$. Rigid rotations of a liquid ${\rm He}^4$ specimen together with the cylindrical vessel where it is contained \cite{Landau52}, involving the group $O(2)$, is an additional construction.} was considered. As the author \cite{Volovik} asserts, the appearance of vortices course violating there the initial $U(1)$ symmetry involves also violating $SO(3)_{\bf L}$ since the direction of the (one) vortex line {\it appears as the axis of spontaneous anisotropy}. This anisotropy line can be chosen as the axis $z$ of the (fixed) reference frame.

\medskip In our case, with the (rigid) $O(2)$ symmetry group, it is a somewhat different situation since we already have  two-dimensional (rigid) rotations about the fixed axis, which, in a natural way,  can be chosen as the axis $z$.
Thus it is now an already ready anisotropy (and simultaneously {\it symmetry}) line, and there are no need for violating $O(2)$ in this case, and the above 'topological' calculations about the liquid helium turning remain legitimate.

\medskip The author  \cite{Volovik} considers a complex scalar $\Psi=\vert\Psi \vert \exp(i\Phi)$ as the order parameter for the superfluid $^4$He. Note that the similar look for the order parameter (coinciding with the helium
Bose condensate wave function) was proposed in the monograph \cite {Landau52}:
\be
\label{Xi11}
 \Xi (t,{\bf r})= \sqrt {N(t,{\bf r})}~ e^{i\Phi(t,{\bf r})},
\ee
with $ N(t,{\bf r})$ being the number of particles in the ground energy state $\epsilon=0$.

The different from zero order parameter $\Psi$ implies the complete breakdown of the $U(1)_{N}$ symmetry. 

For a vortex with the winding number $n_1$, it was set \cite {Volovik} $\Phi(t,{\bf r})\equiv n_1\phi$.

It is natural to guess  \cite {Volovik} that the symmetry-breaking scheme in the presence of  $SO(3)_{\bf L}$ is 
\be \label{scheme}
G^{'} = U(1)_{N} \otimes SO(3)_{\bf L}\rightarrow {\bf H}^{'}=U(1)_Q
\ee Here the remaining symmetry group $U(1)_Q$ is the symmetry with the order parameter given in (\ref{Xi1}). It is the rotation by the angle $\theta$ that transforms $\phi \to \phi +\theta$, accompanied by the global phase rotation $\Phi\to\Phi +\alpha$, with $\alpha=-n_{1}\theta$. The generator of such $U(1)_Q$ transformations is
\be \label{qgen}
Q=L_{z}-n_1 N.
\ee

\medskip From the said in \cite{Volovik} we see a principal difference of the G. Volovik approach to liquid helium ${\rm He}^4$ with that developed in the present study. Our analysis is based upon the assumption  about the initial {\it gauge} $U(1)$ symmetry of the Bogolubov helium Hamiltonian $\hat {\cal H}$, violated then in the (\ref{subgr}) wise.  As for the {\it global} $SO(3)_{\bf L}$ symmetry, which is present naturally in the 3-dimensional flat Euclidian space, we can "forget" about this symmetry which does not change the model. 

On the other hand, global vortices corresponding to the $U(1)_Q$ group ala \cite{Volovik}  {\it also should present}. This can lead to forming composite vortex structures. Also a global vortex may induce a local gauge response. A local vortex may inherit angular momentum from the global sector via the 
$ Q$
 generator.

In the sphere of the stability the discussed theory, the following conclusions can be drawn. Local vortices have finite energy due to gauge screening. Their coexistence leads to nontrivial energy landscapes, possibly stabilizing bound vortex states or vortex lattices.  All this should be the subject of  an additional study!

\medskip Note that the same phenomenological assumptions can be made also for global $O(2)$ vortices in the liquid helium-4 placed in the cylindrical vessel turning around its axis $z$. As it was discussed above, such global $O(2)$ vortices are connected with rotations on  constant angles $\alpha$. Such rotations form a subgroup in $O(2)$.

\bigskip  We have almost finished our discussion about the topological theory of liquid ${\rm He}^4$. Here, as an additional option, we propose our readers our Appendix, where we ground the isomorphism 

$$ \pi_0 (H)\simeq \pi_1 (R)$$ for the (residual) gauge symmetry group $H$ and the degeneration space $R$. Namely this isomorphism guarantees  the existence of thread topological defects inside the degeneration space $R$.  In fact, in our proof we follow the arguments given in the monograph \cite{Al.S.}  (see  \S T17  ibid).

\bigskip  Our readers why does not interested in deep mathematics can go over to our concluding section, where we analyze briefly the perspectives of superconductivity in the framework of discrete $\tilde U$ vacuum geometry developed in the present study. It is very tempting, due to the   striking similarity between the superfluidity and superconductivity phenomena to utilize the same framework for describing superconductivity. However, there are differences between the both models, which we shall discuss.
\section{Discussion.}  The topological theory of helium-4, developed in the present study, is a manifestly gauge model involving the {\it Abelian} $U(1)$ gauge symmetry, violated (spontaneously) in the (\ref{subgr}) wise. It is a quite natural way violating the $U(1)$ gauge symmetry providing, as we argue, the coexistence of superfluid potential motions (corresponding to the zero topology) and local vortices (possessing nontrivial topologies) in the limit of zero absolute temperature $T\to 0$. In this is essence of the first order phase transition taking place. 

\medskip. It turns out that there are related gauge theories involving the spontaneously broken $U(1)$ gauge symmetry, worked out in 20 century. These are the Ginzburg-Landau superconductivity theory \cite{V.L.} and the Abelian Higgs model involving the so-called Nielsen-Olesen (NO) vortices \cite{No} (see also Chapter 10 in \cite{Ryder}) at the $U(1)$ break down. Now we shall attempt to argue that the way (\ref{subgr}) to violate the $U(1)$ gauge symmetry is a quite natural way to do this and to explain the appropriate  phenomenological effects.

\bigskip So, NO vortices \cite{No} are specific solutions to the equations
of motions at violating the initial $U(1)$ gauge symmetry (this involves the nonzero value $|\phi|\neq 0$ for the Higgs field $\phi$ in the ”asymmetric” phase; it is remarkably that $|\phi|$ enters explicitly these equations
of motions, as it was shown in \cite{Ryder}). Herewith  NO vortices are interpreted as {\it thread topological defects} inside the appropriate vacuum manifold, analogous to Halatnikov vortices \cite{ Halatnikov} in a rested liquid helium II specimen.

Since,  from the topological point of view, NO vortices, as a particular case of
thread topological defects, would be naturally associated with the ”discrete” vacuum geometry
should be assumed for the appropriate degeneration space, such degeneration space is similar to the discrete vacuum manifold $R_r$, (\ref{Rr}), has been got in the rested liquid helium II
specimen case.

However there are some principal distinctions between the latter case and the Abelian
Higgs model \cite{No} involving NO vortices. So, for instance, the first-order phase transition occurs primary in that model (see the arguments \cite{Linde}, repeated also in Ref. \cite{ rem1}). Let's look now at the arguments  \cite{Linde}. Suppose, to begin with, the fixed temperature $T$ of environment.  Then it turns out that the plot $\phi (T)$ suffers a gap. More precisely,  it is impossible to link in a continuous wise the false vacuum $\phi(T)=0$ and other, {\it stable}, vacuum. 

If we fix, for example,  the Lorentz gauge $\partial_\mu A^\mu=0$ in the Minkowskian Abelian Higgs model, the first-order
phase transitions occurs when $\lambda\leq e$ (with $e$ being the elementary charge) \cite{Linde}.   In this case it can be shown  that three solutions to appropriate equations of motion exist, including the false vacuum $\phi(T)=0$, in the temperatures interval $T_{c_1}<T< T_{c_2}$.\par

As a result, the false vacuum $\phi(T)=0$ becomes a metastable state at $T> T_{c_1}$, while there exists also the instable   solution $\phi_2$ corresponding to the local maximum of the  appropriate potential $V(\phi,T)$ and $\phi_1$ corresponding to the "true" vacuum, i.e. to the (global) minimum of $V(\phi,T)$. Herewith the phase transition from the $\phi_1$ to the $\phi(T)=0$ state, accompanied by restoring the $U(1)$ gauge symmetry, begins at a temperature $T_c$ at which \cite {Linde}
$$ V(\phi_1(T_c),T_c)\sim V(0,T_c).       $$
Graphically \cite {Linde}, there is however a gap between the both potentials at $T=T_c$, corresponding to the gap $\Delta F(T)=\Delta V(\phi,T)$ (the {\it latent heat}) in the free energy $F(T)$. It is just the sign of the  first-order  phase transition occurring in the  Minkowskian Abelian Higgs model. \par

On the other hand, the metastability of the false vacuum $\phi(T)=0$ implies the {\it supercooling } phenomenon: the system of fields remains in this state, coexisting simultaneously with the "true" vacuum $\phi_1$ even when $T<T_c$, coexisting herewith simultaneously with the "true" vacuum $\phi_1$. Vice verse, when $T>T_c$, the latent  heat is liberated as  $\Delta F(T)$. This phenomenon is referred to as {\it reheating} \cite {Linde,Coles}. Thus the nonzero temperature $T$ plays a crucial role in the NO model. The important discrepancy of the superfluidity model from the NO one is also the fact that the (gauge) $U(1)$ symmetric phase  {\it disappears entirely} upon $U(1)$ 
violating in the former case since the second order phase transition occurs.

\medskip In this context, an one interesting suggestion can be done about the NO model at the fixed environment temperature $T$.  This suggestion assumes retaining the $U(1)$ gauge symmetry and simultaneously forming the stable asymmetric  thermodynamic phase represented by a discrete degeneration space of the $\tilde U$ wise (widely discussed in the present study)- It turns out that this task is quite solvable! 

We must remember that the gauge $U(1)$ group space is isomorphic to the circle $S^1$. In any topological sector of this circle (say, in $n^{\rm -th}$), it is represented, in the "loop framework" \cite{Postn4}, by the appropriate homotopical class $[u]^n$, (\ref{cll}).  Let's speculate what will happen at deleting some loops in several (or even each!) homotopical classes $[u]^n$. Just this promotes forming the stable asymmetric  thermodynamic phase ala $\tilde U$. The topology, i.e. the gauge symmetry, is maintained when we remove some  loops $\gamma^n \in [u]^n$. Herewith we simply restrict the set of available representatives in the class $[u]^n$, the topology of $S^1\cong U(1)$ {\it does not change}! And this is a desired result.

We are also interested in the case when  the gauge $U(1)$ group space is distorted (with forming NO vortices). Let us discuss now   under what circumstances this can occur. This can occur when

 $\bullet$  a single point is removed. Then $S^1/\{p\}\cong {\bf R}$ is contractible. Fundamental group becomes trivial.

$\bullet$ when removing countably many points: still disconnected, possibly totally disconnected depending on density.

$\bullet$ at removing  dense subsets: we may destroy path-connectedness or compactness.

This leads, ultimately, to gradual but complete destroying the symmetric thermodynamic phase  for a finite time interval. This question of coexisting and further destroying/disappearing of the  $U(1)$ gauge symmetric thermodynamic phase in the Abelian Higgs model \cite{No} is very important and interesting. It requires some effort to its investigation- Here we only outlined the problem!

\bigskip 
In connection with our discussion  about the   Abelian Higgs model \cite{No} (implicating NO vortices), it will be relevant to mention the very important analogy (in  the sphere of the topology and phenomenology) between this model and the case \cite{Landau52,V.L.,Lenz} of  superconductors. \par 
First of all, it would be therein  \cite{Landau52,V.L.,Lenz} $\kappa >1/\sqrt 2$ for the Ginzburg-Landau parameter 
$$ \kappa =\frac{\lambda_L}{\xi}, $$ 
with \cite{Lenz} \begin{equation}
   \label{pede}
\lambda_{L} = \frac{1}{M_{\gamma}}   
 \end{equation}
being the {\it penetration or
London depth}, depending on the "photon mass"
\be \label{photomass}
M_\gamma=\sqrt 2~ ea.
\ee and $\xi$ being the coherence length which physical sense is \cite{Landau52} the correlations radius of fluctuations in the order parameter $<\phi^2>$.\par 
The parameter $a \equiv |\phi|$ \cite{Lenz} appears in latter Eq. 
It plays the role of the order  parameter in the { Ginzburg-Landau} superconductivity model \cite{Ryder,V.L.,Lenz}:  $a=0$ for the normal phase, while $a\neq 0$  for the superconducting phase. \par

Generally speaking \cite{Landau52}, the Ginzburg-Landau parameter $\kappa$ is a function of the temperature $T$:  $\kappa\equiv\kappa (T)$. 
As it was demonstrated in the original paper \cite{V.L.}  (see also \cite{Landau52}), the inequality  $\kappa >1/\sqrt 2$, true for Type II superconductors, is associated with the negative surface  tension $\alpha_{ns}<0$ \cite{Landau52}:\be\label{surf}   \alpha_{12}=\int\limits_{-\infty } ^{\infty } (f_1-f_2)~ dx,\ee 
where $f_1$ and $f_2$ are the free energies densities referring to two thermodynamic phases coexisting at the first-order phase transition occurring in the given physical model. Let us denote as (n) the normal phase in a superconductor and, respectively, as (s) the superconducting phase. In this context we shall refer $f_1\equiv f_n$ to the normal phase while $f_2\equiv f_s$ to the superconducting phase.

In the latter case ($\alpha_{ns}<0$)  it becomes advantageous, from the thermodynamic standpoint, to compensate increasing the volume energy by  this negative surface  energy. This implies arising germs of the n-phase inside the s one at the values $H$ of the background magnetic field $H$ exceeding a field  $ H_{c_1}$, called \cite{Landau52} {\it the   lower critical field}. ice verse,  germs of the s-phase inside the n one arise at the values $H$ don't exceeding a field  $ H_{c_2}$, called {\it the   upper critical field}.

Thus in the interval $ H_{c_1}<H< H_{c_2}$, one can observe \cite{Landau52} the {\it mixed state} in Type II superconductors (there are definite alloyed metals and combinations of metals). In this case a Type II superconductor is simultaneously in the n and s states. At $H\leq H_{c_1}$, a specimen is purely in the s-state, while it is purely in the n-state at $H\geq H_{c_2}$.

\bigskip
It turns out that the {\it second-order phase transition} takes place in this situation, and now we would like to clarify this.

First of all, it can be argued (see, for example, the work \cite{Kaj}) that in type II superconductors vortex lines of the n-phase {\it repel each other}; thus the vorteces tend to form a vortex lattice in
the type II region. In fact, due to the periodic lattice boundary conditions, even one vortex forms a
square lattice with its periodic counterparts.

The main argument in favour of one or another kind of phase transitions occurring in a medium is the analysis of second derivatives of thermodynamical potentials by the temperature and pressure (for instance, the heat capacity and compressibility), which change abruptly if the  second-order phase transition takes place. In the same time, the first derivatives (such as the energy and volume of the medium) do not change.

The second case, of the free energy plot, has been analyzed, with the example of the type II superconductor, in the monograph \cite{tinkman}. 
In this monograph the behavior of the magnetization curve $B(H)$ was analyzed in the vicinities of the critical points $H_{c1}$ and  $H_{c2}$. This is in fact the same as the analysis of the Gibbs free energy $G$ in the external magnetic field $H$ switched-on, which contains always the item $(BH)/4\pi$.

Such behavior of the magnetization curve $B(H)$ is closely related to the  picture of vortices in the studied superconductor.

\medskip So, three regimes \cite{tinkman} can be distinguished in the $[H_{c1}, H_{c2}]$ interval.

1. Very near  $H_{c1}$, $\Phi_0/B\gg \lambda_L$ for the elementary flux $\Phi_0=\pi \hbar c/\vert e \vert$. Then the vortices are separated by distances more than $\lambda_L$. In this case only a few neighbors are important.

2. For moderate values of $B$, such that $\xi^2 \ll \Phi_0/B \ll  \lambda_L^2$, many vortices appear within interaction range of any given one; this generate the $\sum _{i>j} F_{ij}$ contribution from the pairwise interactions between vortices  into the Gibbs energy $G$ \footnote{Following \cite{tinkman}, cite the explicit look for $F_{ij}$. This is 
$$F_{ij}\sim \frac{\Phi_0^2}{8\pi^2\lambda_L ^2} \ln ( \frac{\lambda_L}{\xi})$$ It is positive when $\lambda_L > \xi$. But we know already that the type II superconductivity begins when  $\kappa=\lambda_L/\xi >1/\sqrt 2$ ($1/\sqrt 2\approx 0.714$). Herewith the interval $\kappa \in [1/\sqrt 2, 1]$ represents a {\it transitional regime} between type-I and type-II superconductivity. In this interval, the surface energy is small and negative, meaning vortex formation is possible but energetically delicate. Vortices may form, but they are not strongly repulsive.
The interaction between vortices can be weakly attractive or neutral, depending on the exact value of  $\kappa$. This can lead to clustering or nonuniform vortex arrangements, unlike the regular Abrikosov lattice seen in higher $\kappa$ type-II superconductors.
}. However, it is still a good approximation to neglect details of the core. 

3. Near  $H_{c2}$, $\xi^2\approx \Phi_0/B$, so that the cores are almost overlaping.

\medskip Just at the careful analysis of the "regime one" one can make sure that the second order phase transition occurs in an (infinitesimal) neighbourhood of $H_{c1}$. The (long enough) calculations \cite{tinkman} give the following look of the magnetic field $B$ in a neighbourhood of $H_{c1}$:
\be \label{B1}
B=\frac{2\Phi_0}{\sqrt 3  \lambda_L^2} \{ \ln [\frac{3\Phi_0}{4\pi  \lambda_L^2(H-H_{c1})}] \}^{-2}.
\ee
$B$ is {\it continuous} at $H_{c1}$, corresponding to a {\it second order phase transition}  (since $B$ enters the expresiion for the free energy expression of the model studied via the $(BH)/4\pi$  item).

The similar arguments in favor of the  second order phase transition occurring are applied to the regime near $H_{c2}$. More exactly, in \cite{tinkman}
the magnetization curves $4\pi M=f(H)$ for different values of $\kappa$ were considered (see Fig. 5.2. in \cite{tinkman}). This analysis of curves shows that only at $\kappa >1/\sqrt 2$ a  second order phase transition takes place.

\medskip The attraction of vorteces in type one superconductors leads \cite{Kaj} to forming the cylindrical hole (the "broken" s-phase) inside the
magnetic field configuration (the symmetric n-phase). 

For this case, the discontinuity $\Delta \frac{\partial G/V}{\partial y}$ of the canonical free energy $G(x,y,H/e_3^3)$, with \cite{Kaj} $y$ and $x$ being two dimensionless ratios
\be \label{xy}
y=\frac{m_3^2(e_3^2)}{e_3^4}, \quad x=\frac{\lambda_3}{e_3^2}
\ee
(involving the electron mass $m_3$ and the Cooper pair interaction constant $\lambda_3$); $e_3$ is a scale introducing in the model by the authors \cite{Kaj}.

Indeed, 
\be \label{dege}
\Delta \frac{\partial G/V}{\partial y}=e_3^4 \Delta <\phi^\star\phi>; \quad \Delta\frac{\partial G/V}{\partial H/e_3^3}=-e_3^3 \Delta B,
\ee
with $\phi$ being the macroscopic instantaneous wave function associated with the  Cooper pair.

For fixed $x$ the latent heat $L$ of the transition is defined as the discontinuity in the "energy" variable $E$ obtained from $G$ by a Legendre transformation with respect to $y$, $H/e_3^3$:
\be \label{latent}
L=\Delta E= -y \Delta \frac{\partial G}{\partial y} -H_c \Delta \frac{\partial G}{\partial H}= V[-y e_3^4 \Delta <\phi^\star\phi> +H_c \Delta B].
\ee
For fixed x, the identity $\Delta G(x,y, H_C)=0$ leads to the Clausius-Clapeyron equation, relating the different discontinuities: 
\be \label{Claus}
\Delta \frac{\partial G}{\partial y}=-\frac{\partial H_c}{\partial y}\frac{\partial G}{\partial H}\Leftrightarrow e_3^4 \Delta <\phi^\star\phi>=\frac{\partial H_c}{\partial y} \Delta B.
\ee
{\it Thus it is enough to measure one of the discontinuities, and the curve} $H_c(y)$. 

\bigskip   From the topological (geometrical) point of view, the said suggests two ways for breakdown the  $U(1)$ gauge symmetry inherent to the superconductivity model. 

In Type II superconductors, in which the second-order phase transition takes place, the appropriate $U(1)\simeq S^1$ group manifold collapses {\it instantly} in the way when domain walls arise simultaneously between all the topological sectors of the circle $S^1$.

In Type I superconductors, in which the first-order phase transition takes place, collapsing the $U(1)\simeq S^1$ group manifold occurs {\it gradually} in the way similar that in the case of NO vortices, us discussed above.  A distinctive feature of the  first-order phase transition taking place in Type I superconductors thus the coexistence of the mentioned geometrical (topological) structures (at the temperature $T\to 0$). Note herewith that the latent heat depends manifestly on the change in the background magnetic field $H$ (rather on changes in the temperature of environment!) as Eq. (\ref{latent}) reads. Thus we can speak about the gradual "magnetic" distortion of the $U(1)\cong S^1$ group space in this case. This is instead of the thermal distortion in the case of the NO model \cite{No}. Despite the different nature of the gradual "magnetic" and "thermal "distortion of the $U(1)\cong S^1$ group space from the physical point of view, the topological approach to these phenomena is, apparently,   {\it is the same}, us analyzed above!

And one more note is appropriate here. The schemes us described are suitable rather for {\it restoring} the $U(1)$ gauge symmetry, i.e. the normal phase. But these   schemes can be "restarted" in the opposite order: from the n to the s phase.

\bigskip In spite of the said, it is obvious that in the Curie point $T_c$, $ H_{c_1}\to H_{c_2}\to 0$. The said allows, for all that, to construct the plots $H(T)$ for the s, n and mixed states in Type II superconductors (see Fig. 7 in \cite{Landau52}).   

\medskip
It is enough manifest that the case of Type II superconductors may be treated as a nonrelativistic limit of the Abelian Higgs NO model \cite{No} \footnote{It is so at identifying \cite{Ryder, Lenz} the density of  superconducting {a \it Cooper pair} with the Higgs field squared $\vert \phi\vert $ in the NO model \cite{No}. }.

\bigskip Indeed, it is a similarity  between the helium at rest theory \cite{Halatnikov} and the Type II superconductivity \cite{V.L.,Lenz}, namely  that in both the cases the $U(1)\simeq S^1$ group space is  destroyed instantly entirely 
down to the discrete (quite separate) set $\tilde U$ involving domain walls between topologies. \begin{appendix}
\section{Appendix. The topological grounding of the discussed theory.}
\renewcommand{\theequation}{A.\arabic{equation}}

\setcounter{equation}{0}
As we promised above, let us discuss the isomorphism 

\be \label{is1} \pi_0 (H)\simeq \pi_1 (R)\ee playing the decisive role in the presence of (rectilinear)
 thread topological defect inside appropriate degeneration spaces. Recall that in our discussion we have considered in fact two kinds of such degeneration spaces. In the case of a resting liquid helium-4 specimen,  it is the degeneration space $R_r$, (\ref{Rr}).  And for a helium-4 specimen contained in the cylindrical vessel turning around its axis $z$ it is the degeneration space $R_{\rm turn}$, (\ref{rturn}).  In fact, there are the same spaces! It is the direct result of invariability of the Bogolubov Hamiltonian $\hat{\cal H}$ with respect to the global $O(2)$ rotations.

\bigskip  Now we are proceeding to proof of the isomorphism (\ref{is1}). To construct  such an isomorphism, one would consider a way $\phi$ in in the initial gauge symmetry group $G$ and ending in the unit element of $G$. \par 
At  natural mapping $G\to G/H$, the way $\phi$ is mapped in a closed way $\hat \phi$. It is the direct consequence of Eqs.  (\ref{sxema1}), (\ref{sxema}),  (\ref{sxema'}) \cite{Ryder}.

As $\pi_0 (G)=0$ (the initial $SU(2)$ gauge symmetry is just such a case \cite{disc}), two ways joined the $h\in H$ and the unit element of $G$ are homotopical to each  other; therefore the  homotopical class of the way $\hat \phi$, $[\hat \phi]$, does not depend on a choice of the way $\phi$. Thus one can associate  the definite  element of $\pi_1(G/H)$ to the fixed  element $h\in H$. We shall denote this element as $\rho(h)$. \par 
It is easy to check that the  homotopical class $[\rho(h)]$ does not change at a continuous change of $h$. Thus the fixed  element $\pi_1(G/H)$ is associated to the appropriate connection component of $H$. \par 
Also it is easy to show  that 
$$ \rho(h_1 h_2)= \rho(h_1) \rho(h_2):$$  in other words, that $\rho$ is the homomorphism $H\to \pi_1(G/H)$. \par  
Its kernel is the group $H_{\rm con}$, consisting on elements   can be joined in a continuous way with  the unit element of the group $H$. It is the \it maximal connected subgroup \rm  in $H$ (e.g. in the terminology  \cite{Logunov}: see ibid Appendix $\Gamma$.2) \footnote{A group is called \cite{Logunov} connected if each its element can be joined by means of  a continuous curve with  the unit element of this group. Indeed, the authors of the monograph \cite{Logunov} applied the term
"\it linearly connected\rm" in the definition of such groups.  
This term means (see Lecture 11 in \cite{Postn3}) that each two points of the group can be joined by a way. \par  
Actually, the both definitions have the same sense.  
One thinks that a topological space is connected if it has no nontrivial {\it open-closed sets} \cite{Postn3}.  This means that such a topological space  can contain only trivial sets: $\emptyset$ and itself.  
One can show herewith that each linearly connected space is connected \cite{Postn3}.}.

 It turns out that $\pi_1(G/H)$ is isomorphic to $H/H_{\rm con}$.
 To demonstrate this fact, let us consider (see \S T17 in \cite{Al.S.}, Proposition 6 ibid) a principal fiber bundle  $\tilde E\equiv (E,B,F,p)$ with the one-connected total space $E$ ($\pi_1 (E)=0$). In this case  $\pi_1 (B)$ turns out to be isomorphic to $\pi_0 (F)= F/F_{\rm con}$, with $ F_{\rm con}$ being the maximal connected subgroup in the  group $F$ (\it the connected component of the unit  element of \rm $F$).\par 
\medskip
The proof of the latter statement can be reduced to the  case  when the group $F$ is \it discrete\rm.  
Such a group acts free on $E$ and $ F_{\rm con}$ (i.e. for each $p\in E$ the equality $ap=p$ ($a\in F$) is possible only when $a=e$, where $e$  is the unit  element of \rm $F$: see Lecture 1 in \cite{Postn4}). \par  
Thus one can consider the principal fiber bundle
  \be 
\label{E_1} E_1\equiv(E, E/ F_{\rm con}, F_{\rm con}), \ee  
whose base is the space of orbits $ E/ F_{\rm con}$.\par
It is easy to check that $ F_{\rm con}$ is an invariant subgroup in  $F$. Really, each way $a(t)$ ($t\in [0,1]$) in $ F_{\rm con}$ can be replaced by 
\be 
\label {nordel}
  g_0(t) ~ a(t)~ g_0 ^{-1}(t), \quad g_0\in F.
\ee , i.e.

\be 
\label {nordel1}
  a(t)=g_0(t) ~ a(t)~ g_0 ^{-1}(t), \quad g_0\in F.
\ee 
This is correctly since   $a(t)$ is homotopical to the unit subgroup element in $F_{\rm con}$. Thus the commutators $[ g_0(t), a(t)]$ and $[g_0^{-1}(t), a(t)]$ are trivial even in the case of non-Abelian gauge groups!

Since $ F_{\rm con}$ is an invariant subgroup in  $F$, the transformations from $F$ transfer an orbit of $ F_{\rm con}$ into itself (if two elements $e_1, ~e_2\in E$  belong to same orbit of $ F_{\rm con}$, i.e. if $e_2=e_1f$ [$f\in F_{\rm con}$], then $e_2g=e_1 g (g^{-1}fg)$ for any $g\in F$; therefore $e_1g$ and $e_2g$ belong to this of $ F_{\rm con}$).

We see that the group $F$ acts in the space of orbits $ E/ F_{\rm con}$. Herewith  elements of $ F_{\rm con}\subset F$ transfer each orbit of $F$ in itself\footnote{This is the same idea of Faddev-Jackiw \cite{Fadd2} of subdividing gauge transformations onto "small" and "large" in such a wise that "small" (topologically trivial)~transformations form orbits in the subspace of "large" transformations.}; this implies that  the group $ F/ F_{\rm con}$ acts also in $ E/ F_{\rm con}$. 

We see that the group $F$ acts in the space of orbits $ E/ F_{\rm con}$. Herewith  elements of $ F_{\rm con}\subset F$ transfer each orbit of $F$ in itself; this implies that  the group $ F/ F_{\rm con}$ acts also in $ E/ F_{\rm con}$. 

It is easy to check also that the action of the group $ F/ F_{\rm con}$ in $ E/ F_{\rm con}$ is indeed free (for discrete groups  this was demonstrated in the monograph \cite{Postn4}: see ibid Example 5 to Lecture 2) 
nd that the space of orbits of this action is $E/F$.\par
  Thus we have  demonstrated in effect that there exists the principal fiber bundle 
	\be 
\label {E2}
 E_2\equiv (E/ F_{\rm con}, E/F, F/ F_{\rm con})\ee 
with the discrete fiber $ F/F_{\rm con}$. 

Its total space $E/F_{\rm con}$ is (linearly) connected, and $\pi_1 (E/ F_{\rm con})=0$. 
The latter fact follows from Proposition 4  to \S T17 in \cite{Al.S.}.

\medskip This proposition runs as follows: \it if the total space $E$ of a principal fiber bundle is aspheric in dimensions $(k-1)$ and $k$, then $\pi_k (B)= \pi_{k-1} (F)$ \rm \footnote{This result is the particular case of the more general  result, namely that the homotopies sequence 
 $$ \dots \pi _{i+1}(B)\to \pi _i (F)\to \pi _i (E)\to \pi _i (B)\to \pi _{i-1}(F)\dots$$ is exact (see the proof in the monograph \cite{Postn4}, Lecture 26). Such  exact sequence implies the isomorphism $\pi_i (B)= \pi_{i-1} (F)$ if $\pi_{i}(E)=\pi_{i-1}(E)=0$: more precisely, any exact sequence of the shape
  $$0\to G_1\to G_2\to 0,$$
 with $G_1$, $G_2$ being some groups, implies the isomorphism $G_1 \simeq G_2$ of these groups $G_1$ and $G_2$ (see, e.g. \cite{Logunov}); and one should substitute the homotopical groups $\pi _i (E)=\pi _{i-1}(E)=0$ and also $\pi _i (B)$ and $\pi _{i-1}(F)$ in the latter exact sequence.}. In our concrete case we should apply the above proposition for   $\pi_1(B)\equiv \pi_1 (E/ F_{\rm con})$, i.e. for $k=1$, to the  principal fiber bundle $E_1$.

Applying then the same proposition to the  principal fiber bundle $E_2$, we find
\be 
\label{pr.f} \pi_1 (E/F)= \pi_0 (F/F_{\rm con})={\bf Z}.
\ee  
Herewith $E/F$ is the space of 
"large " orbits of the group $ F/F_{\rm con}$ on  $E/F_{\rm con}$, as we have just shown repeating the arguments \cite{Al.S.}.

On the other hand, the space of $F_{\rm con}$-orbits, $ E/F_{\rm con}$, coincides (perhaps, to within a topological equivalence) with $E$, while $E/F$  can be treated as $B$, the base  of the fiber bundle $\tilde E$.

Eq. (\ref{pr.f}) determines the isomorphism between the number of  connection components in the discrete group $ F/F_{\rm con}$  and the fundamental group $\pi_1$ of the base space $E/F$ (see e.g. Proposition 3 to Lecture 4 in \cite{Postn4}). Such construction of the base space in the fiber bundle $E_2$ can result connected as well as  multi-connected spaces as possible bases in this "fashion" fiber bundle. \par

\medskip   Returning again to the proof of the isomorphism (\ref{is1}), now we  substitute $H$ instead of $F$ in  (\ref{E2}). This gives, in particular, 
the group $ H_{\rm con}$ instead of $ F_{\rm con}$ (while $G/H\simeq E\otimes{\bf Z}$ in the role of $B=E/F$)  and thus the quested isomorphism (\ref{is1}). \par 

\medskip   The essential point at this proof is the assumption that group $F$ is discrete. All the said operates very good in the liquid helium-4 case, the basic topic of the present discussion. Here the role of the residual symmetry group $H$ is played by the discrete manifold $\tilde U \cong U_0\otimes {\bf Z}$, given via (\ref{disj}), the disjoint union of the topological
sectors  $U_1^n$  ($n\in {\bf Z}$).  On the other hand, we show that the degeneration space in the  liquid helium-4 (at rest) case, $R_r$,  coincides with $\tilde U$. A decisive argument  here is the fact that any homotopical class $\tau$ (possessing the topological number $n = 1$ and
treated  as the generator of the cyclical group  $\pi_1(U(1), p_0)$) can be treated naturally
as the operator raising the topology: from $n$ to $n + 1$. Inversely,  $\tau^{-1}$ can be treated as operator
lowering topology: from $n$ to $n-1$.  As a result, the
fundamental group of one-dimensional loops, $\pi_1(U(1), p_0)$, acts as an automorphism within the disjoint
union  $\tilde U$. In fact, this is equivalent to say that $\tilde U$ is invariant relatively the (gauge) group
$U(1)$. And this falls under the definition of degeneration space!

\bigskip Namely the nontrivial fundamental group $\pi_1 (G/H)$ is responsible always for line (in particular, thread) topological defects. 
\end{appendix}

\end{document}